\documentclass[acmlarge]{acmart}
\AtBeginDocument{%
  }

\setcopyright{cc}
\setcctype{by}
\acmJournal{IMWUT}
\acmYear{2026} \acmVolume{10} \acmNumber{4} \acmArticle{245}
\acmMonth{12} \acmDOI{10.1145/3857982}

\usepackage{algorithm}
\usepackage{algorithmic}
\usepackage{amsmath}
\usepackage{multirow}
\usepackage{appendix} 
\definecolor{myblue}{RGB}{134,158,255}

\begin{document}

\title{MotionDecipher: LLM-Assisted Clinical Interpretation of Human~Motion Signals for Efficient Gait Analysis}

\author{Mingrui Ma}
\email{mma36@illinois.edu}
\orcid{0000-0002-2073-6424}
\affiliation{%
  \institution{University of Illinois Urbana-Champaign}
  \city{Champaign}
  \state{Illinois}
  \country{USA}
}

\author{Tian Tan}
\email{alanttan@stanford.edu}
\affiliation{%
  \institution{Stanford University}
  \city{Stanford}
  \state{California}
  \country{USA}}

\author{Jinsi Guo}
\email{jinsig2@illinois.edu}
\affiliation{%
  \institution{University of Illinois Urbana-Champaign}
  \city{Champaign}
  \state{Illinois}
  \country{USA}
}

\author{Jessica Rose}
\email{jessica.rose@stanford.edu}
\affiliation{%
  \institution{Stanford University}
  \city{Stanford}
  \state{California}
  \country{USA}}
  
\author{Yiwen Dong}
\email{yiwen@illinois.edu}
\affiliation{%
  \institution{University of Illinois Urbana-Champaign}
  \city{Champaign}
  \state{Illinois}
  \country{USA}}

\renewcommand{\shortauthors}{Ma et al.}


\begin{abstract}
  Clinical gait and motion analysis is widely used to guide targeted interventions for patients with neurological disorders, the outcome of which is typically a report with descriptions of joint movements and forces. Current practices rely on experts’ manual effort, which is labor-intensive and time-consuming. Recent advancements in Large Language Models (LLMs) provide opportunities to address this problem, but translating human motion into clinical gait descriptions remains challenging because the quality of the context directly affects LLMs' reasoning capability over continuous motion signals. To tackle this, we introduce \emph{MotionDecipher}, a novel framework bridging the semantic gap between motion data and clinical interpretations. Specifically, we made 3 key advancements: 1) a \textit{Human Motion Tokenizer} that converts motion signals into domain-specific tokens for time- and frequency-informed gait interpretation; 2) a \textit{Meta Feature Selector} that transforms tokens into \textit{Structured Representations}; and 3) a \textit{Hybrid Context Constructor} to determine the optimal demonstration-based prompt for In-Context Learning. We evaluated \emph{MotionDecipher} on data collected at an accredited gait lab from 122 patients and 46 controls. Results show that \emph{MotionDecipher} achieves 0.91 BERTScore (13\% over GPT-5), while expert evaluation ranks \emph{MotionDecipher} highest in Correctness, Completeness, and Conciseness compared to baselines, highlighting its promise for efficient gait analysis.
\end{abstract}

\ccsdesc[500]{Applied computing~Health informatics}
\ccsdesc[300]{Computing methodologies~Natural language processing}
\ccsdesc[300]{Human-centered computing~Ubiquitous and mobile computing}

\keywords{Gait Analysis, Large Language Model, In-Context Learning, Motion Capture, Report Generation}


\maketitle

\section{Introduction}

Gait analysis \cite{spallone2025step} is an important clinical assessment that observes and analyzes human movement patterns during walking, which plays a crucial role in guiding targeted interventions and supporting clinical decision-making in neurological or musculoskeletal disorders (e.g., cerebral palsy, spina bifida, stroke, Parkinson's disease, and spinal cord injury) \cite{brainsci9020034, na2024kinematic}. Modern gait analysis systems employ diverse sensing modalities such as 3D digital motion capture \cite{Pfister01072014}, force plates \cite{6062376}, video cameras \cite{kidzinski2020deep}, inertial measurement units (IMUs) \cite{RAPP2021110229}, vibration sensors \cite{10.1145/3722570.3726886}, and electromyography (EMG) \cite{Papagiannis02012019} to capture high-dimensional, continuous time-series waveforms of key spatiotemporal, kinematic (e.g., joint angle trajectories), and kinetic (e.g., power, moments) parameters over gait cycles. Such quantitative analysis provides evidence-based insights into patients' mobility and functional status, which can subsequently facilitate individualized treatment planning and objectively evaluate intervention outcomes, ultimately improving patients' functional mobility, autonomy, and quality of life \cite{brainsci9020034}. Given that an individual's walking pattern is closely linked to daily activity participation, fall risk, social engagement, and overall health, gait and motion analysis has become increasingly important in a wide range of medical domains such as rehabilitation, neurology, orthopedics, and gerontology \cite{10.1159/000538270, 10.3389/fmedt.2022.901331}. Gait analysis reports are clinical summaries of quantitative gait parameters (e.g., joint angles) and qualitative descriptions (e.g., abnormal patterns) across different body parts (e.g., trunk, pelvis, hip, knee, and ankle) and gait phases (e.g., stance, swing). Gait reports frequently consist of three parts: the first is a quantitative gait description; the second part links neuromuscular and musculoskeletal impairments to the gait abnormalities; and the third part provides evidence-based treatment recommendations. This paper describes a new LLM-based system that provides the first part of the gait report (i.e., quantitative and clinical gait descriptions). Despite their importance, the generation of these reports still heavily relies on experts' manual effort, which is a labor-intensive process requiring substantial time and effort from clinical experts to mentally parse these complex continuous waveforms, extract critical features, and translate them into clinical text \cite{Gaitanalysisreview, 10.3389/fnhum.2022.907565, Quanthumanmotion}. 

\begin{figure*}[htbp]
\centering
\includegraphics[width=1.0\textwidth]{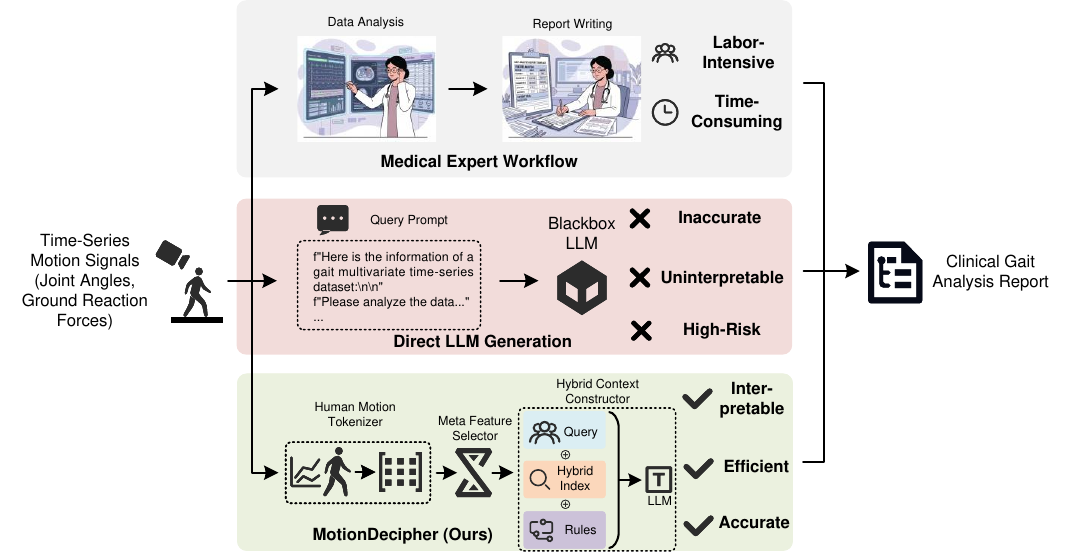}
\caption{Comparison of workflows for time-series motion analysis and clinical report generation. The medical expert workflow requires substantial manual effort across data analysis, result interpretation, and report writing, making it labor-intensive and time-consuming. Directly prompting Large Language Models (LLMs) to generate reports from raw motion signals is unreliable, often producing outputs that are difficult to interpret and prone to inconsistencies or errors, which is undesirable in high-stakes clinical settings (see Section~\ref{3.2}). In contrast, \emph{MotionDecipher} introduces an end-to-end framework consisting of a \textit{Human Motion Tokenizer}, a \textit{Meta Feature Selector}, and a \textit{Hybrid Context Constructor}, which progressively extract and interpret clinically relevant motion features into \textit{Structured Representations} suitable for LLM generation. This design enables more interpretable, efficient, and accurate clinical gait analysis and report generation.}
\label{Fig1}
\end{figure*}

Although prior work has started to develop multiple Artificial Intelligence (AI) models to assist with specific gait analysis tasks (e.g., disease classification, abnormality detection, and outcome prediction) \cite{GaitAbnormalityDetection, Outcomeprediction}, and recent systems have coupled LLMs with biomechanical data for clinical question answering \cite{biomechgpt} and interactive analysis \cite{biomechagent}, these efforts focus on producing a limited scope of outputs (discrete labels, scalar estimates, or interactive visualizations), rather than clinically meaningful gait descriptions. Recent advances in LLMs \cite{10.1145/3744746} offer a promising opportunity to bridge motion data and natural language, due to their strong contextual understanding \cite{DBLP:conf/naacl/ZhaoM0A24}, self-reasoning \cite{NEURIPS2024_e41efb03}, and sequential data modeling capabilities \cite{DBLP:conf/ijcai/0001C0S24, DBLP:conf/ijcai/JiangP0GSNS24}. Accordingly, extensive studies have applied LLMs to various sensor data scenarios, including multimodal sensing, augmented reality, and digital healthcare \cite{DrHouse, ADMarker, TaskSense, MMBind, SocialMind, Otter, SHADE-AD, ARAS, EarlyPrediction}. However, motion signal analysis poses challenges for LLMs that are distinct from discriminative classification \cite{Yang_2025_CVPR}, as current models are not trained to process complex and noisy motion signals and lack the in-depth reasoning capability needed to generate standardized, domain-specific gait interpretations.

To bridge the gap between raw motion signals and clinical language interpretation, we introduce \emph{MotionDecipher}, a novel 3-stage framework that bridges the semantic gap between complex motion signals and clinically interpretable narratives, as shown in Fig.~\ref{Fig1}. In this work, we specifically define the input "motion signals" as the continuous time-series waveforms of ground reaction force (GRF) and joint angle changes over gait cycles. \emph{MotionDecipher} is specifically designed to overcome three fundamental challenges in deciphering continuous sensor data into clinical descriptions. The first challenge is that time-series signals lack human motion semantics, making it difficult to derive clinically meaningful tokens. To tackle the challenge, we develop a \textit{Human Motion Tokenizer} to transform motion signals into four categories of domain-specific tokens for gait interpretation (i.e., Tendency, Comparative, Statistical, Frequency) without an extra training process. However, simply tokenizing motion data is insufficient for gait interpretation because the generated tokens inevitably mix clinically important features with irrelevant details. LLMs struggle to identify important gait features due to the absence of domain-specific pretraining. To address this challenge, the \textit{Meta Feature Selector} identifies and extracts in-depth gait features (e.g., gait abnormalities) from human motion tokens, and reorganizes them into \textit{Structured Representations} according to specific body parts and gait phases. Even with these \textit {Structured Representations}, translating them into precise clinical descriptions remains difficult, as general-purpose LLMs often prioritize vague descriptions over the precise terminology required for motion observations. To handle this problem, we develop \textit{Hybrid Context Constructor} to enable hybrid similarity-based retrieval and automatic In-Context Learning (ICL) prompt construction. Specifically, given a new patient's \textit{Structured Representation} instance, \textit{Hybrid Context Constructor} retrieves the most relevant expert-written descriptions from \textit{Gait Representation-Text Demonstration Corpus (GRTDC)} (Section~\ref{2.3.1}) by jointly calculating textual and numerical similarities through the proposed \textit{Hybrid Index} algorithm (Section~\ref{2.3.2}). Finally, the selected demonstration is embedded in an ICL prompt, enabling \emph{MotionDecipher} to generate clinically aligned gait analysis reports.

To rigorously evaluate the performance of \emph{MotionDecipher}, we deployed \emph{MotionDecipher} on data collected at an accredited gait lab from Jan 2024 to July 2025 on 122 patients and compared it to 46 normal controls using 3D digital motion capture and force plate systems \cite{Pfister01072014, 6062376}. Results showed that \emph{MotionDecipher} outperforms 4 different baseline LLM outputs (GPT-5, Kimi-K2, Google Gemini 3, Qwen3-Max) on 6 different evaluation metrics (\textit{BLEU}, \textit{ROUGE-L}, \textit{METEOR}, \textit{BERTScore}, \textit{MEDCON-Gait}, \textit{GaitTripleF1}), with an average improvement between 9\% and 34\%. In addition, we conducted a clinical evaluation with two medical experts to assess \emph{MotionDecipher} in terms of Correctness, Completeness, and Conciseness (see Section~\ref{3.6}). The results demonstrate that \emph{MotionDecipher} achieves improvements ranging from 71\% to 221\% over Kimi-K2, and can surface clinically relevant findings that are not documented in routine reference reports.

We summarize our contributions as follows:

\begin{itemize}
    \item To the best of our knowledge, \emph{MotionDecipher} is the \textit{first framework} bridging the gap between noisy motion signals and interpretable gait report generation in real-world clinical settings.
    \item To enable semantic abstraction from raw time-series motion signals, we develop a \textit{Human Motion Tokenizer} that automatically transforms complex motion signals into discrete, semantically enriched tokens without an extra training process.
    \item To enable accurate demonstration selection for ICL in clinical settings where textual descriptions are tightly coupled with numerical measurements, we develop the \textit{Hybrid Index} algorithm, which jointly models numerical and textual similarities for retrieving optimal examples in \textit{GRTDC} that allow \emph{MotionDecipher} to generate expert-level gait descriptions. 
    \item We conduct extensive real-world experiments with 122 patients, verifying the superior performance of \emph{MotionDecipher} against various baseline LLM outputs (28\% and 19\% improvement compared to GPT-5 and Google Gemini 3). Moreover, \emph{MotionDecipher} significantly improves efficiency, reducing data analysis and report drafting time from approximately 4-6 hours/patient to an average of 2 minutes/patient.
\end{itemize}

The remainder of the article is organized as follows: Section~\ref{2} elaborates on the algorithm design of each component in \emph{MotionDecipher}. Section~\ref{3} describes the experimental setup, reports the evaluation results, and presents additional analyses of \emph{MotionDecipher}, including ablation, robustness, sensitivity, and medical expert evaluations. Section~\ref{4} discusses the strengths and limitations of \emph{MotionDecipher}, as well as outlines future research directions. Section~\ref{5} categorizes and reviews the relevant research. Section~\ref{6} summarizes the article. 

\section{\emph{MotionDecipher} System Design} \label{2}

\begin{figure*}[htbp]
\centering
\includegraphics[width=1.0\textwidth]{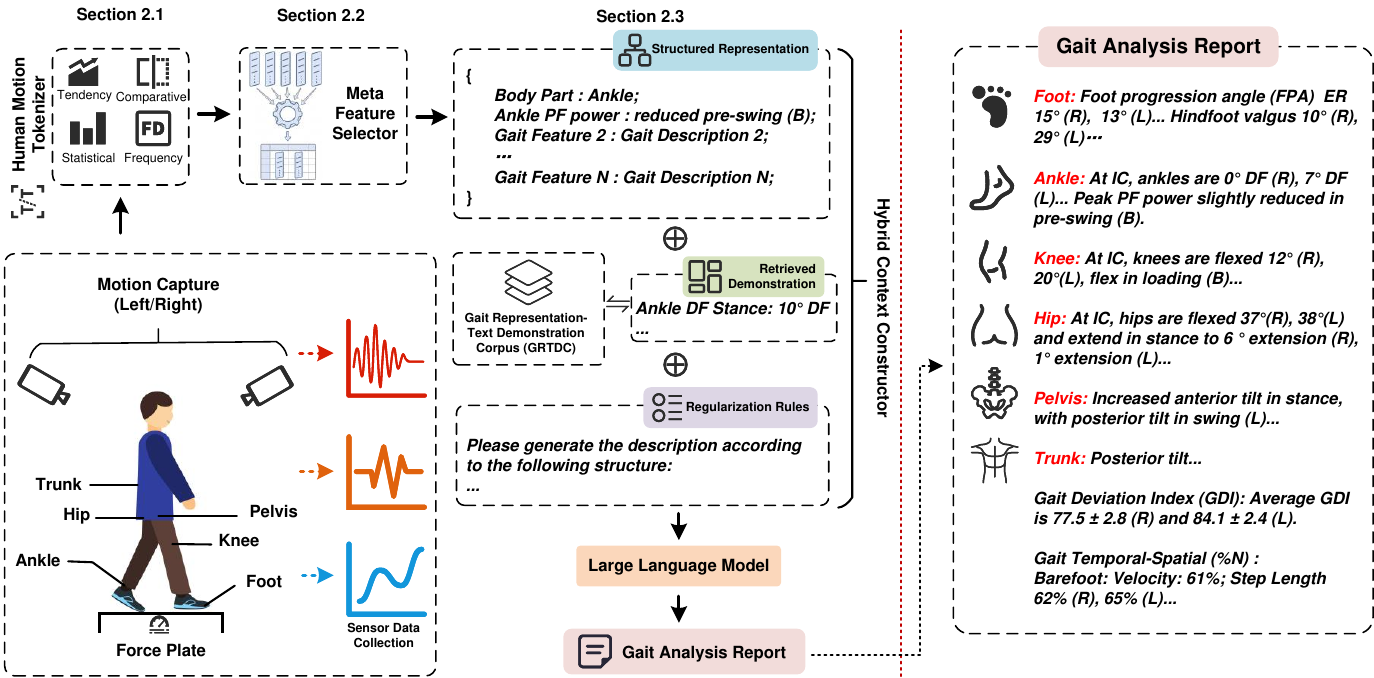}
\caption{System workflow of \emph{MotionDecipher}. \emph{MotionDecipher} is an efficient framework that extracts and analyzes 4 different types of gait features from time-series motion signals, retrieves the optimal demonstration from the \textit{GRTDC}, constructs an ICL prompt based on the retrieved, domain-specific demonstrations, and produces a hierarchical gait analysis report covering multiple body parts (Foot, Ankle, Knee, Hip, Pelvis, and Trunk).}
\label{Fig2}
\end{figure*}

In this section, we present the end-to-end design of \emph{MotionDecipher}, as shown in Fig.~\ref{Fig2}, and introduce each component in detail. We emphasize that the input of \emph{MotionDecipher} is preprocessed motion signals (a visualization of this data format is presented in the lower part of Fig.~\ref{Fig8}), and the corresponding output is a hierarchical gait analysis report covering multiple body parts, as illustrated on the right side of Fig.~\ref{Fig2}.

\subsection{\textit{Human Motion Tokenizer}}

\begin{figure}[htbp]
\centering
\includegraphics[width=1.0\textwidth]{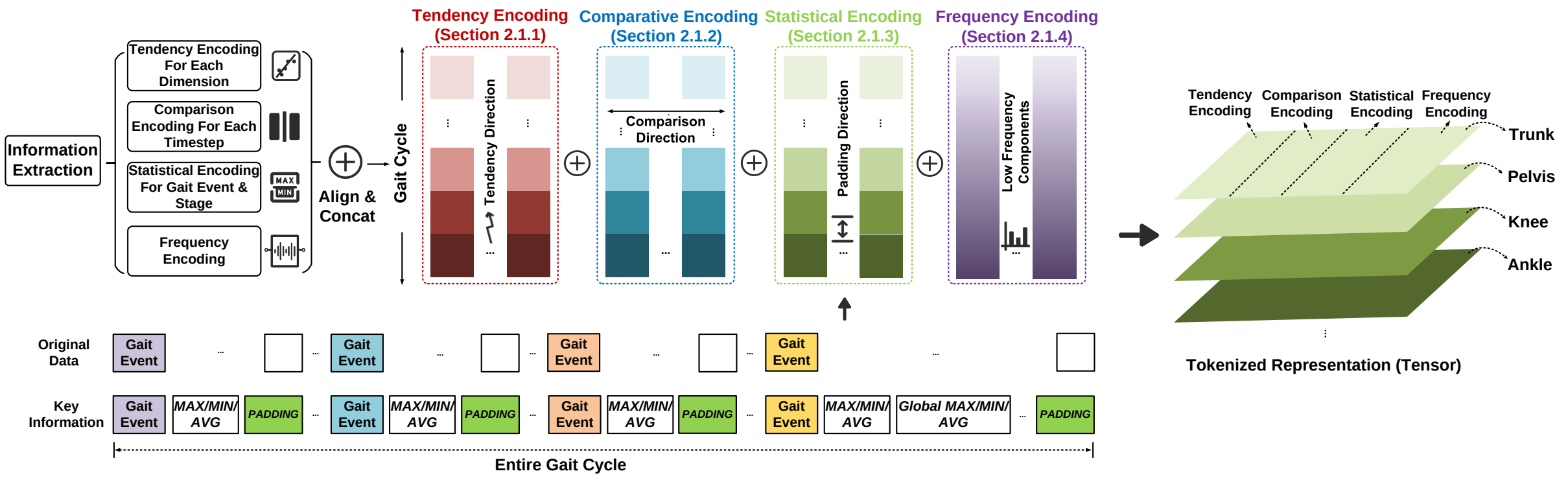}
\caption{The structure of \textit{Human Motion Tokenizer}. For each body part (e.g., Trunk, Pelvis, Hip, Knee, Ankle), \textit{Human Motion Tokenizer} generates 4 types of encodings (Tendency, Comparative, Statistical, Frequency) through different calculation methods to capture both time- and frequency-domain gait characteristics. These encodings are then aligned along the timestep (gait cycle) dimension and concatenated into a 2D token matrix. Finally, matrices from different body parts are stacked to form a 3D tokenized representation in tensor format.}
\label{Fig3}
\end{figure}

To transform continuous human motion time-series data into discrete and interpretable tokens, we introduce \textit{Human Motion Tokenizer}, which provides a multi-feature tokenization of motion data that captures motion patterns for interpretation and reasoning. Conceptually, general human motion data can be sufficiently characterized through four complementary encoding types, as defined below:

\begin{itemize}
    \item \textit{Tendency Encoding}: Captures the trend of the motion signals over time;
    \item \textit{Comparative Encoding}: Calculates the difference of signal magnitudes between the observed gait vs. normative data from the typically developing control population to represent gait deviations;
    \item \textit{Statistical Encoding}: Characterizes critical motion patterns as in basic statistics (e.g., max, min, mean, variance);
    \item \textit{Frequency Encoding}: Represents the frequency spectrum of the motion signals over an average gait cycle.
\end{itemize}

These complementary features provide fundamental discriminative insights for characterizing human motion patterns, as they contain all the necessary information (time, frequency, trend, and statistics) for analyzing human motion. Accordingly, \textit{Human Motion Tokenizer} integrates these encodings into a compact yet informative representation, as shown in Fig.~\ref{Fig3}.

\subsubsection{\textit{Tendency Encoding} For Each Dimension}

\textit{Tendency encoding} is designed to capture the underlying data trend, which is crucial in human motion analysis because it reflects how motion changes over time. For instance, the trend of the trunk tilt effectively informs how much further a person will lean forward or backward in the next timestep.

Capturing reliable tendency features requires extracting motion dynamics rather than signal noise. We formulate the local motion tendency as the smoothed first-order derivative, computed via a configured Savitzky-Golay (SG) operator \cite{Savitzky-Golay}. Unlike linear approaches, this operator preserves the critical waveform trend required for clinical interpretation while reducing noise amplification during differentiation.
Formally, for a time-series channel $\mathbf{y}=[y_1,\dots,y_N]$, we compute the tendency feature \(\hat{y}'(t_i)\) at each timestep $t_i$ as
\begin{equation}
\hat{y}'(t_i) = \frac{1}{\Delta t} \sum_{j=-M}^{M} c^{(1)}_j\, y_{i+j},
\end{equation}
where the convolution with coefficients $c^{(1)}_j$ effectively minimizes noise amplification during differentiation. In our settings, the SG operator is configured with a window size of 5 and a polynomial order of 2 to compute the first-order derivative. These parameters are selected to strike an optimal balance between high-frequency noise attenuation and the preservation of critical motion dynamics. Given that a complete gait cycle is normalized to 100 time steps, a compact 5-frame window with a second-order polynomial effectively prevents over-smoothing. This ensures that the resulting estimated derivative $\hat{y}'(t_i)$ serves as an accurate and continuous proxy of local motion tendency. Therefore, the output for a single channel can be formulated as \(\mathbf{\hat{y}'} = [\hat{y_1}', \hat{y_2}', \dots, \hat{y_N}']\), where \(\mathbf{\hat{y}'}\) directly quantifies the instantaneous direction and intensity of movement.

We generate the \textit{Tendency Encoding} for each channel of the multivariate motion time-series signals along the direction of timesteps (Gait Cycle), and then concatenate the results in channel order to obtain the \textit{Tendency Encoding} component shown in Fig.~\ref{Fig3}.

\subsubsection{\textit{Comparative Encoding} For Each Timestep}

\textit{Comparative Encoding} discretizes every continuous signal value (e.g., knee flexion angle across the entire gait cycle) at each timestep into an interval token by comparing it against normative reference bounds. These bounds are derived from a cohort of 46 typically developing subjects collected at an accredited gait clinic and are consistent with the normal gait kinematic data ranges previously reported~\cite{Normal2006Kinematics}. These data represent the empirical normal variability across the gait cycle~\cite{Referencebound1, Referencebound2}. We emphasize that \emph{MotionDecipher} currently employs a single universal normative boundary: the reference bounds are computed as the \(\text{mean} \pm \text{1SD}\) of each signal channel at every gait-cycle timestep across the entire control cohort, and are applied uniformly to all patients without age- or weight-based stratification. Taking Hip Abduction/Adduction (i.e., lateral movement of the thigh away from/toward the body) as a representative example, the measured hip frontal-plane angle is compared with the normative range derived from the typically developing population to represent gait deviations. To formalize the comparison relationship mathematically, we perform an interval-based projection that maps each continuous feature value \(x\) into its nearest quantized interval determined by a scalable step size \(\theta = 0.5\):

\begin{equation}
f(x) =
\begin{cases}
\theta \lfloor x / \theta \rfloor, & x \ge 0, \\[6pt]
\theta \lceil x / \theta \rceil, & x < 0.
\end{cases}
\label{eq:interval_projection}
\end{equation}

This interval-based quantization suppresses minor fluctuations while preserving clinically meaningful numerical relationships, effectively aligning them with natural language descriptions commonly used in gait and motion analysis (e.g., below, within, or above normal range). We generate the \textit{Comparative Encoding} for each timestep of the motion time-series data, and then concatenate the results in timestep order, as shown in Fig.~\ref{Fig3}.

\subsubsection{\textit{Statistical Encoding} for Gait Event \& Stage}

Human motion analysis often relies on statistical summaries computed over meaningful temporal segments or events, which provide stable and interpretable characterizations of motion patterns. In gait analysis, the magnitudes of motion signals associated with different gait phases (e.g., stance, pre-swing, swing) and specific gait events (e.g., Initial Contact) are crucial because they characterize human motion during critical gait events and gait phases. To accurately acquire such information, we automatically detect gait events of heel strike and toe off for each foot (Right Heel Strike (\textit{RHS}), Right Toe Off (\textit{RTO}), Left Heel Strike (\textit{LHS}), Left Toe Off (\textit{LTO})) according to the turning point detection algorithm, which is widely used to segment walking trials into gait cycles and stance/swing stages \cite{Twosimplemethods}. Using these events, we divide the entire gait cycle into distinct gait phases (taking the right foot as an example): \([RHS, LTO)\), \([LTO, LHS)\), \([LHS, RTO)\), \([RTO, RHS)\), which represent Early stance, Loading, pre-swing, and swing, respectively.

For each gait phase and the gait cycle, we extract the signal magnitude at specific gait events and calculate the maximum (\(MAX\)), minimum (\(MIN\)), and average (\(AVG\)) values. Then, we pad the extracted features along the timestep (gait cycle) dimension using the padding scheme depicted in the lower part of Fig.~\ref{Fig3}, to ensure the uniformity of \textit{Human Motion Tokenizer}, which corresponds to the \textit{Statistical Encoding} of the Tokenized Representation in Fig.~\ref{Fig3}.

\subsubsection{\textit{Frequency Encoding}}

Frequency-domain information captures rhythm-related patterns. We perform a wavelet decomposition on each signal channel in \textit{Human Motion Tokenizer}. Specifically, we apply the Discrete Wavelet Transform (DWT) to handle the non-stationary nature of human motion. In our implementation, we parameterize the DWT using a Daubechies-4 (db4) mother wavelet with a decomposition level of 3, and employ symmetric padding at the signal boundaries. The db4 wavelet is selected because its smooth, asymmetric waveform closely aligns with the biomechanical features of human gait kinematics (e.g., joint flexion curves). A decomposition level of 3 is strategically chosen to isolate the fundamental low-frequency kinematics of human walking while effectively discarding high-frequency sensor noise. Formally, for a temporal sequence \(x_d(t)\), we decompose the signal and selectively reconstruct the denoised representation \(\tilde{x}_d(t)\) by discarding detail coefficients \(D_j\) while keeping the approximation coefficients \(A_L\):

\begin{equation}
\tilde{x}_d(t) = \text{IDWT}(A_L) = \text{IDWT}\left( \mathcal{W}(x_d(t)){\text{approx}} \right),
\end{equation}

where \(\text{IDWT}(\cdot)\) denotes the inverse discrete wavelet transform operation. This process effectively suppresses high-frequency artifacts, yielding a mathematically robust low-frequency representation $X_{\text{Frequency}}$ that aligns with the periodicity of clinical gait cycles.

\subsubsection{Generating Tokenized Representation}

To form a unified and informative representation for human motion tokenization, we concatenate the low-frequency representation $X_{\text{Frequency}}$ with the \textit{Tendency Encoding}, \textit{Comparative Encoding}, and \textit{Statistical Encoding} along the timestep (Gait Cycle) direction:

\begin{equation}
X_{\text{token}} = 
\big[\, X_{\text{Tendency}} \; \| \; 
X_{\text{Comparative}} \; \| \; 
X_{\text{Statistical}} \; \| \; 
X_{\text{Frequency}} \,\big],
\end{equation}

where ``\(\|\)'' denotes feature-wise concatenation. Human motion is naturally organized by anatomical regions, with each body part exhibiting distinct motion patterns while remaining temporally coordinated with others. Accordingly, we stack the tokenized features of different body parts into a 3-dimensional tensor \(\mathcal{X} \in \mathbb{R}^{T \times C \times B}\), as illustrated in Fig.~\ref{Fig3}. This tensor explicitly encodes motion along three orthogonal dimensions:

\begin{itemize}
    \item Temporal Dimension (\(T\)): Represents the timestep of the gait cycle, ensuring temporal alignment across all features;
    \item Feature Dimension (\(C\)): Corresponds to the concatenated domain-specific tokens (Tendency, Comparative, Statistical, Frequency);
    \item Spatial Dimension (\(B\)): Represents different anatomical regions (e.g., Trunk, Pelvis, Knee).
\end{itemize}

This design preserves the independence of body-part representations while enabling coordination-aware reasoning through shared spatial and temporal alignment, and can be naturally extended to additional body parts without altering the representation structure.

\subsection{\textit{Meta Feature Selector}}

\begin{figure*}[htbp]
\centering
\includegraphics[width=1.0\textwidth]{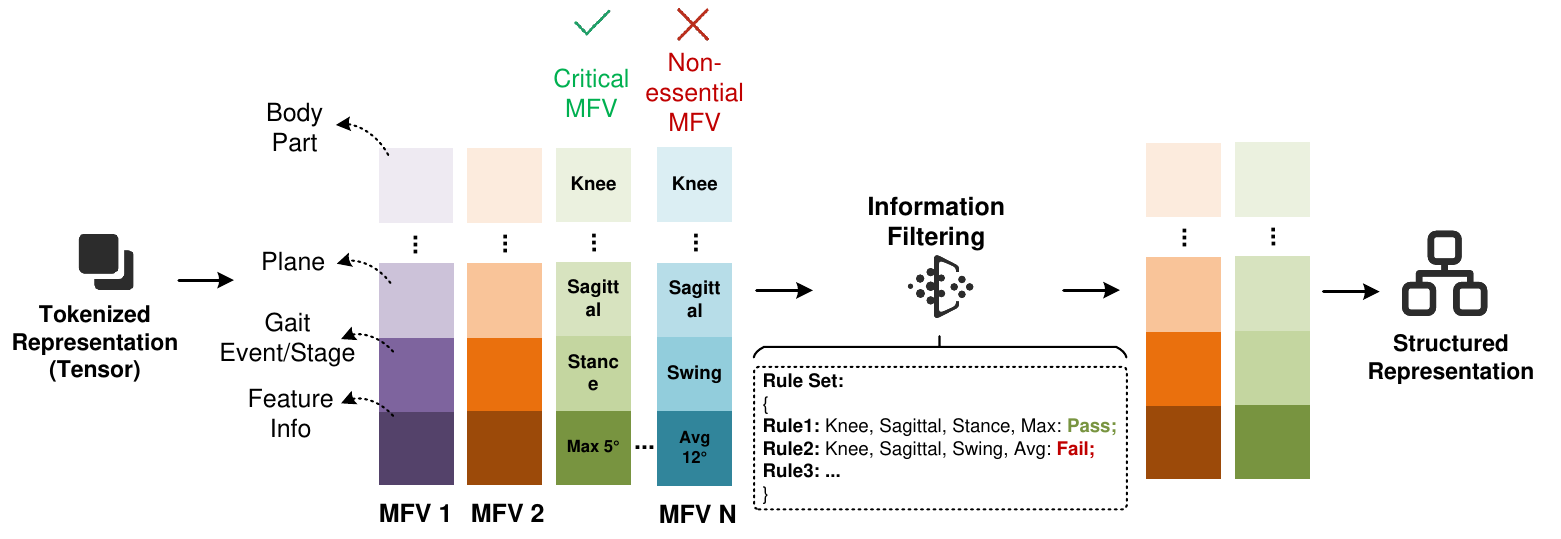}
\caption{The workflow of \textit{Meta Feature Selector}. \textit{Meta Feature Selector} transforms the tokenized representation into Meta Feature Vectors (MFVs), and applies a predefined rule set to select a critical MFV subset while filtering out non-essential ones. The extracted MFVs are then reorganized into \textit{Structured Representations} for subsequent processing.}
\label{Fig4}
\end{figure*}

While the \textit{Human Motion Tokenizer} transforms raw motion signals into a compact tokenized representation, not all encoded features contribute equally to high-quality gait report generation. To reduce information redundancy, we introduce the \textit{Meta Feature Selector}, a crucial component that identifies critical gait features relevant to clinical decision-making (e.g., maximum knee joint angle at the stance phase) while filtering out details that are less significant for gait health inference (e.g., average knee joint angle during the swing phase) from domain-specific token representations and reorganizes them into \textit{Structured Representations}, as shown in Fig.~\ref{Fig4}.

\subsubsection{Meta Feature Vector (MFV)}

To facilitate the selection of important information, we first split and reorganize the tokenized representation to generate multiple Meta Feature Vectors (MFVs). An MFV serves as an informative feature vector that groups all the necessary details (e.g., Trend, Gait Event/Stage, Body part, etc.) needed to generate a complete description in clinical gait reports. By organizing the tokenized representation into MFVs, we enable systematic rule-based filtering, composition, and restructuring of gait features, which serves as a crucial intermediate representation for constructing structured outputs and downstream report generation within the \emph{MotionDecipher} framework. 

\subsubsection{Information Filtering}

In the filtering stage, our objective is to highlight clinically important gait features while filtering out less informative content. Unlike traditional noise filtering, this discarded information (e.g., “minimal knee extension in swing”) does not correspond to measurement noise, but rather to details that clinical experts deliberately omit because they are unlikely to influence diagnosis or clinical decision-making, and would otherwise detract from a concise and focused report.

To automate this procedure, we define a set of filtering rules denoted as 
\begin{equation}
\mathcal{S} = \{ R_1, R_2, R_3, \ldots, R_N \},
\end{equation}

where each rule \( R_j,\quad 1 \le j \le N\) represents a programmatic formalization of a clinician’s selection criterion. To improve filtering efficiency, each rule \(R_j, \quad 1 \le j \le N\) is designed as a set of discrete keywords aligned with the dimensions of an MFV. For example, Rule 1 in Fig.~\ref{Fig4} consists of keywords such as Knee (body part), Sagittal (plane), stance (gait event/stage), and Max (feature information). During filtering, matching is performed via direct keyword comparison, eliminating the need for additional fuzzy matching algorithms. Every MFV, denoted as \( \mathrm{MFV}_i \), is sequentially passed through this set of rules. For each \( i \), if 
\begin{equation}
\mathrm{MFV}_i \not\models R_j, \quad 1 \le j \le N,
\end{equation}
That is, if the feature vector does not satisfy the condition specified by rule \( R_j \), it will be filtered out from further processing and excluded from subsequent analytical stages. The visualization of this process is shown in Fig.~\ref{Fig4}.

\subsubsection{Structured Representation} \label{2.2.3}

The final step of the \textit{Meta Feature Selector} is to convert the filtered MFVs into \textit{Structured Representations}. We define a \textit{Structured Representation} as a collection of selected MFVs categorized by anatomical region, where each body part is associated with a set of clinically meaningful gait descriptions (Fig.~\ref{Fig2} provides a specific example). Compared to unstructured textual descriptions, this representation explicitly groups human motion characteristics under semantic units (i.e., body parts), reducing cross-context interference and ambiguity. Such structuring provides a clear and consistent context for subsequent ICL, allowing LLMs to focus on motion semantics rather than searching for relevant details across long, unstructured text. For a single individual, we generate \textit{Structured Representations} corresponding to the key anatomical regions essential for report generation (i.e., Foot, Ankle, Knee, Hip, Pelvis, and Trunk), which serve as fundamental indexing units in the \textit{Hybrid Index} algorithm (Section~\ref{2.3.2}).

\subsection{Hybrid Context Constructor} \label{2.3}

\begin{figure*}[htbp]
\centering
\includegraphics[width=1.0\textwidth]{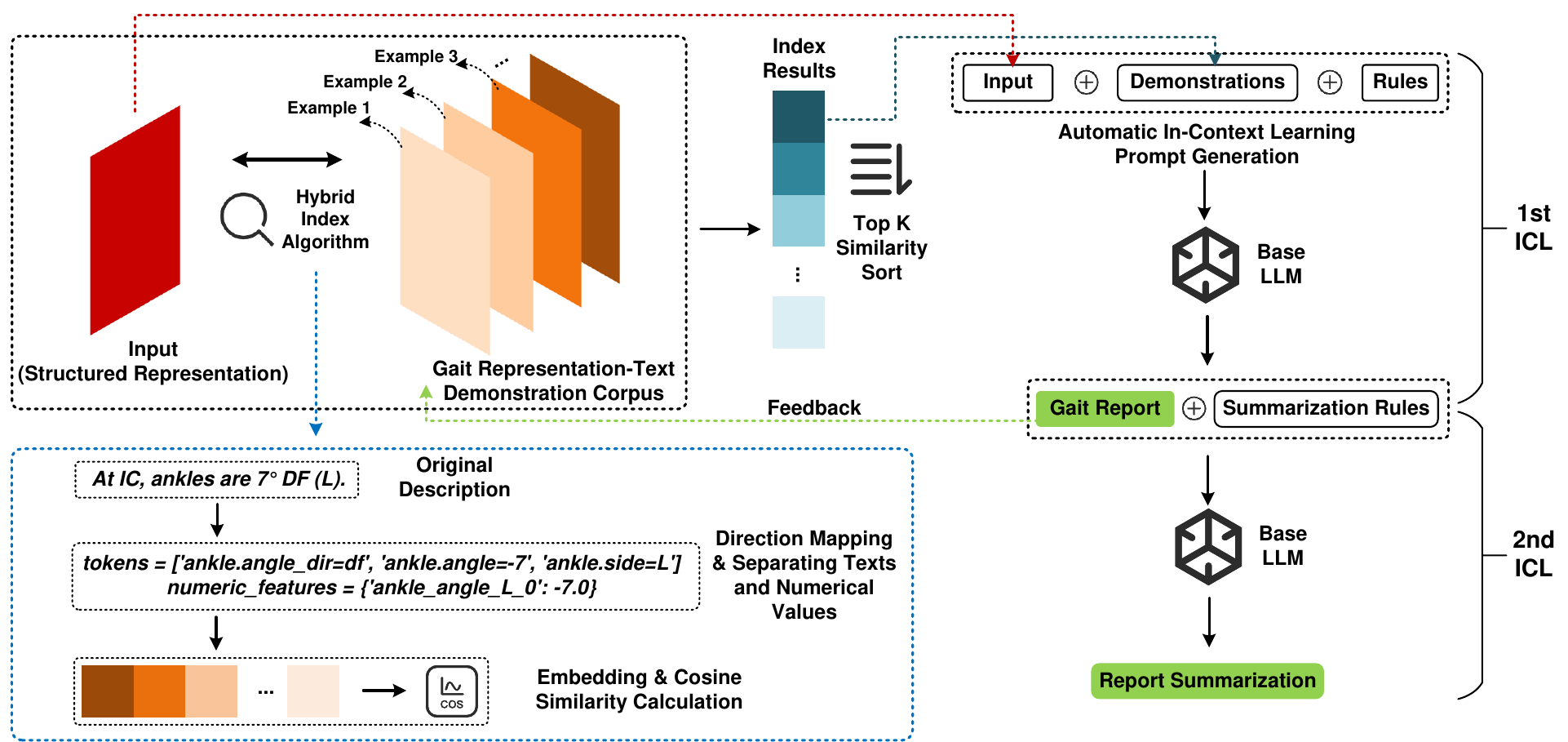}
\caption{The detailed design of \textit{Hybrid Context Constructor}. We compute the hybrid (\(textual + numerical\)) similarity between the input and examples in the Gait Representation-Text Demonstration Corpus (GRTDC) using the \textit{Hybrid Index} algorithm, select the Top-K examples (In practice, we set K=1) as demonstrations, and automatically construct the ICL prompt for 2-stage reasoning in the base LLM to generate gait analysis reports and summaries. Expert-refined benchmark reports are then fed back into the GRTDC to continuously enrich example diversity.}
\label{Fig5}
\end{figure*}

To provide the LLM with optimal demonstrations for clinical reasoning, we propose the \textit{Hybrid Context Constructor} component that retrieves the most relevant expert-written descriptions and automatically constructs precise ICL prompts, enabling LLMs to perform gait analysis report generation effectively, as shown in Fig.~\ref{Fig5}.

\subsubsection{Gait Representation-Text Demonstration Corpus} \label{2.3.1}

\begin{figure}[htbp]
\centering
\includegraphics[width=0.71\textwidth]{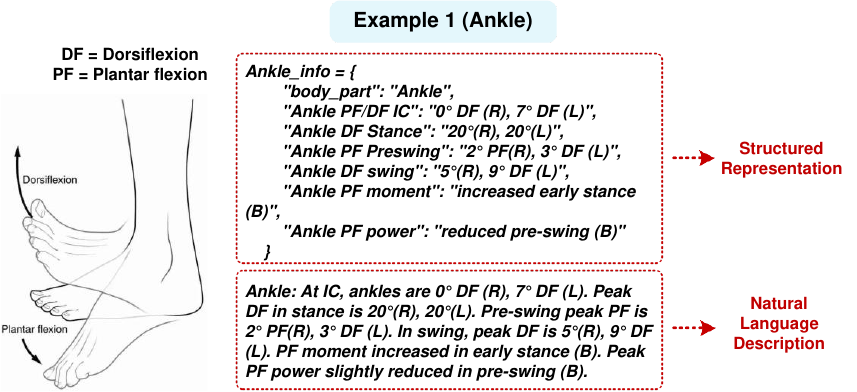}
\caption{A detailed example in GRTDC. The \textit{Structured Representation} is used for precise similarity computation in the \textit{Hybrid Index} algorithm, while the \textit{Natural Language Description} provides clinically grounded expressions with accurate terminology.}
\label{Fig6}
\end{figure}

Ensuring that LLM-generated narratives strictly adhere to professional clinical conventions is challenging due to the lack of clinical demonstrations. General-purpose prompts often lead to generic descriptions that fail to emphasize critical gait abnormalities and associations with neuromuscular impairments. To bridge this gap, we introduce the Gait Representation-Text Demonstration Corpus (GRTDC) for retrieval-based ICL. Rather than treating demonstrations as unstructured text, each demonstration in the set is explicitly constructed as a paired exemplar consisting of (i) a \textit{Structured Representation} of a specific body part (e.g., ankle), and (ii) a medical expert-written gait description, as shown in Fig.~\ref{Fig6}. This paired design decouples semantic similarity estimation from language generation: \textit{Structured Representations} are used for indexing and retrieval, while the corresponding natural language descriptions serve as ICL demonstrations for guiding LLM generation. 

This method is inspired by advances in NLP, where researchers construct a corpus (a set of words) to assist mainstream machine learning models in performing various downstream tasks, such as machine translation and natural language generation \cite{CorpusDownstream}.

\subsubsection{Hybrid Index Algorithm} \label{2.3.2}

To precisely retrieve expert-written demonstrations that are most similar to an input query, we propose the \textit{Hybrid Index} algorithm, which explicitly separates the input (\textit{Structured Representation}) into linguistic descriptors and quantitative values. This unique design addresses a critical limitation of standard text embedding models: the tendency to conflate clinically distinct measurements. For instance, standard models often prioritize the identical numeric token "10" while overlooking minor directional differences, leading to the misinterpretation of distinct clinical phrases like "10$^{\circ}$ DF" and "10$^{\circ}$ PF." Unlike traditional methods that compress raw text into a single vector, \textit{Hybrid Index} independently embeds textual semantics and normalizes numerical magnitudes. By concatenating these features into a joint \(Textual + Numerical\) representation, the algorithm preserves both the quantitative information (e.g., 10$^{\circ}$) and the precise physical constraints (e.g., DF/PF) of motion data. The pseudocode is detailed in Algorithm \ref{alg:hybrid_index}.

\begin{algorithm}[t]
\caption{Hybrid Index Algorithm}
\label{alg:hybrid_index}
\begin{algorithmic}[1]
\REQUIRE
Query gait sample $Q$; Demonstration Set $\mathcal{E}=\{E_i\}_{i=1}^{N}$; Scaling $\alpha$; Top-$K$ parameter $K$.
\ENSURE
Ranked list $\mathcal{O}=\{(E_{(k)}, s_{(k)})\}_{k=1}^{K}$.
    \FOR{each $E_i \in \mathcal{E} \cup \{Q\}$}
        \STATE $(\mathcal{T}_i, \mathcal{N}_i) \gets \textsc{ParseText}(E_i)$
        \STATE $\mathcal{N}_i \gets \textsc{MapDir}(\mathcal{N}_i, D)$
    \ENDFOR
    \STATE $\mathbf{T}, \mathbf{t}_q \gets \textsc{Vectorize}(\{\mathcal{T}_i\}, \mathcal{T}_Q)$
    \STATE $\mathbf{N}, \mathbf{n}_q \gets \textsc{AlignNum}(\{\mathcal{N}_i\}, \mathcal{N}_Q)$
    \STATE $(\tilde{\mathbf{N}}, \tilde{\mathbf{n}}_q) \gets \textsc{Normalize}(\mathbf{N}, \mathbf{n}_q)$
    \STATE $\tilde{\mathbf{N}} \gets \alpha \tilde{\mathbf{N}}, \quad \tilde{\mathbf{n}}_q \gets \alpha \tilde{\mathbf{n}}_q$
    \STATE $\mathbf{X}_i = [\mathbf{T}_i; \tilde{\mathbf{N}}_i], \quad \mathbf{x}_q = [\mathbf{t}_q; \tilde{\mathbf{n}}_q]$
    \STATE $s_i = \textsc{CosSim}(\mathbf{X}_i, \mathbf{x}_q), \quad i=1,\dots,N$
    \STATE $\mathcal{O} = \textsc{TopK}(\{(E_i, s_i)\}, K)$
    \STATE \RETURN $\mathcal{O}$
\end{algorithmic}
\end{algorithm}

Specifically, we first build a \textit{Direction Map}, which defines the symbol mapping corresponding to directional terms (e.g., PF/DF, Internal/External, etc.), converting angular directions in free text into signed numerical values. Subsequently, the \textit{PARSETEXT} function extracts and separates textual tokens from numeric features. The numerical features are then passed through the \textit{MAPDIR} function to perform directional mapping. Then, the textual tokens and the signed numeric features are independently vectorized, aligned, normalized, and concatenated into a sparse feature matrix. Eventually, we calculate and sort the \textit{Cosine Similarity} for the feature matrix pairs, and return the \textit{Top-K} most similar examples. 

\subsubsection{2-Step In-Context Learning}

To ensure the LLM adheres to clinical standards, we construct the final prompt by grounding the input in retrieved expert knowledge. Specifically, we concatenate the query's \textit{Structured Representation} with the retrieved demonstrations and explicit formatting constraints, then pass the resulting prompts into the base LLM (we select Kimi-K2 \cite{Kimi-k2} as the base LLM for \emph{MotionDecipher}) to obtain gait descriptions for corresponding body parts. The entire process can be formalized as:

\[
\mathrm{LLM}\biggl(
  Constraints,\;
  \underbrace{f(x_1,y_1),\ldots,f(x_k,y_k)}_{\text{demonstrations}},\;
  \underbrace{f(x_{k+1},}_{\text{input}}
  \underbrace{\ \_\_\ }_{\text{answer}})
\biggr)\;\rightarrow\;\hat{y}_{k+1}.
\]

Here, \(\hat{y}_{k+1}\) denotes the LLM output of the input query \(x_{k+1}\). Once the LLM generates descriptions for each body part, we aggregate them to form the final gait analysis report. To further highlight gait abnormalities for personalized treatment planning, we perform the second-step ICL based on the generated report, guiding the LLM to emphasize abnormal gait features, thereby generating a condensed summary. For example, given a patient exhibiting knee-related gait abnormalities, the second-stage ICL condenses the content into a focused summary that highlights key abnormal patterns (e.g., excessive knee valgus during stance). The complete process is shown in Fig.~\ref{Fig5}.

\subsection{\textit{GaitTripleF1}: A New Metric for Gait Description Evaluation} \label{2.4}

\begin{figure}[htbp]
\centering
\includegraphics[width=0.38\textwidth]{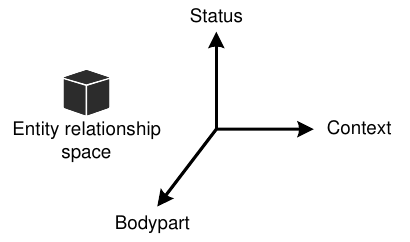}
\caption{The orthogonal relationship of \textit{GaitTripleF1}. Each entity in the triple occupies a distinct feature dimension in the relationship space.}
\label{Fig7}
\end{figure}

To rigorously evaluate the semantic accuracy of clinical gait reports, we introduce \textit{GaitTripleF1}, a new evaluation metric developed based on standardized clinical gait report templates \cite{ClinicalGENI} with clinical experts, in order to overcome the lack of specificity in existing metrics that are overly generic for medical report evaluations. Clinical descriptions consistently organize motion information along three orthogonal key dimensions: anatomical location (Bodypart), movement characterization (Status), and temporal context within the gait cycle (Context). Therefore, we interpret each report not as a collection of loose observations, but as a set of structured semantic triples \((Bodypart, Status, Context)\) (Fig.~\ref{Fig7}), which captures the essential associations between anatomical regions and motion states. On this basis, \textit{GaitTripleF1} measures triple similarity between model-generated and ground truth reports.

Formally, the three components are defined as:

\[
\begin{aligned}
\textit{Bodypart} &\in \{\text{Foot},~\text{Ankle},~\text{Knee},~\text{Hip},~\text{Pelvis},~\text{Trunk},~\text{etc.}\} \\
\textit{Status} &\in \{\text{flexion},~\text{extension},~\text{abducted},~\text{adducted},~\text{etc.}\} \\
\textit{Context} &\in \{\text{stance},~\text{swing},~\text{loading},~\text{pre-swing},~\text{etc.}\}
\end{aligned}
\]

This triple structure addresses a unique challenge in gait analysis: the necessity of temporal and anatomical synchronization. \textit{GaitTripleF1} factorizes gait semantics into the three-dimensional orthogonal space. This ensures that a specific \textit{Status} (e.g., flexion) is accurately credited only when correctly associated with both its corresponding \textit{Bodypart} and \textit{Context}. By transforming natural-language descriptions into these standardized relationships, \textit{GaitTripleF1} enables an objective comparison of the method's ability to synchronize "where" (\textit{Bodypart}), "how" (\textit{Status}), and "when" (\textit{Context}) a motion occurs. 

Given an input report, we first transform the free-text description into a set of triples \((Bodypart, Status, Context)\) using a rule-guided parsing procedure. Specifically, we maintain predefined vocabularies for each dimension derived from standardized clinical gait report templates and expert annotations. Then we further identify candidate words in the description via keyword matching and synonym normalization, and then associate them into triples based on sentence-level co-occurrence and dependency patterns. We discard partial matches to ensure each valid triple contains all the elements. We then compute \textit{Precision}, \textit{Recall}, and \textit{GaitTripleF1} score for the generated report \(T_{gen}\) and the ground truth \(T_{gt}\) based on exact triple matching:

\[
\text{Precision} = \frac{|T_{\text{gen}} \cap T_{\text{gt}}|}{|T_{\text{gen}}|}, 
\quad
\text{Recall} = \frac{|T_{\text{gen}} \cap T_{\text{gt}}|}{|T_{\text{gt}}|}
\]

\[
\text{GaitTripleF1} = \frac{2 \cdot \text{Precision} \cdot \text{Recall}}{\text{Precision} + \text{Recall}}
\]

Therefore, if a generated report fails to produce any valid triples (e.g., due to irrelevant, incoherent, or non-clinical descriptions), then \(T_{gen} = \emptyset\), leading to a \textit{GaitTripleF1} score of 0. \textit{GaitTripleF1} is used to evaluate \emph{MotionDecipher} and several baseline LLMs in Section~\ref{3}.
\section{Real-world Evaluation of \emph{MotionDecipher}} \label{3}

We deployed \emph{MotionDecipher} with 122 patients to evaluate its effectiveness, robustness, and clinical relevance. Our results show that \emph{MotionDecipher} consistently outperforms multiple baseline LLMs across quantitative metrics, remains robust across different patient age and BMI groups, and achieves the highest ratings from medical experts in terms of Correctness, Completeness, and Conciseness.

\subsection{Experimental Setup} \label{3.1}

Data were collected in real-world clinical settings and analyzed with quantitative evaluation metrics.

\begin{figure*}[htbp]
\centering
\includegraphics[width=0.815\textwidth]{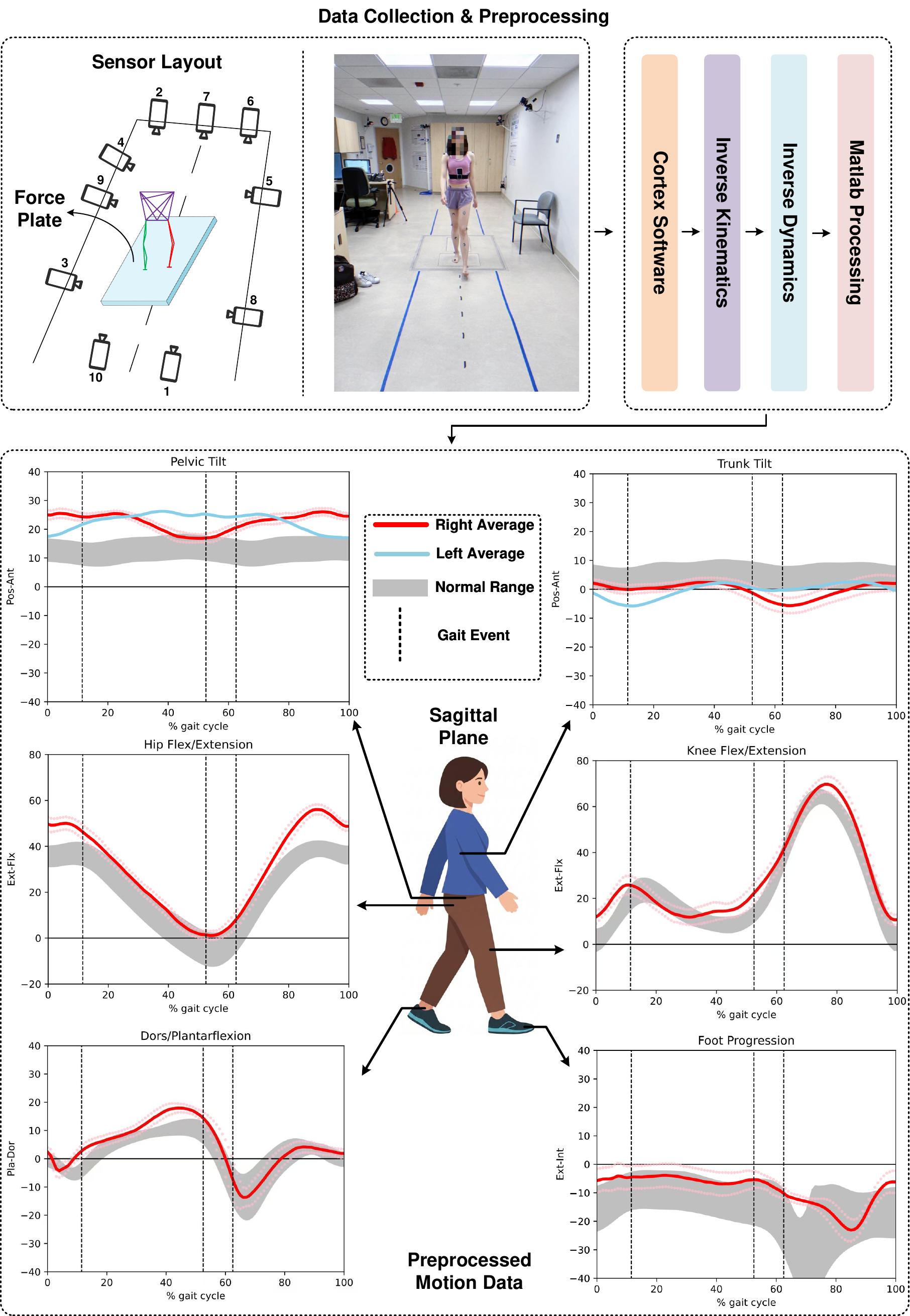}
\caption{Data collection \& Preprocessing process for \emph{MotionDecipher}: We use motion capture devices and force plates in an accredited gait lab to collect the data and transform the data into a visualized form through multiple preprocessing steps.}
\label{Fig8}
\end{figure*}

\textbf{Data Collection}. To provide clinically relevant gait analysis and ensure high-fidelity motion measurements, data were collected at an accredited gait lab from 122 patients between January 2024 and July 2025 and 46 controls following standardized clinical gait analysis protocols. The overall data collection process is shown in Fig.~\ref{Fig8}.

The main equipment for data collection includes clinical-grade 3D motion capture and force plate systems. The motion capture system is a 10-camera optical 3D motion capture system from Motion Analysis with a sampling rate of \(100Hz\), accompanied by the Cortex software to record data and track marker trajectories.  A modified Helen Hayes marker set is used for gait analysis following the standard practice of the gait lab. The modified Helen Hayes marker set is a widely used marker set in clinical gait analysis, and its variants form the basis of many standard marker configurations \cite{COLLINS2009173}. A total of 18 reflective markers are placed to comprehensively capture walking posture, including the sacrum, the left and right anterior superior iliac spines, the mid-shaft femurs, the lateral knee joint axes, the mid-shaft tibias (shanks), the lateral malleoli, the calcanei, the dorsum of the foot between the second and third metatarsal heads, the left and right acromia, and the left scapula respectively. We place 4 additional markers on the right and left medial knee joint axes and malleoli. For the force plate, we use a microcomputer-controlled force plate measurement device (Sampling rate: \(50Hz\)) with the Cortex software to collect the 3-dimensional ground reaction forces to obtain gait kinetics.

For each participant, the standard data collection procedure was performed, certified by the Commission for Motion Laboratory Accreditation (CMLA). The procedure is typically conducted through the collaboration of two to three on-site medical experts, including a biomechanist, a licensed physical therapist, and a specialized technician. The standard clinical gait analysis process typically takes 2–3 hours per patient. 

Before the patient arrives, the clinical team reviews the patient’s history, assesses potential challenges and solutions during the assessment, calibrates the motion capture (MoCap) and force plate systems, and prepares markers and documentation. Upon arrival, the technician and physical therapist first communicate with the patient and their family members to explain the assessment process and introduce the study, obtaining signed data disclosure and consent forms. Marker placement is conducted collaboratively by the physical therapist and biomechanist to ensure anatomical accuracy and proper alignment with the patient’s musculoskeletal structure, accounting for existing musculoskeletal deformities. Following marker placement, the biomechanist collects a static trial with the patient standing still in the center of the motion capture area to fit the digital skeletal template to the individual’s profile, followed by subsequent walking trials. During the walking trials, the patient is instructed by the physical therapist to walk along a 7-meter path marked on the floor using their typical gait. The physical therapist walks alongside the patient to ensure safety; the biomechanist records the data, logs trial numbers, and identifies clear force plate impacts. Each patient walks back and forth along the path 10-20 times, with unlimited rest periods provided upon request. Data collection continues until the biomechanist obtains at least 3-5 valid left and right foot impacts on the force plates, respectively. Upon completion of the assessment, the markers are removed. Following the gait analysis, the medical team conducts standard physical examinations and administrative procedures; manual measurements from these subsequent steps are not used for report generation in this study.

After the patient leaves, the biomechanist performs data preprocessing following the standardized protocol, which typically takes 1–2 hours per participant. The motion capture and force plate data are first filtered, segmented, and smoothed using the Cortex software KinTools RT package (Motion Analysis Inc. ~\cite{KinRT}), and then analyzed using built-in algorithms for inverse kinematics and inverse dynamics to compute joint angles, moments, and powers from marker trajectories and ground reaction forces. Critical gait events (e.g., foot strikes, foot off) are initially detected using the built-in algorithms in Cortex software and are subsequently verified or manually corrected by the biomechanist if any deviations from motion observations are identified. The output of the Cortex software includes timestamps of gait events, kinematics (joint angles), and kinetics (force, moment, power) in the form of time-series data. The time-series data are then processed through a series of MATLAB scripts to segment the walking trials into gait cycles based on the timestamps of gait events and to normalize the gait cycle time to standardized percentages (0–100\%). Finally, multiple streams of time-series motion data from the frontal, sagittal, and transverse planes are visualized by plotting the averaged gait cycles along with their corresponding standard deviations (see examples in Fig.~\ref{Fig8}). This processed data serves as the input to \emph{MotionDecipher} and other comparative LLM baselines.

Similarly, the processed data also serves as the input for medical experts to manually analyze gait and generate clinical reports. This process typically involves collaborative assessment by the two on-site clinicians, along with the director of the gait lab, who oversees the process and reviews the report to ensure accurate interpretation. The full reporting workflow requires approximately 5-10 cumulative hours per participant, which can be decomposed into 3 steps: 

\begin{enumerate}
    \item Data analysis and interpretation: 2-3 hours;
    \item Report drafting: 2-3 hours;
    \item Consensus discussion: 1-4 hours.
\end{enumerate}

The resulting manually generated reports are used as the ground truth in this study. In collaboration with the clinical team, we compiled a dataset of 122 patients during the data collection period through routine appointments conducted in the gait laboratory. This study was reviewed and approved by the Institutional Review Board (IRB) of [Stanford University] under protocol number [11910: Kinematic, Kinetic, and Electromyographic Measures of Human Movement]. All participants provided informed consent before participation. In all subsequent experiments, unless otherwise specified, we use the expert-written reports from 2 representative participants as examples in the GRTDC and evaluate \emph{MotionDecipher} on the remaining 120 participants. This split is intentionally designed to maximize the size of the unseen test set to enhance the rigor of our evaluation. 

\vspace{0.5em}
\noindent \textbf{Quantitative Evaluation Metrics}. To evaluate the report generation quality of \emph{MotionDecipher} and other baseline LLMs, we use 4 general Natural Language Generation (NLG) evaluation metrics (\textit{BLEU}, \textit{ROUGE-L}, \textit{METEOR}, and \textit{BERTScore}) and 2 medical-specific evaluation metrics (\textit{MEDCON-Gait} and the proposed \textbf{\textit{GaitTripleF1}} (Section~\ref{2.4})) to quantify the similarity between each method's output and the ground truth reports. The definitions and explanations of each metric are as follows:

\begin{itemize}
    \item \textbf{\textit{BLEU}} \cite{BLEU}: It measures the \textit{N-Gram} level overlap between the generated text and the reference text (ground truth).
    \item \textbf{\textit{ROUGE-L}} \cite{ROUGE}: It measures the similarity of the Longest Common Subsequence (LCS) between the generated text and the reference text.
    \item \textbf{\textit{METEOR}} \cite{METEOR}: A similarity metric that jointly accounts for \textit{Precision}, \textit{Recall}, lemmatization, and synonym matching.
    \item \textbf{\textit{BERTScore}} \cite{BERTScore}: It computes sentence-level semantic similarity with BERT embeddings, aligning better with semantic coherence than \textit{N-Gram} metrics.
    \item \textbf{\textit{MEDCON-Gait}}: It measures the overlap rate of medical concepts between the generated text and the reference text. \textit{MEDCON-Gait} is an adapted version of the \textit{MEDCON} \cite{MEDCON} metric specifically optimized to assess the quality of generated gait analysis reports through a domain-specific gait term library. This adaptation is necessary because the original \textit{MEDCON} provides limited coverage of gait- and motion-specific terminology, which can lead to underestimation of semantic overlap in clinical gait descriptions. To address this, we construct a gait term library from standardized clinical gait analysis report templates \cite{ClinicalGENI}, spanning multiple dimensions, including anatomical structures, motion attributes, gait phases and events, clinical status descriptors, and abbreviations (Appendix ~\ref{appendix:gaitterms}). 
    \item \textbf{\textit{GaitTripleF1}} (Section~\ref{2.4}): It captures the associations between anatomical regions and motion states.
\end{itemize}

For gait analysis report generation, the descriptions contain a mixture of quantitative information (e.g., joint angles) and qualitative information (e.g., gait tendencies), allowing substantial linguistic variability even when conveying the same clinical meaning. As a result, NLG evaluation metrics such as \textit{BLEU}, \textit{ROUGE-L}, and \textit{METEOR}, which are sensitive to surface-form differences (e.g., "anterior tilt in stance" vs. "anterior tilt at stance") and numerical variations (e.g., 10$^{\circ}$ vs. 11$^{\circ}$), may yield moderate scores despite clinically consistent outputs. Therefore, these metrics should be interpreted in conjunction with medical-specific evaluation metrics (\textit{MEDCON-Gait} and \textit{GaitTripleF1}) to provide a more comprehensive assessment of model performance. At the same time, extremely low scores generally indicate substantial divergence from the ground truth reports in both content and form.

\vspace{0.5em}
\noindent \textbf{Comparison Models}. We select 4 mainstream LLMs (GPT-5 \cite{GPT-5}, Kimi-K2 \cite{Kimi-k2}, Google Gemini 3 \cite{gemini}, and Qwen3-Max \cite{Qwen3}) as 4 general LLM comparison baselines and set the hyperparameter \(Temperature = 0.7\). To ensure a fair comparison, we employ a standardized Chain-of-Thought \cite{CoTPrompt} prompt template across all baselines and enable the file upload functionality (Appendix ~\ref{appendix:CoT}). This setup guides the LLMs to perform step-by-step reasoning based on the preprocessed data (Fig.~\ref{Fig8}) and subsequently generate the gait analysis reports. We further include a strong non-LLM baseline (\textit{Rule-based method}) in Section \ref{3.3}. Essentially, the \textit{Rule-based method} is derived from the complete \emph{MotionDecipher} pipeline by removing the \textit{Hybrid Context Constructor} component. It concatenates the \textit{Structured Representation} into natural language descriptions according to corpus concatenation rules (e.g., concatenating "body\_part": "Ankle", "Ankle PF/DF IC": "0° DF (R), 7° DF (L)", into a natural language description: "Ankle PF/DF at Initial Contact: 0° DF (R), 7° DF (L)"), and ultimately generates the complete gait analysis report.

\subsection{Overall Performance Compared to Baseline LLMs} \label{3.2}

Using our self-collected dataset, we adopt \emph{MotionDecipher} and other baseline LLMs to generate gait analysis reports for each participant individually, then compute the scores and corresponding standard deviations (SD) between each generated report and the ground truth using the metrics described in Section~\ref{3.1}. Subsequently, we average the scores of 120 participants, and the results are presented in Table \ref{Table1}.

\begin{table}[htbp]
\centering
\caption{Comparison Results between \textit{\emph{MotionDecipher}} and baseline LLMs on 6 different evaluation metrics}
\label{Table1}
\begin{tabular}{cccc}
\toprule
\multirow{2}{*}{\textbf{Model}} & \multicolumn{3}{c}{Evaluation Metric (\(Avg \pm 1SD\))} \\
\cmidrule(r){2-4}
 & \textit{BLEU} & \textit{ROUGE-L} & \textit{METEOR} \\
\midrule
\textit{\emph{MotionDecipher}} & \textbf {0.40 $\pm$ 0.04} \textcolor[RGB]{0,100,0} {(\(+34\%\))} &   \textbf {0.43 $\pm$ 0.07} \textcolor[RGB]{0,100,0} {(\(+29\%\))} & \textbf{0.49 $\pm$ 0.06} \textcolor[RGB]{0,100,0} {(\(+34\%\))} \\
\textit{GPT-5} & 0.02 $\pm$ 0.01 & 0.13 $\pm$ 0.02 & 0.12 $\pm$ 0.04 \\
\textit{Kimi-K2} & 0.01 $\pm$ 0.00 & 0.11 $\pm$ 0.01 & 0.09 $\pm$ 0.02 \\ 
\textit{Google Gemini 3} & 0.06 $\pm$ 0.02 & 0.14 $\pm$ 0.02 & 0.15 $\pm$ 0.03 \\
\textit{Qwen3-Max} & 0.01 $\pm$ 0.00 & 0.11 $\pm$ 0.02 & 0.11 $\pm$ 0.02 \\
\bottomrule
\addlinespace[3pt]
\multirow{2}{*}{\textbf{Model}} & \multicolumn{3}{c}{Evaluation Metric (\(Avg \pm 1SD\))} \\
\cmidrule(r){2-4}
 & \textit{BERTScore} & \textit{MEDCON-Gait} & \textit{GaitTripleF1} \\
\midrule
\textit{\emph{MotionDecipher}} & \textbf{0.91 $\pm$ 0.01} \textcolor[RGB]{0,100,0} {(\(+9\%\))} & \textbf{0.76 $\pm$ 0.05} \textcolor[RGB]{0,100,0} {(\(+16\%\))} & \textbf{0.91 $\pm$ 0.07} \textcolor[RGB]{0,100,0} {(\(+19\%\))} \\
\textit{GPT-5}  & 0.78 $\pm$ 0.03 & 0.60 $\pm$ 0.06 & 0.63 $\pm$ 0.10 \\
\textit{Kimi-K2} & 0.82 $\pm$ 0.01 & 0.45 $\pm$ 0.05 & 0.62 $\pm$ 0.12 \\ 
\textit{Google Gemini 3} & 0.71 $\pm$ 0.02 & 0.58 $\pm$ 0.06 & 0.72 $\pm$ 0.16 \\
\textit{Qwen3-Max} & 0.81 $\pm$ 0.03 & 0.38 $\pm$ 0.05 & 0.60 $\pm$ 0.14 \\
\bottomrule
\end{tabular}%
\end{table}

\begin{figure}[htbp]
\centering
\includegraphics[width=0.98\textwidth]{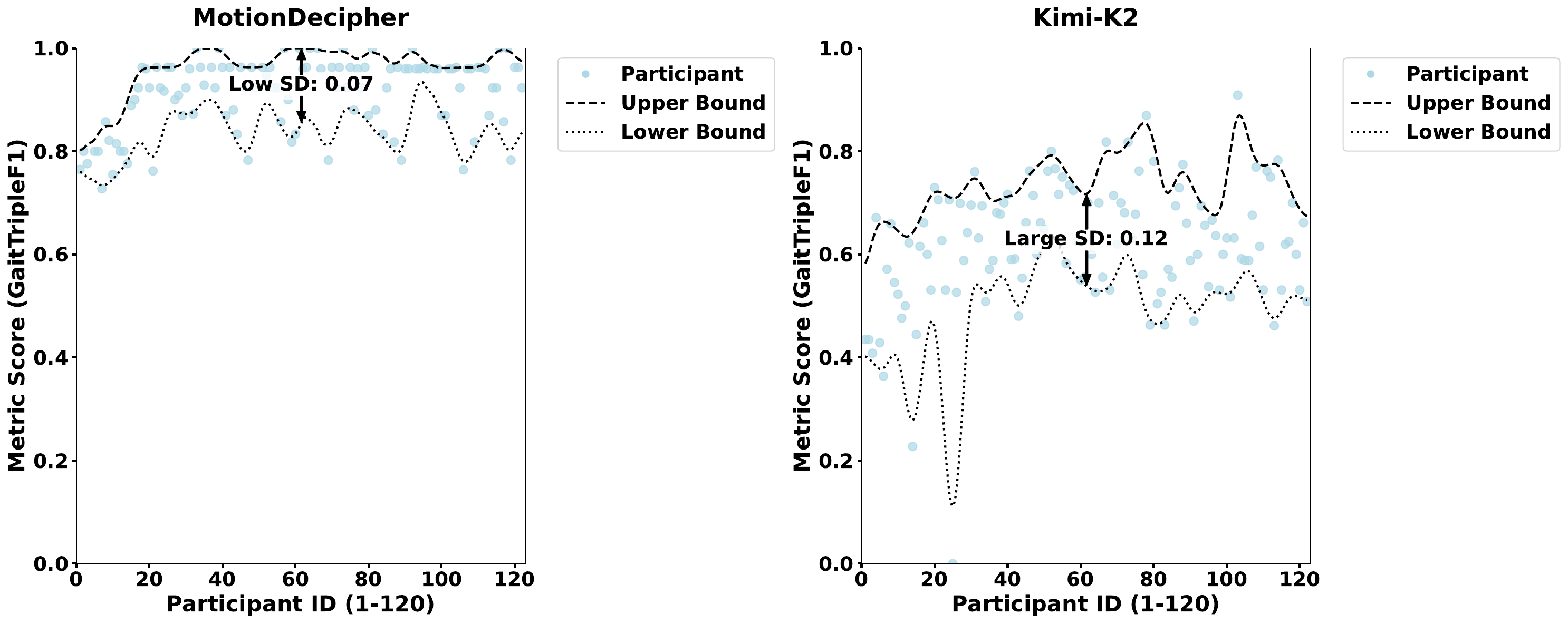}
\caption{A visualization example of SD results: \emph{MotionDecipher} outperforms the Kimi-K2 baseline in average performance while exhibiting a lower SD (0.07 vs. 0.12).}
\label{Fig9}
\end{figure}

From Table \ref{Table1}, \emph{MotionDecipher} achieves the best overall performance, outperforming other baseline LLMs on all 6 metrics, with an improvement from 9\% to 34\%. In terms of NLG evaluation, \emph{MotionDecipher}'s \textit{BERTScore} exceeds 90\%, indicating its capability of generating semantically meaningful gait analysis reports. \emph{MotionDecipher} also shows significant advantages in \textit{BLEU}, \textit{ROUGE-L}, and \textit{METEOR} metrics, demonstrating that the reports generated by \emph{MotionDecipher} are more consistent with the ground truth at the textual level. Regarding medical-specific metrics, \emph{MotionDecipher}'s \textit{GaitTripleF1} score also surpasses 90\%, showing its ability to correctly link motion states (e.g., flexion, extension, etc.) to the corresponding anatomical regions in gait descriptions. The SD score further supports the superior performance of \emph{MotionDecipher}. Since the baseline LLMs perform poorly in \textit{BLEU}, \textit{ROUGE-L}, and \textit{METEOR} metrics, their lower SD in these categories is of little significance. In contrast, for \textit{BERTScore}, \textit{MEDCON-Gait}, and \textit{GaitTripleF1} metrics, \emph{MotionDecipher} not only maintains a clear performance advantage but also achieves the lowest (or equivalent) SD, indicating stronger consistency in report generation when handling different participants. Fig.~\ref{Fig9} provides an intuitive comparison of the SD on the \textit{GaitTripleF1} metric between \emph{MotionDecipher} and the Kimi-K2 baseline across 120 participants. While achieving superior average performance, \emph{MotionDecipher} also exhibits a substantially lower SD, as evidenced by the smaller distance between the upper and lower bounds.

\begin{figure}[htbp]
\centering
\includegraphics[width=0.98\textwidth]{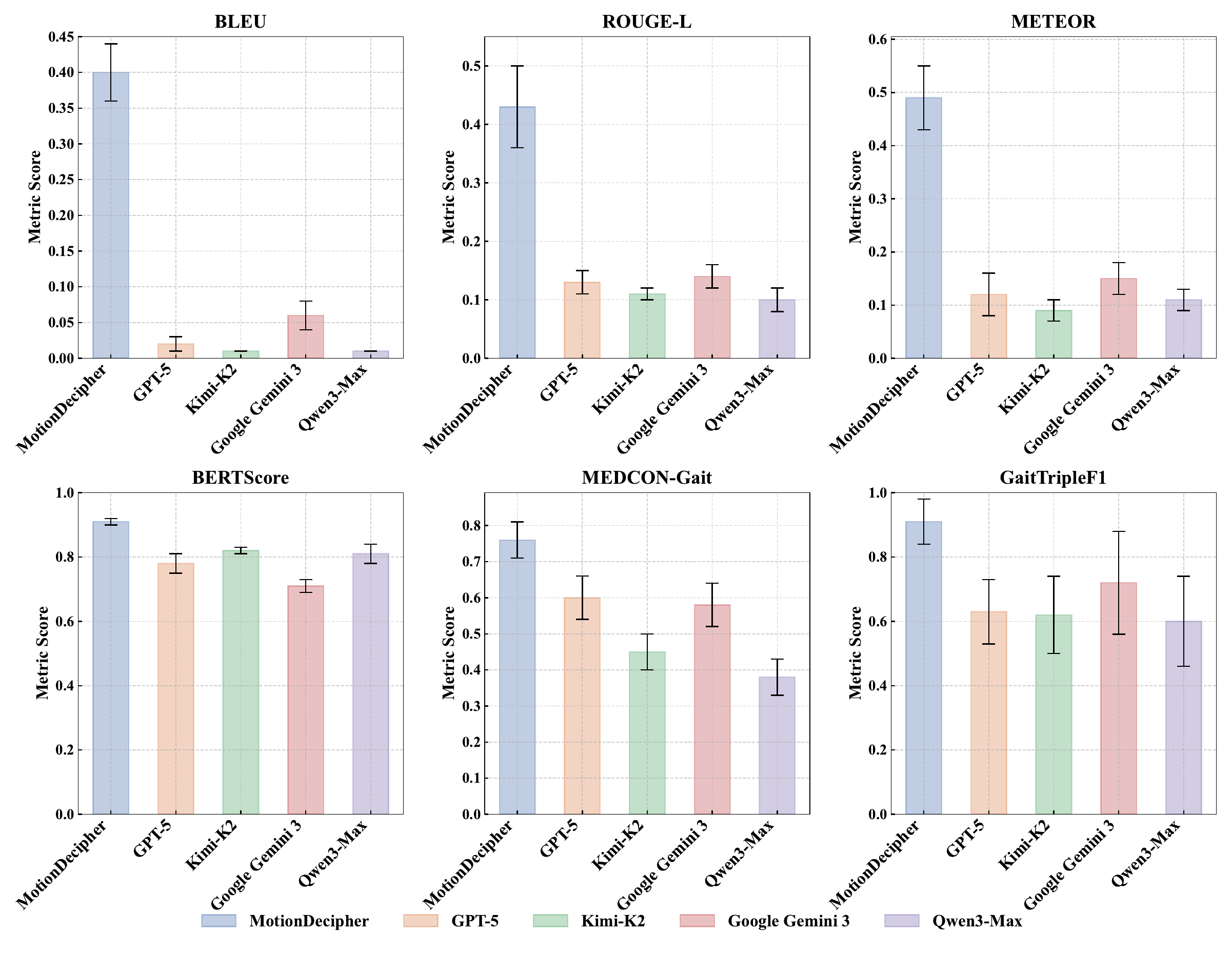}
\caption{Visualization of Comparative Results: \emph{MotionDecipher} shows significant performance advantages on all evaluation metrics.}
\label{Fig10}
\end{figure}

Horizontally comparing each baseline, no specific LLM demonstrates an obvious performance advantage. For example, \textit{Google Gemini 3} and \textit{GPT-5} perform better on medical-specific evaluation metrics, indicating that they incorporate more domain-specific terms and background knowledge related to gait analysis during the model pre-training stage. Kimi-K2 achieves the optimal performance in \textit{BERTScore}, suggesting that it can better understand the semantic information in motion time-series data. However, all these baselines exhibit a performance gap compared to \emph{MotionDecipher}, which reflects that LLMs lack analytical expertise in the gait and motion analysis domain and cannot generate gait analysis reports that integrate both quantitative information (e.g., 10° DF) and qualitative description (e.g., increased anterior tilt). Fig.~\ref{Fig10} visualizes the information shown in Table \ref{Table1}.

\subsection{Ablation Analysis} \label{3.3}

We conduct ablation analysis on \emph{MotionDecipher} to verify the effectiveness of its different components (\textit{Human Motion Tokenizer}, \textit{Meta Feature Selector}, and \textit{Hybrid Context Constructor}). Specifically, we first remove the GRTDC, enabling the LLM to generate gait reports based solely on a single fixed demonstration (A1). We further remove the entire \textit{Hybrid Context Constructor} module, which is the Rule-based method (A2). Next, we remove the \textit{Meta Feature Selector}, directly prompting unfiltered \textit{Structured Representations} into the LLM to assess the impact of feature selection (A3). Finally, we remove the \textit{Human Motion Tokenizer}, replacing the tokenized representation with preprocessed data, to evaluate the necessity of feature embedding for downstream reasoning and report generation (A4). The experimental results are presented in Table \ref{Table2}.

\begin{table}[htbp]
\centering
\caption{Ablation Analysis}
\label{Table2}
\begin{tabular}{cccc}
\toprule
\multirow{2}{*}{\textbf{Model}} & \multicolumn{3}{c}{Evaluation Metric (\(Avg \pm 1SD\))} \\
\cmidrule(r){2-4}
 & \textit{BLEU} & \textit{ROUGE-L} & \textit{METEOR} \\
\midrule
\textit{\emph{MotionDecipher}} & \textbf{0.40 $\pm$ 0.04} & \textbf{0.43 $\pm$ 0.07} & \textbf{0.49 $\pm$ 0.06} \\
\textit{\emph{MotionDecipher} w/o GRTDC (A1)} & 0.34 $\pm$ 0.11 \textcolor[RGB]{150,0,0}{(\(-6\%\))} & 0.42 $\pm$ 0.16 \textcolor[RGB]{150,0,0}{(\(-1\%\))} & 0.44 $\pm$ 0.13 \textcolor[RGB]{150,0,0}{(\(-5\%\))} \\
\textit{Rule-based method (A2)} & 0.30 $\pm$ 0.05 \textcolor[RGB]{150,0,0}{(\(-10\%\))} & 0.35 $\pm$ 0.06 \textcolor[RGB]{150,0,0}{(\(-8\%\))} & 0.41 $\pm$ 0.08 \textcolor[RGB]{150,0,0}{(\(-8\%\))} \\ 
\textit{\emph{MotionDecipher} w/o \text{Meta Feature Selector} (A3)} & 0.24 $\pm$ 0.07 \textcolor[RGB]{150,0,0}{(\(-16\%\))} & 0.32 $\pm$ 0.09 \textcolor[RGB]{150,0,0}{(\(-11\%\))} & 0.34 $\pm$ 0.12 \textcolor[RGB]{150,0,0}{(\(-15\%\))} \\
\textit{\emph{MotionDecipher} w/o \text{Human Motion Tokenizer} (A4)} & 0.15 $\pm$ 0.06 \textcolor[RGB]{150,0,0}{(\(-25\%\))} & 0.21 $\pm$ 0.08 \textcolor[RGB]{150,0,0}{(\(-22\%\))} & 0.23 $\pm$ 0.15 \textcolor[RGB]{150,0,0}{(\(-26\%\))} \\
\bottomrule
\addlinespace[3pt]
\multirow{2}{*}{\textbf{Model}} & \multicolumn{3}{c}{Evaluation Metric (\(Avg \pm 1SD\))} \\
\cmidrule(r){2-4}
 & \textit{BERTScore} & \textit{MEDCON-Gait} & \textit{GaitTripleF1} \\
\midrule
\textit{\emph{MotionDecipher}} & \textbf{0.91 $\pm$ 0.01} & \textbf{0.76 $\pm$ 0.05} & \textbf{0.91 $\pm$ 0.07} \\
\textit{\emph{MotionDecipher} w/o GRTDC (A1)} & 0.89 $\pm$ 0.06 \textcolor[RGB]{150,0,0}{(\(-2\%\))} & 0.68 $\pm$ 0.11 \textcolor[RGB]{150,0,0}{(\(-8\%\))} & 0.85 $\pm$ 0.14 \textcolor[RGB]{150,0,0}{(\(-6\%\))} \\
\textit{Rule-based method (A2)} & 0.87 $\pm$ 0.02 \textcolor[RGB]{150,0,0}{(\(-4\%\))} & 0.63 $\pm$ 0.06 \textcolor[RGB]{150,0,0}{(\(-13\%\))} & 0.74 $\pm$ 0.06 \textcolor[RGB]{150,0,0}{(\(-17\%\))} \\ 
\textit{\emph{MotionDecipher} w/o \text{Meta Feature Selector} (A3)} & 0.86 $\pm$ 0.02 \textcolor[RGB]{150,0,0}{(\(-5\%\))} & 0.61 $\pm$ 0.08 \textcolor[RGB]{150,0,0}{(\(-15\%\))} & 0.71 $\pm$ 0.09 \textcolor[RGB]{150,0,0}{(\(-20\%\))} \\
\textit{\emph{MotionDecipher} w/o \text{Human Motion Tokenizer} (A4)} & 0.81 $\pm$ 0.01 \textcolor[RGB]{150,0,0}{(\(-10\%\))} & 0.57 $\pm$ 0.07 \textcolor[RGB]{150,0,0}{(\(-19\%\))} & 0.68 $\pm$ 0.11 \textcolor[RGB]{150,0,0}{(\(-23\%\))} \\
\bottomrule
\end{tabular}%
\end{table}

\begin{figure}[htbp]
\centering
\includegraphics[width=0.98\textwidth]{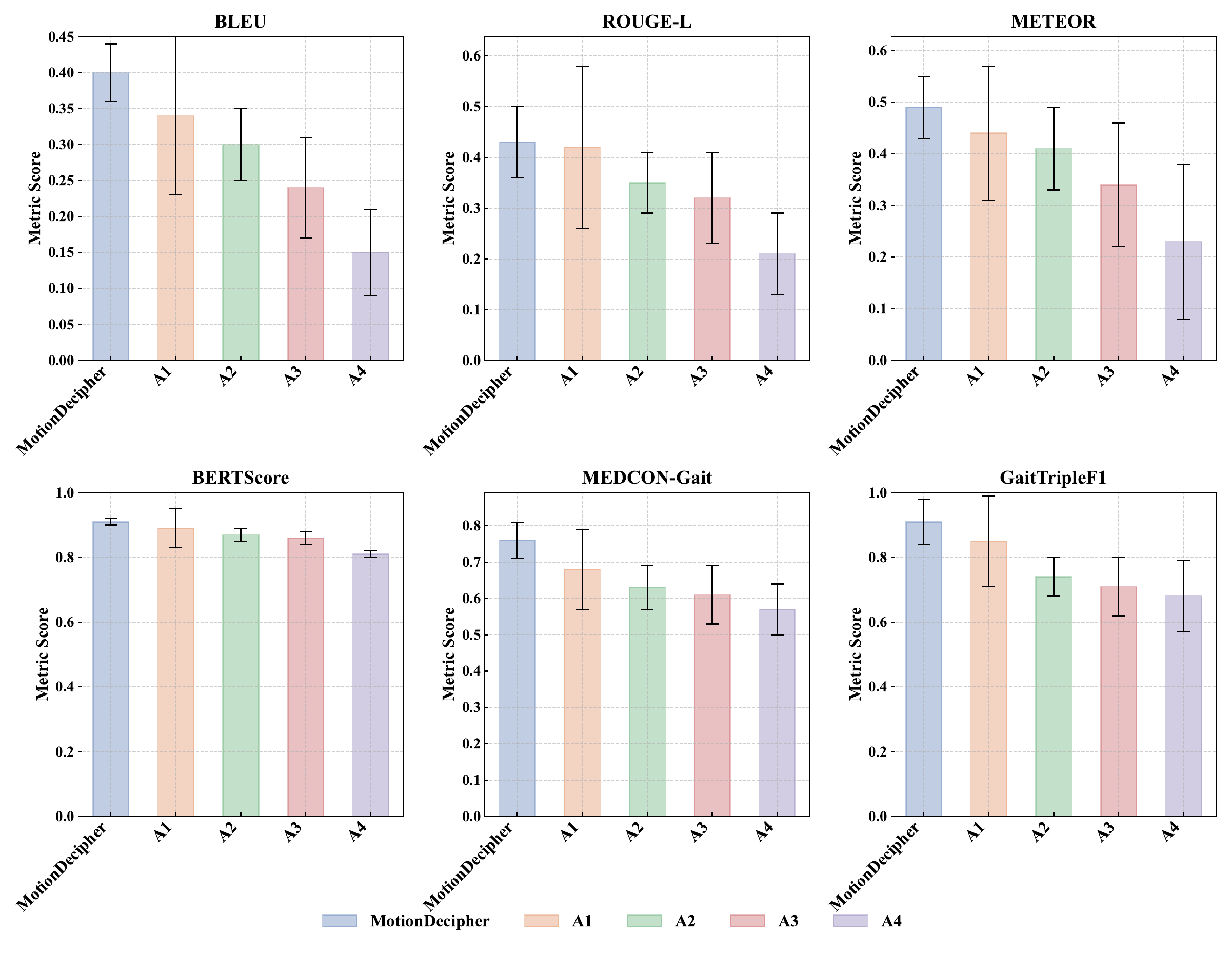}
\caption{Ablation Analysis of \emph{MotionDecipher}: Removing different components leads to varying degrees of performance degradation.}
\label{Fig11}
\end{figure}

After removing the GRTDC, \emph{MotionDecipher} exhibits varying degrees of performance degradation (1\%-8\%) across all metrics. This is because, in some cases, the testing gait data is highly similar to the fixed demonstration in GRTDC (e.g., both the input and the demonstration have "Trunk" as the body part, and both feature "posterior tilt"), where the demonstration serves as a description template for the LLM. However, in other cases, there is almost no similarity between the input and the demonstration (e.g., the body part of the input (Pelvis) differs from that of the demonstration (Trunk)), and such mismatched demonstrations may increase the risk of hallucinations by anchoring the LLM to irrelevant context. In contrast, the GRTDC can automatically retrieve and select the demonstration with the optimal similarity based on the input, ensuring the LLM generates gait analysis descriptions with the optimal example, thus improving the consistency of report generation. For the Rule-based method, its standard deviation is similar to that of \emph{MotionDecipher} due to fixed generation rules. However, lacking the LLM's reasoning and analysis processes, it shows a noticeable performance degradation in all metrics (4\%-17\%). After removing the \textit{Meta Feature Selector} component, \emph{MotionDecipher} exhibits a further performance decline due to the interference of substantial redundant information and the absence of \textit{Structured Representations}. When the \textit{Human Motion Tokenizer} is removed, the downstream components receive inputs that lack explicit gait feature information, leading to further performance degradation. Nevertheless, it still outperforms the direct outputs of Kimi-K2, as the examples retrieved by the \textit{Hybrid Index} algorithm continue to provide useful guidance to the LLM generation. In summary, the results of the ablation analysis indicate that every component of \emph{MotionDecipher} is indispensable. Fig.~\ref{Fig11} visualizes the information presented in Table \ref{Table2}.

\subsection{Robustness Analysis} \label{3.4}

\subsubsection{Robustness to different participant groups}

To further verify the robustness and generalizability of \emph{MotionDecipher} across different participant groups, we divide 120 participants into 5 intervals based on 2 dimensions: Age and BMI score. Regarding age, we classify participants as: Children (0-12) (55 participants, 45.8\%), Adolescents (13-17) (38 participants, 31.7\%), Young Adults (18-35) (8 participants, 6.7\%), Middle-aged Adults (36-59) (7 participants, 5.8\%), and Older Adults (60 and above) (12 participants, 10\%). For BMI, it synthesizes 2 factors (Weight and Height) simultaneously. Similarly, we classify participants' BMI ranges as: BMI Range I (Below 18.5) (48 participants, 40.0\%), BMI Range II (18.5-24.9) (40 participants, 33.3\%), BMI Range III (25.0-27.9) (17 participants, 14.2\%), BMI Range IV (28.0-31.9) (9 participants, 7.5\%), BMI Range V (32.0 or higher) (6 participants, 5\%).

\begin{figure}[htbp]
\centering
\includegraphics[width=1.0\textwidth]{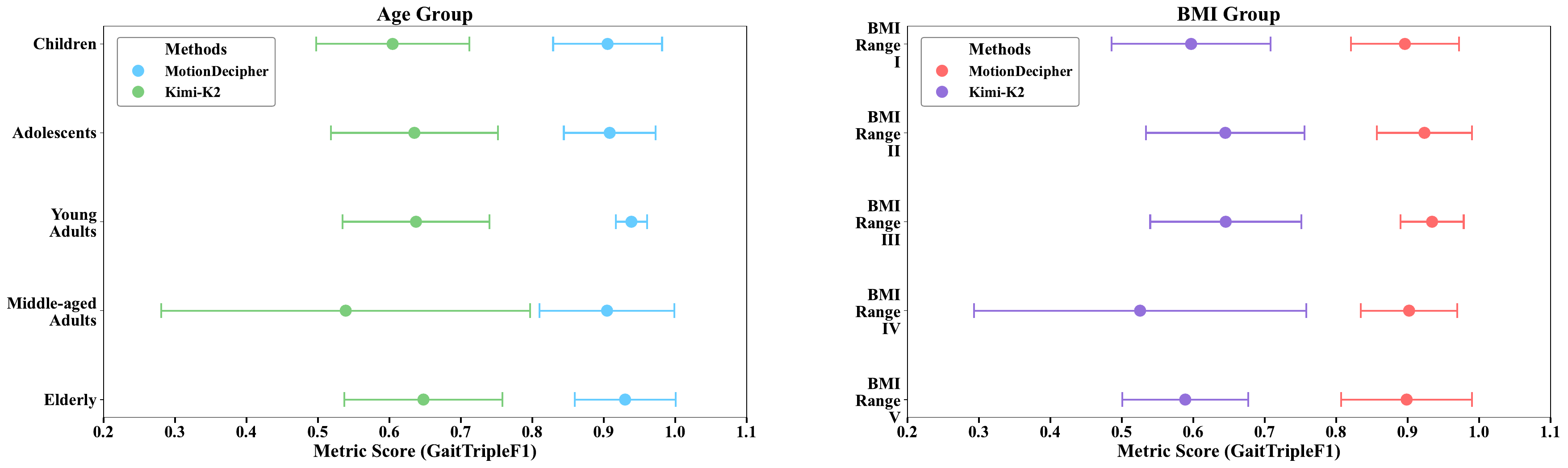}
\caption{Robustness analysis on Age and BMI Group: \emph{MotionDecipher} demonstrates stronger robustness across different participant groups (Age and BMI), as evidenced by the increased score of data points and lower error compared with the Kimi-K2 baseline.}
\label{Fig12}
\end{figure}

We adopt the proposed \textit{GaitTripleF1} evaluation metric because it can effectively reflect the model's ability to capture the relationships between body parts and joint motions, thereby enabling a better assessment of the model's adaptability when applied to different age/BMI groups. Kimi-K2 is selected as the comparative baseline. The results are shown in Fig.~\ref{Fig12}. \emph{MotionDecipher} exhibits consistently low error/high score results across different age and BMI groups, with a low SD, making it widely applicable to gait and motion analysis tasks for various patients. In contrast, Kimi-K2 not only demonstrates inferior performance across age and BMI groups but also shows significant score fluctuations, with a large SD.

\subsubsection{Robustness to Gait Deviation Index Score}

\begin{figure}[htbp]
\centering
\includegraphics[width=0.9\textwidth]{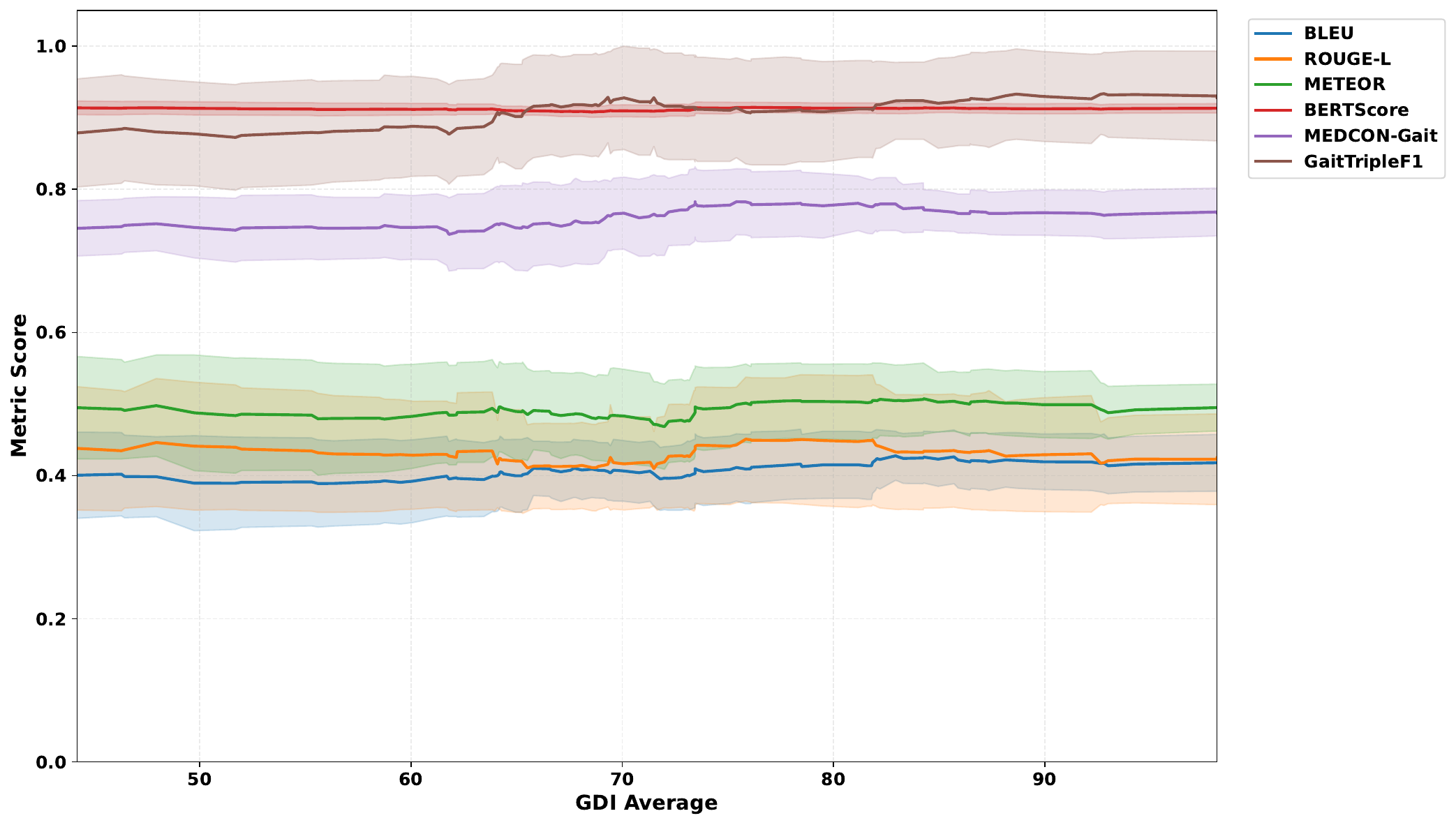}
\caption{Robustness analysis on GDI Score: \emph{MotionDecipher} demonstrates stable performance on 6 evaluation metrics across different GDI scores, exhibiting minimal fluctuations in the line plots and a relatively uniform distribution of the shaded regions.}
\label{Fig13}
\end{figure}

Gait Deviation Index (GDI) \cite{GDI} is a clinically established summary metric that quantifies overall gait pathology by comparing an individual’s kinematic gait patterns to those of a normative reference population, enabling quantitative assessment of gait abnormality severity and changes following intervention. To investigate the consistency of reports generated by \emph{MotionDecipher} across different GDI scores, we calculate the mean of the left and right leg GDI for each participant, and compute a sliding-window moving average as the observed value for each metric score data point (with the window size set to 5). Based on these observations, we further plot the average trend line and the \(Normal Mean \pm 1SD\) shaded region of each evaluation metric, as illustrated in Fig.~\ref{Fig13}.

The consistently stable trend and shaded bands across 6 evaluation metrics over the GDI range indicate \emph{MotionDecipher} is robust to gait variability, demonstrating its ability to generate accurate gait analysis reports regardless of the participant's gait abnormality status.

\subsection{Sensitivity Analysis}

\begin{figure}[htbp]
\centering
\includegraphics[width=0.99\textwidth]{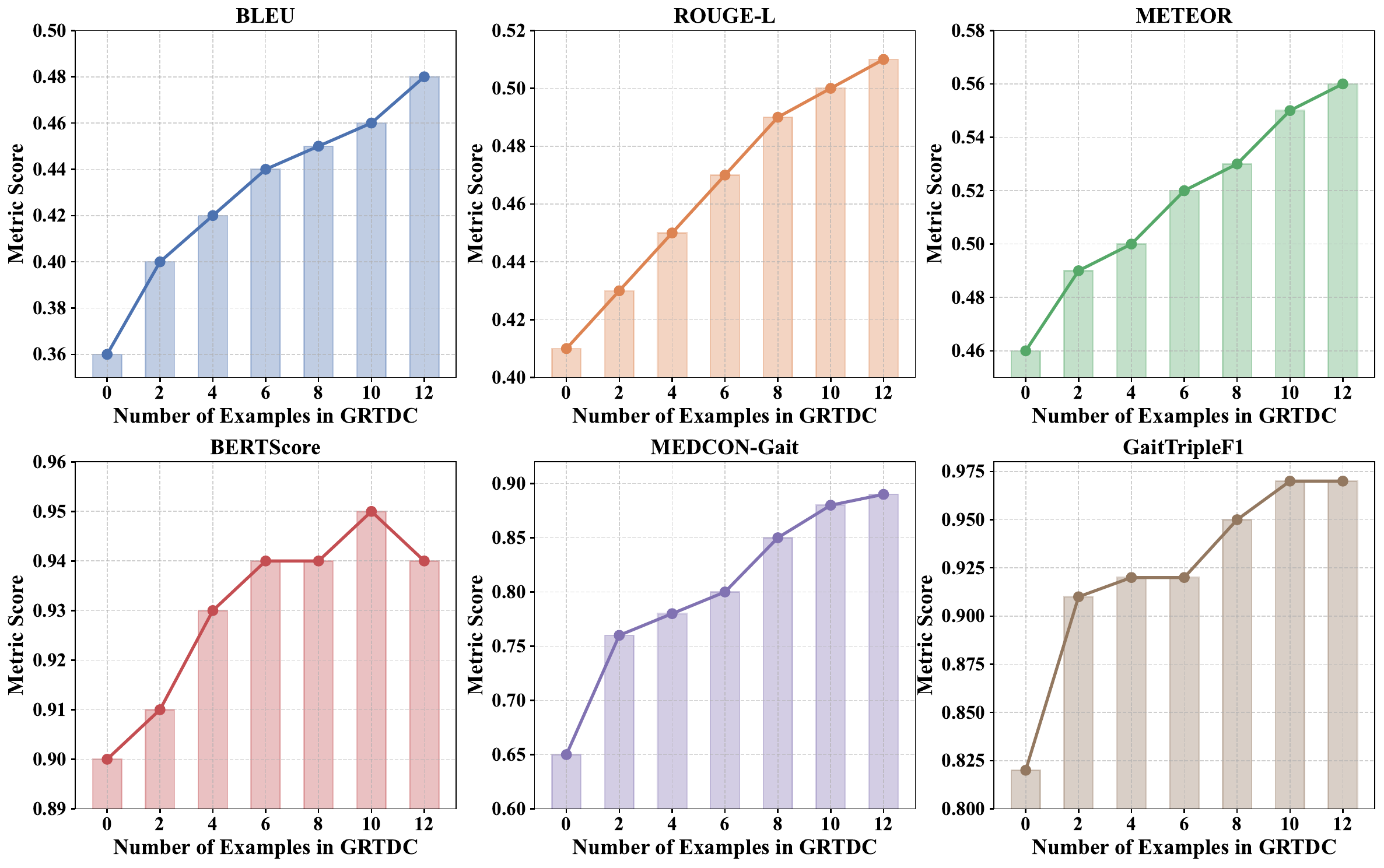}
\caption{The sensitivity of \emph{MotionDecipher} to the number of examples in GRTDC: As the number of examples in the GRTDC increases, all 6 metrics show varying degrees of performance improvement.}
\label{Fig14}
\end{figure}

We further investigate how the number of examples in the GRTDC influences the performance of \emph{MotionDecipher}. Since the total number of participants is fixed, in the sensitivity analysis, the number of examples in the GRTDC is gradually increased (0, 2, 4, 6, 8, 10, 12), and correspondingly, the number of participants used for testing \emph{MotionDecipher} is gradually decreased, with the sum of the two remaining constant at 122. Note that 0 examples in the GRTDC represent the \textit{without-retrieval} process (i.e., no retrieved demonstrations are included in the ICL prompt). This is distinct from method A1 in Section~\ref{3.3}, where the GRTDC module is removed from \emph{MotionDecipher} but specific demonstrations are still retained. For each test set, we calculate the average performance of \emph{MotionDecipher} across the 6 evaluation metrics, and the results are presented in Fig.~\ref{Fig14}.

Overall, as the number of examples increases, the performance of \emph{MotionDecipher} across the 6 different evaluation metrics also improves. Notably, when the number of examples increases from 0 to 2, \emph{MotionDecipher} achieves considerable performance improvement across all six evaluation metrics. This is because, with 0 examples, the LLM's zero-shot inference amplifies output uncertainty, which highlights the necessity of the GRTDC design. With the enhanced diversity of the example set in the GRTDC, \emph{MotionDecipher} can retrieve and utilize more accurate examples as ICL demonstrations when generating gait analysis reports. Since \emph{MotionDecipher} already demonstrates excellent performance on these metrics, both \textit{BERTScore} and \textit{GaitTripleF1} metrics stabilize at a threshold of approximately 0.95 when the number of examples continues to increase. We additionally conducted a toy experiment where all example types in the GRTDC were restricted to a specific body part (e.g., Knee). In this scenario, no matter how the number of examples is increased, these results indicate that performance improvements cannot be achieved by simply increasing the number of examples; instead, effective gains require diverse and clinically relevant demonstrations, which motivates the design of the GRTDC. Therefore, in real-world practice, we recommend prioritizing the diversity of examples rather than the number of examples in the GRTDC.

\subsection{Medical Expert Evaluation} \label{3.6}

The Medical Expert Evaluation aims to assess and analyze the clinical utility of \emph{MotionDecipher} from the perspective of human experts. Although Section~\ref{3.2} to Section~\ref{3.4} comprehensively quantify the similarity between reports generated by \emph{MotionDecipher}, various baseline LLMs, non-LLM baselines, and the ground truth reports using evaluation metrics (e.g., \textit{BLEU}, \textit{ROUGE-L}, \textit{BERTScore}, \textit{GaitTripleF1}, etc.) across different levels (e.g., subsequence, semantic, N-gram, relation, term, etc.), these metrics overlook the fact that, in certain cases, \emph{MotionDecipher} and other baselines may provide clinically relevant findings that are not documented in reference reports. Therefore, we submit the analysis reports generated by \emph{MotionDecipher}, the baseline LLM (Kimi-K2), and the non-LLM baseline (Rule-based method) for evaluation by two certified gait clinic experts through a blinded review process, where model identities are anonymized to prevent bias. The evaluation is conducted across 3 orthogonal indices: \textbf{Correctness}, \textbf{Completeness}, and \textbf{Conciseness}~\cite{van2024adapted}. The definitions and scoring rubrics of each index are shown in Fig.~\ref{Fig15}:

\begin{figure}[htbp]
\centering
\includegraphics[width=0.99\textwidth]{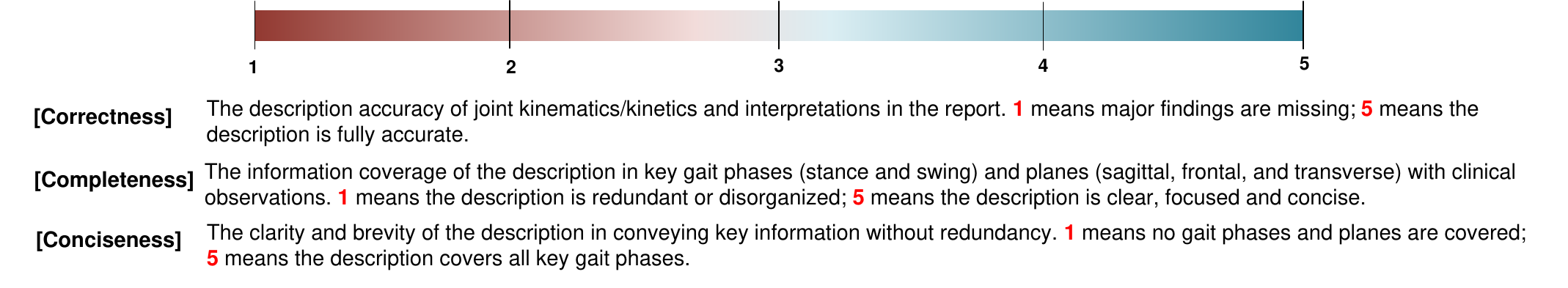}
\caption{Scoring rubrics for medical expert evaluation indices. Each index has its own definition and scoring criteria.}
\label{Fig15}
\end{figure}

\begin{figure}[htbp]
\centering
\includegraphics[width=0.99\textwidth]{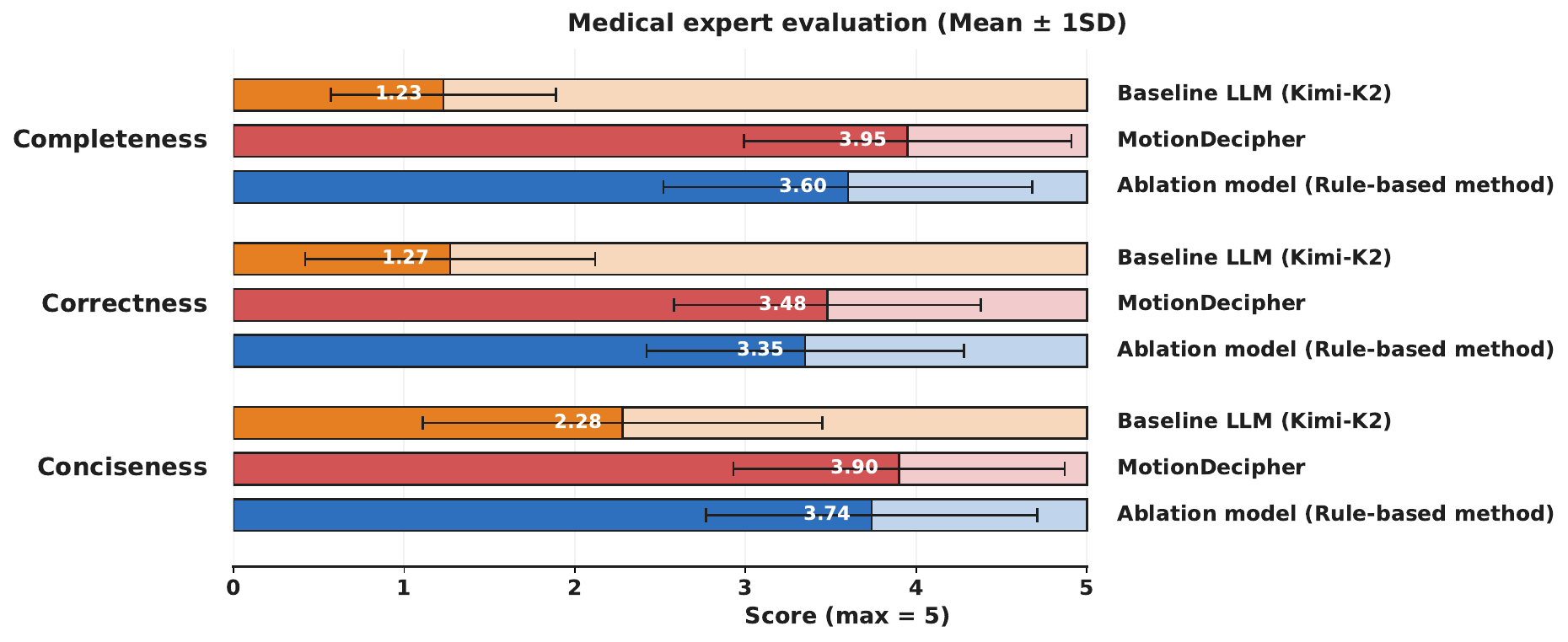}
\caption{Medical expert evaluation results of different methods. \emph{MotionDecipher} achieves the optimal score across different indices.}
\label{Fig16}
\end{figure}

For Medical Expert Evaluation, we decompose each gait analysis report into components corresponding to different body parts. We focus on the knee, hip, and ankle as representative regions. These joints capture both quantitative kinematic and kinetic measurements (e.g., joint angles at specific gait events) and qualitative descriptions (e.g., increased trends during stance). During the evaluation, each expert not only scored the reports based on 3 orthogonal indices but also provided qualitative comments on the descriptions generated by the various methods. Given the time-consuming nature of this process (experts were required to jointly analyze the preprocessed motion signals and the visualization plots to evaluate the quality of the analysis report), we assigned different sets of reports to different experts. This distribution strategy was employed to maximize the total number of evaluated reports, thereby ensuring statistical significance.
Fig.~\ref{Fig16} shows the average scores of all comparison methods on the 3 orthogonal indices.


\emph{MotionDecipher} received the highest scores from medical experts, with mean ratings exceeding 3.4 across all three indices (Correctness, Conciseness, and Completeness) for each of the three body parts (Knee, Hip, and Ankle). The \textit{Rule-based method} demonstrates inferior overall performance compared with \emph{MotionDecipher} (with mean scores ranging from 3.35 to 3.74); however, it shows a significant performance advantage over the baseline LLM (Kimi-K2). Expert feedback further indicates that Kimi-K2 performs very poorly, producing reports with almost entirely incorrect numerical information and qualitative descriptions, and generating partially correct clinical interpretations only in very rare cases.

\begin{figure}[htbp]
\centering
\includegraphics[width=0.99\textwidth]{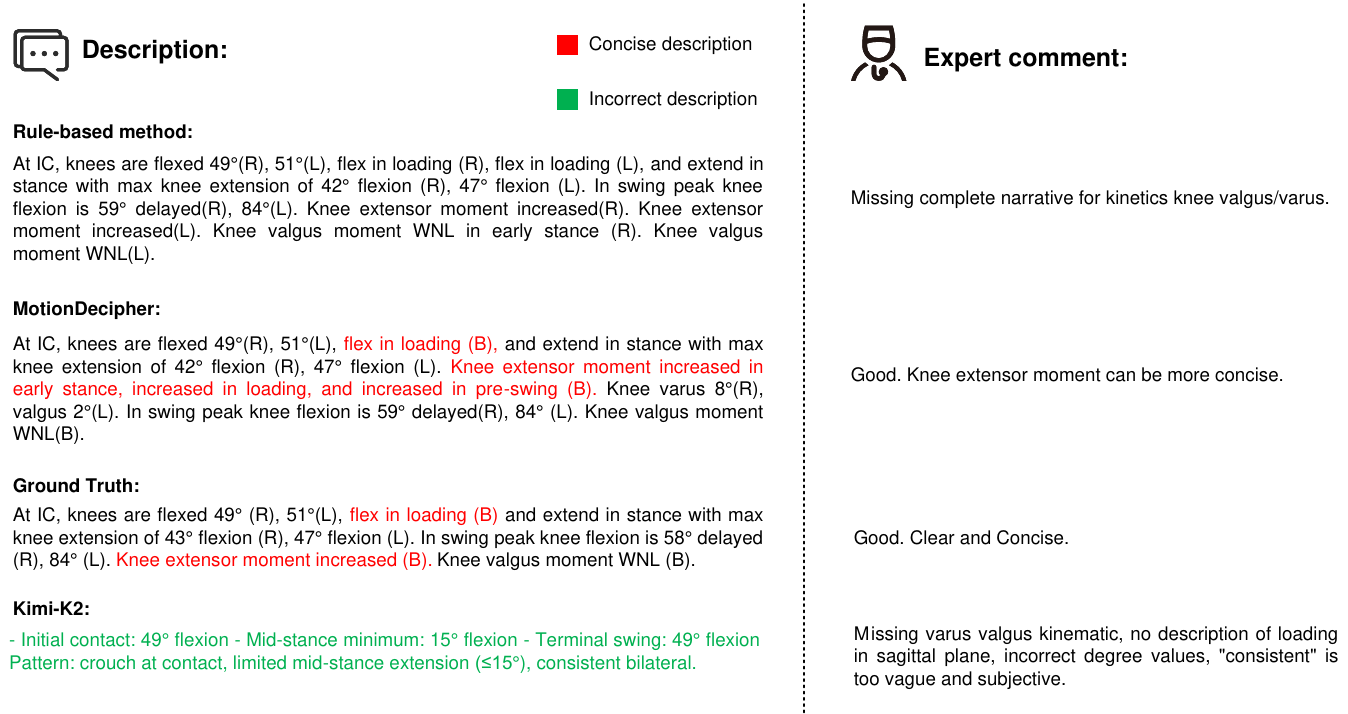}
\caption{A specific example in which \emph{MotionDecipher} outperforms the ground truth description. \emph{MotionDecipher} is the only method that contains knee valgus/varus information. }
\label{Fig17}
\end{figure}

Analysis of expert feedback reveals that occasionally, \emph{MotionDecipher} may capture clinically relevant information (e.g., knee valgus/varus) that is not documented in the ground truth reports, offering opportunities for further clinical review, as illustrated in Fig.~\ref{Fig17}. \emph{MotionDecipher} includes additional information on knee valgus/varus. Compared with the rule-based method, the ground truth description exhibits more concise language expression (red text). The Kimi-K2 baseline performs the worst: it not only generates incorrect numerical information but also lacks feature descriptions across different gait stages, with wording that is highly vague and overly subjective.

Taken together, the results in Fig.~\ref{Fig16} and the example in Fig.~\ref{Fig17} show that \emph{MotionDecipher} exhibits competitive overall performance and captures critical gait features (e.g., knee valgus/varus) overlooked by medical experts, underscoring its potential to assist medical experts in clinical practice.

\subsection{Error Analysis}

Although \emph{MotionDecipher} has demonstrated significant performance advantages over the comparative baselines, to further investigate the errors that remain in the generated analysis reports and the performance gap between \emph{MotionDecipher} and the ground truth reports, we invited experts to annotate the issues present in each description during the scoring process. The results indicate that the existing errors can be categorized into the following 2 types:

\begin{enumerate}
    \item Type 1: Missing analytical statements;
    \item Type 2: Imprecise Phrasing.
\end{enumerate}

\begin{figure}[htbp]
\centering
\includegraphics[width=0.99\textwidth]{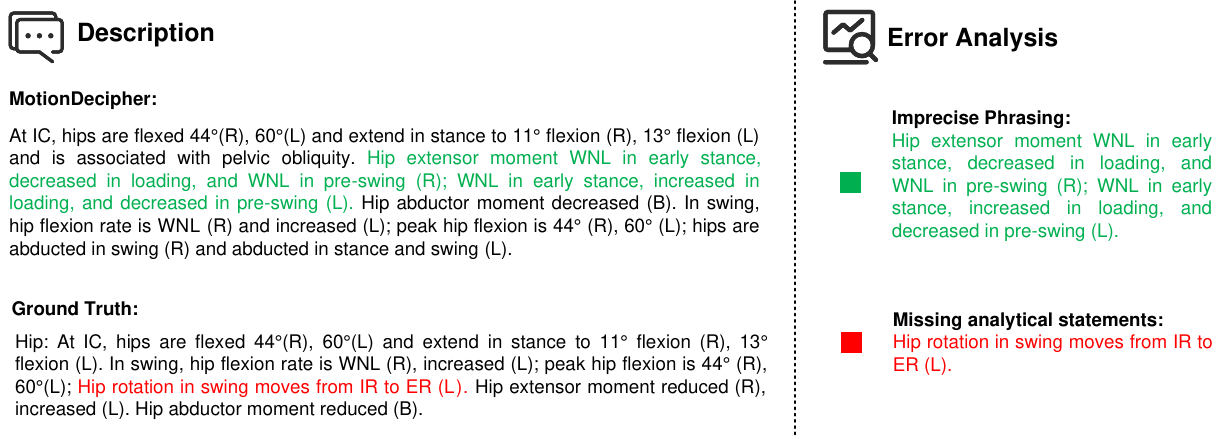}
\caption{A concrete example containing both types of errors. The green text denotes the "Imprecise Phrasing" errors present in \emph{MotionDecipher}'s output, while the red text denotes the analytical statements that are missing from \emph{MotionDecipher}'s output but present in the ground truth report.}
\label{Fig18}
\end{figure}

For both types of errors, one typical example is shown in Fig.~\ref{Fig18}. 

Fig.~\ref{Fig18} provides an intuitive example of the two types of errors that arise when \emph{MotionDecipher} generates reports. The green parts correspond to "Imprecise Phrasing" errors. Compared with the ground truth statement regarding hip extensor moment ("Hip extensor moment reduced (R), increased (L)"), \emph{MotionDecipher} describes it separately according to different gait phases (the green parts) but lacks an analysis of the overall status across the entire gait cycle. The red part of the ground truth, which describes the changing trend of hip rotation during the swing phase, represents an analytical statement that goes beyond direct feature analysis. However, \emph{MotionDecipher} lacks such information. The root cause of both types of errors is \emph{MotionDecipher}'s lack of secondary, in-depth analysis of the available information; conducting such an analysis is precisely the future research objective of our work (see Section~\ref{4}). It is worth noting that, in the example presented in Fig.~\ref{Fig18}, \emph{MotionDecipher} maintains a high degree of consistency with the ground truth in both qualitative and quantitative statements. This is because gait analysis report generation is a high-fidelity scenario, in which we tend to prioritize information accuracy over completeness to prevent potential medical incidents caused by error propagation within the clinical workflow.

Statistical analysis shows that, among the 60 clinical descriptions generated by \emph{MotionDecipher}, 4 exhibit one or both of the two aforementioned types of errors, corresponding to an error rate of 6.7\%. This indirectly supports that \emph{MotionDecipher} can generate accurate gait analysis reports in most cases.

\section{Discussion and Future Work} \label{4}

\emph{MotionDecipher} demonstrates strong potential for human-AI collaboration in clinical gait analysis. Our experiments (both quantitative metrics and medical expert evaluation) show that its generated reports achieve performance very close to expert-written ground truth and significantly outperform the remaining baseline LLMs. This suggests that \emph{MotionDecipher} can help clinicians rapidly obtain structured and interpretable gait assessments. In practice, this efficiency gain is substantial: manual data analysis and report drafting typically require approximately 4-6hr/patient, whereas \emph{MotionDecipher} produces a complete gait analysis report in about 2min/patient. This reduction in reporting time significantly lowers clinician workload and enables more consistent and scalable gait evaluation in routine clinical settings.

\textit{Adaptability of Fixed Hyperparameters to Heterogeneous Populations.} \emph{MotionDecipher} employs fixed hyperparameters across all participants: the SG operator (window size 5, polynomial order 2) in \textit{Tendency Encoding}; the quantization step \(\theta = 0.5\) in \textit{Comparative Encoding}; the db4 wavelet with decomposition level 3 in \textit{Frequency Encoding}, and the scaling factor \(\alpha\) and Top-K in the \textit{Hybrid Index} algorithm. Several properties make these fixed settings adaptable to heterogeneous populations. First, all hyperparameters parameterize deterministic signal-processing operations, which cannot overfit the demographic distribution of any specific cohort. Second, the time-domain parameters are defined relative to the normalized gait cycle (100 timesteps): the 5-frame SG window corresponds to 5\% of a gait cycle and the level-3 DWT isolates stride-frequency-related bands, making both invariant to inter-group differences in cadence, height, and age. The quantization step \(\theta\) preserves clinically meaningful deviations in any subgroup while suppressing subclinical fluctuations. Furthermore, \emph{MotionDecipher} maintains stable, low-variance report quality across different age groups, BMI groups, and GDI Score (Section~\ref{3.4}) under the fixed parameters, demonstrating the adaptability of \emph{MotionDecipher} to heterogeneous populations. 

Despite these efficiency gains, several limitations remain. While \emph{MotionDecipher} effectively bridges the gap between human motion signals and clinical descriptions, its current scope is primarily restricted to interpreting existing gait patterns without inferring their long-term clinical implications. Furthermore, our current deployment relies on marker-based motion capture and force plates in an accredited gait laboratory. We want to clarify that this lab-first strategy is a stepping stone toward ubiquitous deployment rather than the target setting of \emph{MotionDecipher}: developing and rigorously validating a clinical analysis and interpretation framework requires high-fidelity motion measurements and expert-written reference reports. Importantly, the core contribution of \emph{MotionDecipher} lies in the interpretation layer rather than the sensing hardware, as all of its components (\textit{Human Motion Tokenizer}, \textit{Meta Feature Selector}, and \textit{Hybrid Context Constructor}) operate on normalized motion signals and are thus decoupled from specific sensing modalities. Therefore, our validation proves the core interpretation capability of \emph{MotionDecipher}, making it readily adaptable to wearable and mobile sensors, as outlined in our future work below.

\textit{Equipment Cost and Deployment Feasibility in Resource-Limited Settings.} A practical consideration for clinical translation is the cost of the sensing infrastructure. Our current validation relies on an accredited gait laboratory, including a multi-camera optical motion capture system (10 Kestral Digital Cameras from Motion Analysis Corporation), force plates (FP4060-10 Force Plate from Bertec Corporation), and licensed processing software, together with dedicated space, periodic calibration, and a trained multi-person team (Section~\ref{3.1}). The cost of the motion capture system is \$100k, and the cost of the force plate is \$36k. The total cost, including both hardware devices and software tools, is approximately \$150k. Such facilities are typically available in specialized clinical centers, and the associated personnel costs limit the accessibility of instrumented gait analysis in resource-limited settings. We emphasize that this constraint stems from the current sensing platform rather than from \emph{MotionDecipher} itself. As discussed above, all components of \emph{MotionDecipher} operate on normalized motion signals and are decoupled from the specific hardware. Therefore, a feasible deployment pathway for resource-limited settings is to pair the interpretation layer of \emph{MotionDecipher} with low-cost sensing front-ends: single-camera markerless video pose estimation \cite{kidzinski2020deep} and wearable IMUs \cite{RAPP2021110229} can provide kinematic information after standard gait-event segmentation and gait-cycle normalization. We have initiated a follow-up effort that replaces the laboratory motion capture system with single-camera markerless video. This ongoing work directly targets the deployment scenario of resource-limited clinics, where the sensing infrastructure can be reduced from a \$150k laboratory setup to a single camera and standard computing hardware.

These limitations highlight several promising directions for future work. First, \emph{MotionDecipher} currently focuses on generating gait analysis reports (descriptive factual interpretation) and is not yet capable of producing personalized treatment recommendations. In real clinical workflows, such recommendations are essential for guiding patient care and rehabilitation. As such, a key direction of future work is to explore the recommendation generation capability of LLMs on the basis of gait analysis reports. Second, we plan to extend \emph{MotionDecipher} from laboratory motion capture to wearable and mobile sensing modalities, including wearable IMUs and smartphone-based markerless video pose estimation. Since its interpretation layer operates on normalized motion signals rather than sensor-specific raw streams, this extension will reuse the framework validated in this work and focus on the sensing front-end. Concretely, we plan to (1) integrate IMU- and video-derived kinematics into the \textit{Human Motion Tokenizer} after standard gait-event segmentation and gait-cycle normalization, and (2) systematically evaluate the accuracy and robustness of the generated reports under such low-cost inputs across diverse clinical and free-living settings, quantifying the trade-off between sensing fidelity and report quality. We believe this extension will move \emph{MotionDecipher} toward ubiquitous and scalable gait assessment beyond the gait laboratory.

\section{Related Work} \label{5}

\textbf{Clinical Gait Analysis and Automated Gait Reporting.}
Clinical gait analysis traditionally relies on expert interpretation of kinematic and kinetic measurements, where clinicians synthesize 3D joint trajectories, angles, and spatiotemporal parameters into structured clinical reports \cite{GaitReview1}. Prior computational research has focused primarily on gait classification, abnormality detection, and person identification using wearable sensors or motion-capture data \cite{GaitAbnormalityDetection, GaitIdentification1, WirelessSensingGait}. Although these methods demonstrate strong modeling capabilities, no existing work addresses the generation of clinical gait analysis reports, which require interpreting complex motion features into precise, domain-specific language.

\vspace{0.5em}
\noindent \textbf{LLM-Assisted Biomechanical Analysis.}
The biomechanics community has recently begun coupling LLMs with human movement data. BiomechGPT \cite{biomechgpt} fine-tunes an LLM on tokenized biomechanical trajectories to answer clinically relevant questions, such as detecting gait impairments, inferring their etiologies, and estimating walking parameters. BiomechAgent \cite{biomechagent} equips an LLM with code-generation and database tools for interactive biomechanical analysis and visualization. Notably, BiomechAgent's finding that domain-specific tools substantially improve the base agent supports our observation (Section~\ref{3.2}) that general-purpose LLMs require domain-specific scaffolding to reason reliably over motion data. However, both systems produce short-form outputs (classification labels, scalar estimates, or visualizations) based on markerless-derived kinematics, rather than standardized clinical documentation. In contrast, \emph{MotionDecipher} generates clinical gait analysis reports grounded in gold-standard kinematics and kinetic measurements. 

\vspace{0.5em}
\noindent \textbf{LLMs for Sensor and Time-Series Interpretation.}
Recent studies have begun exploring how LLMs can interface with sensor modalities \cite{MentalHealthAI}. Sensor-language alignment frameworks \cite{SensorLM} and multimodal time-series methods \cite{TimeGPT, UniTime} show that LLMs can interpret or describe physiological and wearable signals when paired with high-quality textual supervision. Similar trends appear in mobile and ubiquitous computing, where LLMs are used to infer human states from IMU, inertial, and environmental sensors \cite{OneModeltoFit, LeveragingLLMs}. However, these systems are typically designed for question answering, captioning, or state recognition, not for generating structured, clinically meaningful gait analysis reports.

\vspace{0.5em}
\noindent \textbf{Structure-Aware Language Models and Domain-Specific In-Context Learning.}
Recent advances in ICL demonstrate that structured knowledge, retrieval-based prompting, and domain-guided demonstrations can significantly enhance the reasoning ability of LLMs \cite{StructGPT, GraphPrompt}. Retrieval-augmented prompting methods \cite{SelfadaptiveICL} show that selecting domain-relevant examples is crucial for generating accurate outputs in specialized fields. In clinical NLP, models such as ClinicalGPT \cite{ClinicalGPT} and MedPrompt \cite{MedPrompt} leverage curated corpora or structured clinical abstractions to guide medical text generation. Nevertheless, no prior work introduces a structured gait-specific knowledge corpus to support ICL for gait and motion analysis tasks.

\section{Conclusion} \label{6}

We present \emph{MotionDecipher}, a pioneering framework that bridges the semantic gap between continuous motion signals and clinical interpretation to generate gait analysis reports. By jointly designing and integrating three core components (\textit{Human Motion Tokenizer}, \textit{Meta Feature Selector}, \textit{Hybrid Context Constructor}), \emph{MotionDecipher} outperforms several baseline LLMs in both quantitative metrics (34\% improvement over Google Gemini 3 in the \textit{METEOR} metric) and medical expert evaluation indexes (71\%-221\% average improvement over Kimi-K2). By embedding clinical expertise in LLMs, \emph{MotionDecipher} enables efficient gait and motion analysis in clinics, unlocking the potential for human-AI collaboration in the future of healthcare.

\begin{acks}
We sincerely thank the medical experts from [Stanford Medicine Children's Health] for their support in the evaluation of gait analysis reports.
\end{acks}

\bibliographystyle{ACM-Reference-Format}
\bibliography{reference}


\begin{thebibliography}{68}


\ifx \showCODEN    \undefined \def \showCODEN     #1{\unskip}     \fi
\ifx \showISBNx    \undefined \def \showISBNx     #1{\unskip}     \fi
\ifx \showISBNxiii \undefined \def \showISBNxiii  #1{\unskip}     \fi
\ifx \showISSN     \undefined \def \showISSN      #1{\unskip}     \fi
\ifx \showLCCN     \undefined \def \showLCCN      #1{\unskip}     \fi
\ifx \shownote     \undefined \def \shownote      #1{#1}          \fi
\ifx \showarticletitle \undefined \def \showarticletitle #1{#1}   \fi
\ifx \showURL      \undefined \def \showURL       {\relax}        \fi
\providecommand\bibfield[2]{#2}
\providecommand\bibinfo[2]{#2}
\providecommand\natexlab[1]{#1}
\providecommand\showeprint[2][]{arXiv:#2}

\bibitem[Aoyama et~al\mbox{.}(2023)]%
        {CorpusDownstream}
\bibfield{author}{\bibinfo{person}{Tatsuya Aoyama}, \bibinfo{person}{Shabnam
  Behzad}, \bibinfo{person}{Luke Gessler}, \bibinfo{person}{Lauren Levine},
  \bibinfo{person}{Jessica Lin}, \bibinfo{person}{Yang~Janet Liu},
  \bibinfo{person}{Siyao Peng}, \bibinfo{person}{Yilun Zhu}, {and}
  \bibinfo{person}{Amir Zeldes}.} \bibinfo{year}{2023}\natexlab{}.
\newblock \showarticletitle{{GENTLE}: A Genre-Diverse Multilayer Challenge Set
  for {E}nglish {NLP} and Linguistic Evaluation}. In
  \bibinfo{booktitle}{\emph{Proceedings of the 17th Linguistic Annotation
  Workshop (LAW-XVII)}}, \bibfield{editor}{\bibinfo{person}{Jakob Prange} {and}
  \bibinfo{person}{Annemarie Friedrich}} (Eds.).
  \bibinfo{publisher}{Association for Computational Linguistics},
  \bibinfo{address}{Toronto, Canada}, \bibinfo{pages}{166--178}.
\newblock
\href{https://doi.org/10.18653/v1/2023.law-1.17}{doi:\nolinkurl{10.18653/v1/2023.law-1.17}}


\bibitem[Banerjee and Lavie(2005)]%
        {METEOR}
\bibfield{author}{\bibinfo{person}{Satanjeev Banerjee} {and}
  \bibinfo{person}{Alon Lavie}.} \bibinfo{year}{2005}\natexlab{}.
\newblock \showarticletitle{{METEOR}: An Automatic Metric for {MT} Evaluation
  with Improved Correlation with Human Judgments}. In
  \bibinfo{booktitle}{\emph{Proceedings of the {ACL} Workshop on Intrinsic and
  Extrinsic Evaluation Measures for Machine Translation and/or Summarization}},
  \bibfield{editor}{\bibinfo{person}{Jade Goldstein}, \bibinfo{person}{Alon
  Lavie}, \bibinfo{person}{Chin-Yew Lin}, {and} \bibinfo{person}{Clare Voss}}
  (Eds.). \bibinfo{publisher}{Association for Computational Linguistics},
  \bibinfo{address}{Ann Arbor, Michigan}, \bibinfo{pages}{65--72}.
\newblock
\urldef\tempurl%
\url{https://aclanthology.org/W05-0909/}
\showURL{%
\tempurl}


\bibitem[Bennett et~al\mbox{.}(2018)]%
        {Referencebound1}
\bibfield{author}{\bibinfo{person}{Hunter~J. Bennett},
  \bibinfo{person}{Kristina Fleenor}, {and} \bibinfo{person}{Joshua~T.
  Weinhandl}.} \bibinfo{year}{2018}\natexlab{}.
\newblock \showarticletitle{A normative database of hip and knee joint
  biomechanics during dynamic tasks using anatomical regression prediction
  methods}.
\newblock \bibinfo{journal}{\emph{Journal of Biomechanics}}
  \bibinfo{volume}{81} (\bibinfo{year}{2018}), \bibinfo{pages}{122--131}.
\newblock
\showISSN{0021-9290}
\href{https://doi.org/10.1016/j.jbiomech.2018.10.003}{doi:\nolinkurl{10.1016/j.jbiomech.2018.10.003}}


\bibitem[Buckley et~al\mbox{.}(2019)]%
        {brainsci9020034}
\bibfield{author}{\bibinfo{person}{Christopher Buckley}, \bibinfo{person}{Lisa
  Alcock}, \bibinfo{person}{Ríona McArdle}, \bibinfo{person}{Rana Zia~Ur
  Rehman}, \bibinfo{person}{Silvia Del~Din}, \bibinfo{person}{Claudia Mazzà},
  \bibinfo{person}{Alison~J. Yarnall}, {and} \bibinfo{person}{Lynn Rochester}.}
  \bibinfo{year}{2019}\natexlab{}.
\newblock \showarticletitle{The Role of Movement Analysis in Diagnosing and
  Monitoring Neurodegenerative Conditions: Insights from Gait and Postural
  Control}.
\newblock \bibinfo{journal}{\emph{Brain Sciences}} \bibinfo{volume}{9},
  \bibinfo{number}{2} (\bibinfo{year}{2019}).
\newblock
\showISSN{2076-3425}
\href{https://doi.org/10.3390/brainsci9020034}{doi:\nolinkurl{10.3390/brainsci9020034}}


\bibitem[Chen et~al\mbox{.}(2024)]%
        {MedPrompt}
\bibfield{author}{\bibinfo{person}{Xuhang Chen}, \bibinfo{person}{Shenghong
  Luo}, \bibinfo{person}{Chi{-}Man Pun}, {and} \bibinfo{person}{Shuqiang
  Wang}.} \bibinfo{year}{2024}\natexlab{}.
\newblock \showarticletitle{MedPrompt: Cross-modal Prompting for Multi-task
  Medical Image Translation}. In \bibinfo{booktitle}{\emph{Pattern Recognition
  and Computer Vision - 7th Chinese Conference, {PRCV} 2024, Urumqi, China,
  October 18-20, 2024, Proceedings, Part {XIV}}}
  \emph{(\bibinfo{series}{Lecture Notes in Computer Science},
  Vol.~\bibinfo{volume}{15044})}, \bibfield{editor}{\bibinfo{person}{Zhouchen
  Lin}, \bibinfo{person}{Ming{-}Ming Cheng}, \bibinfo{person}{Ran He},
  \bibinfo{person}{Kurban Ubul}, \bibinfo{person}{Wushouer Silamu},
  \bibinfo{person}{Hongbin Zha}, \bibinfo{person}{Jie Zhou}, {and}
  \bibinfo{person}{Cheng{-}Lin Liu}} (Eds.). \bibinfo{publisher}{Springer},
  \bibinfo{pages}{61--75}.
\newblock
\href{https://doi.org/10.1007/978-981-97-8496-7\_5}{doi:\nolinkurl{10.1007/978-981-97-8496-7\_5}}


\bibitem[Choi et~al\mbox{.}(2024)]%
        {EarlyPrediction}
\bibfield{author}{\bibinfo{person}{Yoonseon Choi}, \bibinfo{person}{Dayoung
  Jeong}, \bibinfo{person}{Bogoan Kim}, {and} \bibinfo{person}{Kyungsik Han}.}
  \bibinfo{year}{2024}\natexlab{}.
\newblock \showarticletitle{Early Prediction of Cybersickness in Virtual
  Reality Using a Large Language Model for Multimodal Time Series Data}. In
  \bibinfo{booktitle}{\emph{Companion of the 2024 on ACM International Joint
  Conference on Pervasive and Ubiquitous Computing}} (Melbourne VIC, Australia)
  \emph{(\bibinfo{series}{UbiComp '24})}. \bibinfo{publisher}{Association for
  Computing Machinery}, \bibinfo{address}{New York, NY, USA},
  \bibinfo{pages}{25–29}.
\newblock
\showISBNx{9798400710582}
\href{https://doi.org/10.1145/3675094.3677578}{doi:\nolinkurl{10.1145/3675094.3677578}}


\bibitem[Clewes et~al\mbox{.}(2025)]%
        {ClinicalGENI}
\bibfield{author}{\bibinfo{person}{Kylie Clewes}, \bibinfo{person}{Yiwen Dong},
  \bibinfo{person}{Mary Meyer}, \bibinfo{person}{Evan Lowe},
  \bibinfo{person}{Kornél Schadl}, {and} \bibinfo{person}{Jessica Rose}.}
  \bibinfo{year}{2025}\natexlab{}.
\newblock \showarticletitle{Clinical Gait Evaluation with Neuromuscular
  Impairments (Clinical GENI) for spastic cerebral palsy}.
\newblock \bibinfo{journal}{\emph{Frontiers in Human Neuroscience}}
  \bibinfo{volume}{Volume 19 - 2025} (\bibinfo{year}{2025}).
\newblock
\showISSN{1662-5161}
\href{https://doi.org/10.3389/fnhum.2025.1637164}{doi:\nolinkurl{10.3389/fnhum.2025.1637164}}


\bibitem[Collins et~al\mbox{.}(2009)]%
        {COLLINS2009173}
\bibfield{author}{\bibinfo{person}{Thomas~D. Collins},
  \bibinfo{person}{Salim~N. Ghoussayni}, \bibinfo{person}{David~J. Ewins},
  {and} \bibinfo{person}{Jenny~A. Kent}.} \bibinfo{year}{2009}\natexlab{}.
\newblock \showarticletitle{A six degrees-of-freedom marker set for gait
  analysis: Repeatability and comparison with a modified Helen Hayes set}.
\newblock \bibinfo{journal}{\emph{Gait \& Posture}} \bibinfo{volume}{30},
  \bibinfo{number}{2} (\bibinfo{year}{2009}), \bibinfo{pages}{173--180}.
\newblock
\showISSN{0966-6362}
\href{https://doi.org/10.1016/j.gaitpost.2009.04.004}{doi:\nolinkurl{10.1016/j.gaitpost.2009.04.004}}


\bibitem[Cotton and Leonard(2026)]%
        {biomechagent}
\bibfield{author}{\bibinfo{person}{R.~James Cotton} {and}
  \bibinfo{person}{Thomas Leonard}.} \bibinfo{year}{2026}\natexlab{}.
\newblock \bibinfo{title}{BiomechAgent: AI-Assisted Biomechanical Analysis
  Through Code-Generating Agents}.
\newblock
\showeprint[arxiv]{2602.06975}~[cs.CL]
\urldef\tempurl%
\url{https://arxiv.org/abs/2602.06975}
\showURL{%
\tempurl}


\bibitem[Fan et~al\mbox{.}(2023)]%
        {Outcomeprediction}
\bibfield{author}{\bibinfo{person}{Shaoyi Fan}, \bibinfo{person}{Jieshun Ye},
  \bibinfo{person}{Qing Xu}, \bibinfo{person}{Runxin Peng},
  \bibinfo{person}{Bin Hu}, \bibinfo{person}{Zhong Pei},
  \bibinfo{person}{Zhimin Yang}, {and} \bibinfo{person}{Fuping Xu}.}
  \bibinfo{year}{2023}\natexlab{}.
\newblock \showarticletitle{Digital health technology combining wearable gait
  sensors and machine learning improve the accuracy in prediction of frailty}.
\newblock \bibinfo{journal}{\emph{Frontiers in Public Health}}
  \bibinfo{volume}{Volume 11 - 2023} (\bibinfo{year}{2023}).
\newblock
\showISSN{2296-2565}
\href{https://doi.org/10.3389/fpubh.2023.1169083}{doi:\nolinkurl{10.3389/fpubh.2023.1169083}}


\bibitem[Fu et~al\mbox{.}(2025)]%
        {SHADE-AD}
\bibfield{author}{\bibinfo{person}{Heming Fu}, \bibinfo{person}{Hongkai Chen},
  \bibinfo{person}{Shan Lin}, {and} \bibinfo{person}{Guoliang Xing}.}
  \bibinfo{year}{2025}\natexlab{}.
\newblock \showarticletitle{SHADE-AD: An LLM-Based Framework for Synthesizing
  Activity Data of Alzheimer’s Patients}. In
  \bibinfo{booktitle}{\emph{Proceedings of the 23rd ACM Conference on Embedded
  Networked Sensor Systems}} (UC Irvine Student Center., Irvine, CA, USA)
  \emph{(\bibinfo{series}{SenSys '25})}. \bibinfo{publisher}{Association for
  Computing Machinery}, \bibinfo{address}{New York, NY, USA},
  \bibinfo{pages}{290–296}.
\newblock
\showISBNx{9798400714795}
\href{https://doi.org/10.1145/3715014.3722062}{doi:\nolinkurl{10.1145/3715014.3722062}}


\bibitem[Garza and Canseco(2023)]%
        {TimeGPT}
\bibfield{author}{\bibinfo{person}{Azul Garza} {and}
  \bibinfo{person}{Max~Mergenthaler Canseco}.} \bibinfo{year}{2023}\natexlab{}.
\newblock \showarticletitle{TimeGPT-1}.
\newblock \bibinfo{journal}{\emph{CoRR}}  \bibinfo{volume}{abs/2310.03589}
  (\bibinfo{year}{2023}).
\newblock
\showeprint[arXiv]{2310.03589}
\href{https://doi.org/10.48550/ARXIV.2310.03589}{doi:\nolinkurl{10.48550/ARXIV.2310.03589}}


\bibitem[Hulleck et~al\mbox{.}(2022)]%
        {10.3389/fmedt.2022.901331}
\bibfield{author}{\bibinfo{person}{Abdul~Aziz Hulleck}, \bibinfo{person}{Dhanya
  Menoth~Mohan}, \bibinfo{person}{Nada Abdallah}, \bibinfo{person}{Marwan
  El~Rich}, {and} \bibinfo{person}{Kinda Khalaf}.}
  \bibinfo{year}{2022}\natexlab{}.
\newblock \showarticletitle{Present and future of gait assessment in clinical
  practice: Towards the application of novel trends and technologies}.
\newblock \bibinfo{journal}{\emph{Frontiers in Medical Technology}}
  \bibinfo{volume}{Volume 4 - 2022} (\bibinfo{year}{2022}).
\newblock
\showISSN{2673-3129}
\href{https://doi.org/10.3389/fmedt.2022.901331}{doi:\nolinkurl{10.3389/fmedt.2022.901331}}


\bibitem[Hummel et~al\mbox{.}(2024)]%
        {10.1159/000538270}
\bibfield{author}{\bibinfo{person}{Jonas Hummel}, \bibinfo{person}{Michael
  Schwenk}, \bibinfo{person}{Daniel Seebacher}, \bibinfo{person}{Philipp
  Barzyk}, \bibinfo{person}{Joachim Liepert}, {and} \bibinfo{person}{Manuel
  Stein}.} \bibinfo{year}{2024}\natexlab{}.
\newblock \showarticletitle{Clustering Approaches for Gait Analysis within
  Neurological Disorders: A Narrative Review}.
\newblock \bibinfo{journal}{\emph{Digital Biomarkers}} \bibinfo{volume}{8},
  \bibinfo{number}{1} (\bibinfo{date}{05} \bibinfo{year}{2024}),
  \bibinfo{pages}{93--101}.
\newblock
\showISSN{2504-110X}
\showeprint{https://karger.com/dib/article-pdf/8/1/93/4225500/000538270.pdf}
\href{https://doi.org/10.1159/000538270}{doi:\nolinkurl{10.1159/000538270}}


\bibitem[Javaheri et~al\mbox{.}(2024)]%
        {ARAS}
\bibfield{author}{\bibinfo{person}{Hamraz Javaheri}, \bibinfo{person}{Omid
  Ghamarnejad}, \bibinfo{person}{Paul Lukowicz},
  \bibinfo{person}{Gregor~Alexander Stavrou}, {and} \bibinfo{person}{Jakob
  Karolus}.} \bibinfo{year}{2024}\natexlab{}.
\newblock \showarticletitle{ARAS: LLM-Supported Augmented Reality Assistance
  System for Pancreatic Surgery}. In \bibinfo{booktitle}{\emph{Companion of the
  2024 on ACM International Joint Conference on Pervasive and Ubiquitous
  Computing}} (Melbourne VIC, Australia) \emph{(\bibinfo{series}{UbiComp
  '24})}. \bibinfo{publisher}{Association for Computing Machinery},
  \bibinfo{address}{New York, NY, USA}, \bibinfo{pages}{176–180}.
\newblock
\showISBNx{9798400710582}
\href{https://doi.org/10.1145/3675094.3677543}{doi:\nolinkurl{10.1145/3675094.3677543}}


\bibitem[Jiang et~al\mbox{.}(2023)]%
        {StructGPT}
\bibfield{author}{\bibinfo{person}{Jinhao Jiang}, \bibinfo{person}{Kun Zhou},
  \bibinfo{person}{Zican Dong}, \bibinfo{person}{Keming Ye},
  \bibinfo{person}{Xin Zhao}, {and} \bibinfo{person}{Ji{-}Rong Wen}.}
  \bibinfo{year}{2023}\natexlab{}.
\newblock \showarticletitle{StructGPT: {A} General Framework for Large Language
  Model to Reason over Structured Data}. In
  \bibinfo{booktitle}{\emph{Proceedings of the 2023 Conference on Empirical
  Methods in Natural Language Processing, {EMNLP} 2023, Singapore, December
  6-10, 2023}}, \bibfield{editor}{\bibinfo{person}{Houda Bouamor},
  \bibinfo{person}{Juan Pino}, {and} \bibinfo{person}{Kalika Bali}} (Eds.).
  \bibinfo{publisher}{Association for Computational Linguistics},
  \bibinfo{pages}{9237--9251}.
\newblock
\href{https://doi.org/10.18653/V1/2023.EMNLP-MAIN.574}{doi:\nolinkurl{10.18653/V1/2023.EMNLP-MAIN.574}}


\bibitem[Jiang et~al\mbox{.}(2024)]%
        {DBLP:conf/ijcai/JiangP0GSNS24}
\bibfield{author}{\bibinfo{person}{Yushan Jiang}, \bibinfo{person}{Zijie Pan},
  \bibinfo{person}{Xikun Zhang}, \bibinfo{person}{Sahil Garg},
  \bibinfo{person}{Anderson Schneider}, \bibinfo{person}{Yuriy Nevmyvaka},
  {and} \bibinfo{person}{Dongjin Song}.} \bibinfo{year}{2024}\natexlab{}.
\newblock \showarticletitle{Empowering Time Series Analysis with Large Language
  Models: {A} Survey}. In \bibinfo{booktitle}{\emph{Proceedings of the
  Thirty-Third International Joint Conference on Artificial Intelligence,
  {IJCAI} 2024, Jeju, South Korea, August 3-9, 2024}}.
  \bibinfo{publisher}{ijcai.org}, \bibinfo{pages}{8095--8103}.
\newblock
\urldef\tempurl%
\url{https://www.ijcai.org/proceedings/2024/895}
\showURL{%
\tempurl}


\bibitem[John et~al\mbox{.}(2021)]%
        {Savitzky-Golay}
\bibfield{author}{\bibinfo{person}{Arlene John}, \bibinfo{person}{Jishnu
  Sadasivan}, {and} \bibinfo{person}{Chandra~Sekhar Seelamantula}.}
  \bibinfo{year}{2021}\natexlab{}.
\newblock \showarticletitle{Adaptive Savitzky-Golay Filtering in Non-Gaussian
  Noise}.
\newblock \bibinfo{journal}{\emph{IEEE Transactions on Signal Processing}}
  \bibinfo{volume}{69} (\bibinfo{year}{2021}), \bibinfo{pages}{5021--5036}.
\newblock
\href{https://doi.org/10.1109/TSP.2021.3106450}{doi:\nolinkurl{10.1109/TSP.2021.3106450}}


\bibitem[Kaufman and Sutherland(2006)]%
        {Normal2006Kinematics}
\bibfield{author}{\bibinfo{person}{Kenton~R. Kaufman} {and}
  \bibinfo{person}{David~H. Sutherland}.} \bibinfo{year}{2006}\natexlab{}.
\newblock \showarticletitle{Kinematics of Normal Walking}.
\newblock In \bibinfo{booktitle}{\emph{Human Walking} (\bibinfo{edition}{3rd}
  ed.)}, \bibfield{editor}{\bibinfo{person}{Jessica Rose} {and}
  \bibinfo{person}{James~G. Gamble}} (Eds.). \bibinfo{publisher}{Lippincott
  Williams \& Wilkins}, \bibinfo{address}{Philadelphia, PA}, Chapter~3,
  \bibinfo{pages}{55--74}.
\newblock
\showISBNx{978-0-7817-5954-0}


\bibitem[Kidzi{\'n}ski et~al\mbox{.}(2020)]%
        {kidzinski2020deep}
\bibfield{author}{\bibinfo{person}{{\L}ukasz Kidzi{\'n}ski},
  \bibinfo{person}{Bryan Yang}, \bibinfo{person}{Jennifer~L Hicks},
  \bibinfo{person}{Apoorva Rajagopal}, \bibinfo{person}{Scott~L Delp}, {and}
  \bibinfo{person}{Michael~H Schwartz}.} \bibinfo{year}{2020}\natexlab{}.
\newblock \showarticletitle{Deep neural networks enable quantitative movement
  analysis using single-camera videos}.
\newblock \bibinfo{journal}{\emph{Nature communications}} \bibinfo{volume}{11},
  \bibinfo{number}{1} (\bibinfo{year}{2020}), \bibinfo{pages}{4054}.
\newblock
\href{https://doi.org/10.1038/s41467-020-17807-z}{doi:\nolinkurl{10.1038/s41467-020-17807-z}}


\bibitem[Li and Dong(2025)]%
        {10.1145/3722570.3726886}
\bibfield{author}{\bibinfo{person}{Jiaxin Li} {and} \bibinfo{person}{Yiwen
  Dong}.} \bibinfo{year}{2025}\natexlab{}.
\newblock \showarticletitle{Human-Centered Gait Balance Estimation Using
  Footstep-Induced Floor Vibrations}. In \bibinfo{booktitle}{\emph{Proceedings
  of the 3rd International Workshop on Human-Centered Sensing, Modeling, and
  Intelligent Systems}} (Irvine, CA, USA) \emph{(\bibinfo{series}{HumanSys
  '25})}. \bibinfo{publisher}{Association for Computing Machinery},
  \bibinfo{address}{New York, NY, USA}, \bibinfo{pages}{80–85}.
\newblock
\showISBNx{9798400716096}
\href{https://doi.org/10.1145/3722570.3726886}{doi:\nolinkurl{10.1145/3722570.3726886}}


\bibitem[Lin(2004)]%
        {ROUGE}
\bibfield{author}{\bibinfo{person}{Chin-Yew Lin}.}
  \bibinfo{year}{2004}\natexlab{}.
\newblock \showarticletitle{{ROUGE}: A Package for Automatic Evaluation of
  Summaries}. In \bibinfo{booktitle}{\emph{Text Summarization Branches Out}}.
  \bibinfo{publisher}{Association for Computational Linguistics},
  \bibinfo{address}{Barcelona, Spain}, \bibinfo{pages}{74--81}.
\newblock
\urldef\tempurl%
\url{https://aclanthology.org/W04-1013/}
\showURL{%
\tempurl}


\bibitem[Liu et~al\mbox{.}(2025)]%
        {TaskSense}
\bibfield{author}{\bibinfo{person}{Kaiwei Liu}, \bibinfo{person}{Bufang Yang},
  \bibinfo{person}{Lilin Xu}, \bibinfo{person}{Yunqi Guo},
  \bibinfo{person}{Guoliang Xing}, \bibinfo{person}{Xian Shuai},
  \bibinfo{person}{Xiaozhe Ren}, \bibinfo{person}{Xin Jiang}, {and}
  \bibinfo{person}{Zhenyu Yan}.} \bibinfo{year}{2025}\natexlab{}.
\newblock \bibinfo{booktitle}{\emph{TaskSense: A Translation-like Approach for
  Tasking Heterogeneous Sensor Systems with LLMs}}.
\newblock \bibinfo{publisher}{Association for Computing Machinery},
  \bibinfo{address}{New York, NY, USA}, \bibinfo{pages}{213–225}.
\newblock
\showISBNx{9798400714795}
\urldef\tempurl%
\url{https://doi.org/10.1145/3715014.3722070}
\showURL{%
\tempurl}


\bibitem[Liu et~al\mbox{.}(2012)]%
        {6062376}
\bibfield{author}{\bibinfo{person}{Tao Liu}, \bibinfo{person}{Yoshio Inoue},
  \bibinfo{person}{Kyoko Shibata}, {and} \bibinfo{person}{K. Shiojima}.}
  \bibinfo{year}{2012}\natexlab{}.
\newblock \showarticletitle{A Mobile Force Plate and Three-Dimensional Motion
  Analysis System for Three-Dimensional Gait Assessment}.
\newblock \bibinfo{journal}{\emph{IEEE Sensors Journal}} \bibinfo{volume}{12},
  \bibinfo{number}{5} (\bibinfo{year}{2012}), \bibinfo{pages}{1461--1467}.
\newblock
\href{https://doi.org/10.1109/JSEN.2011.2173763}{doi:\nolinkurl{10.1109/JSEN.2011.2173763}}


\bibitem[Liu et~al\mbox{.}(2024)]%
        {UniTime}
\bibfield{author}{\bibinfo{person}{Xu Liu}, \bibinfo{person}{Junfeng Hu},
  \bibinfo{person}{Yuan Li}, \bibinfo{person}{Shizhe Diao},
  \bibinfo{person}{Yuxuan Liang}, \bibinfo{person}{Bryan Hooi}, {and}
  \bibinfo{person}{Roger Zimmermann}.} \bibinfo{year}{2024}\natexlab{}.
\newblock \showarticletitle{UniTime: A Language-Empowered Unified Model for
  Cross-Domain Time Series Forecasting}. In
  \bibinfo{booktitle}{\emph{Proceedings of the ACM Web Conference 2024}}
  (Singapore, Singapore) \emph{(\bibinfo{series}{WWW '24})}.
  \bibinfo{publisher}{Association for Computing Machinery},
  \bibinfo{address}{New York, NY, USA}, \bibinfo{pages}{4095–4106}.
\newblock
\showISBNx{9798400701719}
\href{https://doi.org/10.1145/3589334.3645434}{doi:\nolinkurl{10.1145/3589334.3645434}}


\bibitem[Liu et~al\mbox{.}(2023)]%
        {GraphPrompt}
\bibfield{author}{\bibinfo{person}{Zemin Liu}, \bibinfo{person}{Xingtong Yu},
  \bibinfo{person}{Yuan Fang}, {and} \bibinfo{person}{Xinming Zhang}.}
  \bibinfo{year}{2023}\natexlab{}.
\newblock \showarticletitle{GraphPrompt: Unifying Pre-Training and Downstream
  Tasks for Graph Neural Networks}. In \bibinfo{booktitle}{\emph{Proceedings of
  the ACM Web Conference 2023}} (Austin, TX, USA) \emph{(\bibinfo{series}{WWW
  '23})}. \bibinfo{publisher}{Association for Computing Machinery},
  \bibinfo{address}{New York, NY, USA}, \bibinfo{pages}{417–428}.
\newblock
\showISBNx{9781450394161}
\href{https://doi.org/10.1145/3543507.3583386}{doi:\nolinkurl{10.1145/3543507.3583386}}


\bibitem[Na et~al\mbox{.}(2024)]%
        {na2024kinematic}
\bibfield{author}{\bibinfo{person}{Chuh-Hyoun Na}, \bibinfo{person}{Hannah~Lena
  Siebers}, \bibinfo{person}{Julia Reim}, \bibinfo{person}{J{\"o}rg
  Eschweiler}, \bibinfo{person}{Frank Hildebrand}, \bibinfo{person}{Hans
  Clusmann}, {and} \bibinfo{person}{Marcel Betsch}.}
  \bibinfo{year}{2024}\natexlab{}.
\newblock \showarticletitle{Kinematic movement and balance parameter analysis
  in neurological gait disorders}.
\newblock \bibinfo{journal}{\emph{Journal of Biological Engineering}}
  \bibinfo{volume}{18}, \bibinfo{number}{1} (\bibinfo{year}{2024}),
  \bibinfo{pages}{6}.
\newblock
\href{https://doi.org/10.1186/s13036-023-00398-w}{doi:\nolinkurl{10.1186/s13036-023-00398-w}}


\bibitem[Naveed et~al\mbox{.}(2025)]%
        {10.1145/3744746}
\bibfield{author}{\bibinfo{person}{Humza Naveed}, \bibinfo{person}{Asad~Ullah
  Khan}, \bibinfo{person}{Shi Qiu}, \bibinfo{person}{Muhammad Saqib},
  \bibinfo{person}{Saeed Anwar}, \bibinfo{person}{Muhammad Usman},
  \bibinfo{person}{Naveed Akhtar}, \bibinfo{person}{Nick Barnes}, {and}
  \bibinfo{person}{Ajmal Mian}.} \bibinfo{year}{2025}\natexlab{}.
\newblock \showarticletitle{A Comprehensive Overview of Large Language Models}.
\newblock \bibinfo{journal}{\emph{ACM Trans. Intell. Syst. Technol.}}
  \bibinfo{volume}{16}, \bibinfo{number}{5}, Article \bibinfo{articleno}{106}
  (\bibinfo{date}{Aug.} \bibinfo{year}{2025}), \bibinfo{numpages}{72}~pages.
\newblock
\showISSN{2157-6904}
\href{https://doi.org/10.1145/3744746}{doi:\nolinkurl{10.1145/3744746}}


\bibitem[Orekhov et~al\mbox{.}(2019)]%
        {KinRT}
\bibfield{author}{\bibinfo{person}{Greg Orekhov}, \bibinfo{person}{A.~Matt
  Robinson}, \bibinfo{person}{Scott~J. Hazelwood}, {and}
  \bibinfo{person}{Stephen~M. Klisch}.} \bibinfo{year}{2019}\natexlab{}.
\newblock \showarticletitle{Knee joint biomechanics in transtibial amputees in
  gait, cycling, and elliptical training}.
\newblock \bibinfo{journal}{\emph{PLOS ONE}} \bibinfo{volume}{14},
  \bibinfo{number}{12} (\bibinfo{date}{12} \bibinfo{year}{2019}),
  \bibinfo{pages}{1--16}.
\newblock
\href{https://doi.org/10.1371/journal.pone.0226060}{doi:\nolinkurl{10.1371/journal.pone.0226060}}


\bibitem[Ouyang et~al\mbox{.}(2024)]%
        {ADMarker}
\bibfield{author}{\bibinfo{person}{Xiaomin Ouyang}, \bibinfo{person}{Xian
  Shuai}, \bibinfo{person}{Yang Li}, \bibinfo{person}{Li Pan},
  \bibinfo{person}{Xifan Zhang}, \bibinfo{person}{Heming Fu},
  \bibinfo{person}{Sitong Cheng}, \bibinfo{person}{Xinyan Wang},
  \bibinfo{person}{Shihua Cao}, \bibinfo{person}{Jiang Xin},
  \bibinfo{person}{Hazel Mok}, \bibinfo{person}{Zhenyu Yan},
  \bibinfo{person}{Doris Sau~Fung Yu}, \bibinfo{person}{Timothy Kwok}, {and}
  \bibinfo{person}{Guoliang Xing}.} \bibinfo{year}{2024}\natexlab{}.
\newblock \showarticletitle{ADMarker: A Multi-Modal Federated Learning System
  for Monitoring Digital Biomarkers of Alzheimer's Disease}. In
  \bibinfo{booktitle}{\emph{Proceedings of the 30th Annual International
  Conference on Mobile Computing and Networking}} (Washington D.C., DC, USA)
  \emph{(\bibinfo{series}{ACM MobiCom '24})}. \bibinfo{publisher}{Association
  for Computing Machinery}, \bibinfo{address}{New York, NY, USA},
  \bibinfo{pages}{404–419}.
\newblock
\showISBNx{9798400704895}
\href{https://doi.org/10.1145/3636534.3649370}{doi:\nolinkurl{10.1145/3636534.3649370}}


\bibitem[Ouyang et~al\mbox{.}(2025)]%
        {MMBind}
\bibfield{author}{\bibinfo{person}{Xiaomin Ouyang}, \bibinfo{person}{Jason Wu},
  \bibinfo{person}{Tomoyoshi Kimura}, \bibinfo{person}{Yihan Lin},
  \bibinfo{person}{Gunjan Verma}, \bibinfo{person}{Tarek Abdelzaher}, {and}
  \bibinfo{person}{Mani Srivastava}.} \bibinfo{year}{2025}\natexlab{}.
\newblock \bibinfo{booktitle}{\emph{MMBind: Unleashing the Potential of
  Distributed and Heterogeneous Data for Multimodal Learning in IoT}}.
\newblock \bibinfo{publisher}{Association for Computing Machinery},
  \bibinfo{address}{New York, NY, USA}, \bibinfo{pages}{491–503}.
\newblock
\showISBNx{9798400714795}
\urldef\tempurl%
\url{https://doi.org/10.1145/3715014.3722053}
\showURL{%
\tempurl}


\bibitem[Papagiannis et~al\mbox{.}(2019)]%
        {Papagiannis02012019}
\bibfield{author}{\bibinfo{person}{Georgios~I. Papagiannis},
  \bibinfo{person}{Athanasios~I. Triantafyllou}, \bibinfo{person}{Ilias~M.
  Roumpelakis}, \bibinfo{person}{Frantzeska Zampeli},
  \bibinfo{person}{Pinioti~Garyfallia Eleni}, \bibinfo{person}{Panayiotis
  Koulouvaris}, \bibinfo{person}{Elias~C. Papadopoulos},
  \bibinfo{person}{Panayiotis~J. Papagelopoulos}, {and}
  \bibinfo{person}{George~C. Babis}.} \bibinfo{year}{2019}\natexlab{}.
\newblock \showarticletitle{Methodology of surface electromyography in gait
  analysis: review of the literature}.
\newblock \bibinfo{journal}{\emph{Journal of Medical Engineering \&
  Technology}} \bibinfo{volume}{43}, \bibinfo{number}{1}
  (\bibinfo{year}{2019}), \bibinfo{pages}{59--65}.
\newblock
\showeprint{https://doi.org/10.1080/03091902.2019.1609610}
\href{https://doi.org/10.1080/03091902.2019.1609610}{doi:\nolinkurl{10.1080/03091902.2019.1609610}}
\newblock
\shownote{PMID: 31074312}.


\bibitem[Papineni et~al\mbox{.}(2002)]%
        {BLEU}
\bibfield{author}{\bibinfo{person}{Kishore Papineni}, \bibinfo{person}{Salim
  Roukos}, \bibinfo{person}{Todd Ward}, {and} \bibinfo{person}{Wei-Jing Zhu}.}
  \bibinfo{year}{2002}\natexlab{}.
\newblock \showarticletitle{BLEU: a method for automatic evaluation of machine
  translation}. In \bibinfo{booktitle}{\emph{Proceedings of the 40th Annual
  Meeting on Association for Computational Linguistics}} (Philadelphia,
  Pennsylvania) \emph{(\bibinfo{series}{ACL '02})}.
  \bibinfo{publisher}{Association for Computational Linguistics},
  \bibinfo{address}{USA}, \bibinfo{pages}{311–318}.
\newblock
\href{https://doi.org/10.3115/1073083.1073135}{doi:\nolinkurl{10.3115/1073083.1073135}}


\bibitem[Pfister et~al\mbox{.}(2014)]%
        {Pfister01072014}
\bibfield{author}{\bibinfo{person}{Alexandra Pfister},
  \bibinfo{person}{Alexandre~M. West}, \bibinfo{person}{Shaw Bronner}, {and}
  \bibinfo{person}{Jack~Adam Noah}.} \bibinfo{year}{2014}\natexlab{}.
\newblock \showarticletitle{Comparative abilities of Microsoft Kinect and Vicon
  3D motion capture for gait analysis}.
\newblock \bibinfo{journal}{\emph{Journal of Medical Engineering \&
  Technology}} \bibinfo{volume}{38}, \bibinfo{number}{5}
  (\bibinfo{year}{2014}), \bibinfo{pages}{274--280}.
\newblock
\showeprint{https://doi.org/10.3109/03091902.2014.909540}
\href{https://doi.org/10.3109/03091902.2014.909540}{doi:\nolinkurl{10.3109/03091902.2014.909540}}
\newblock
\shownote{PMID: 24878252}.


\bibitem[Potluri et~al\mbox{.}(2019)]%
        {GaitAbnormalityDetection}
\bibfield{author}{\bibinfo{person}{Sasanka Potluri}, \bibinfo{person}{Srinivas
  Ravuri}, \bibinfo{person}{Christian Diedrich}, {and} \bibinfo{person}{Lutz
  Schega}.} \bibinfo{year}{2019}\natexlab{}.
\newblock \showarticletitle{Deep Learning based Gait Abnormality Detection
  using Wearable Sensor System}. In \bibinfo{booktitle}{\emph{2019 41st Annual
  International Conference of the IEEE Engineering in Medicine and Biology
  Society (EMBC)}}. \bibinfo{pages}{3613--3619}.
\newblock
\href{https://doi.org/10.1109/EMBC.2019.8856454}{doi:\nolinkurl{10.1109/EMBC.2019.8856454}}


\bibitem[Rapp et~al\mbox{.}(2021)]%
        {RAPP2021110229}
\bibfield{author}{\bibinfo{person}{Eric Rapp}, \bibinfo{person}{Soyong Shin},
  \bibinfo{person}{Wolf Thomsen}, \bibinfo{person}{Reed Ferber}, {and}
  \bibinfo{person}{Eni Halilaj}.} \bibinfo{year}{2021}\natexlab{}.
\newblock \showarticletitle{Estimation of kinematics from inertial measurement
  units using a combined deep learning and optimization framework}.
\newblock \bibinfo{journal}{\emph{Journal of Biomechanics}}
  \bibinfo{volume}{116} (\bibinfo{year}{2021}), \bibinfo{pages}{110229}.
\newblock
\showISSN{0021-9290}
\href{https://doi.org/10.1016/j.jbiomech.2021.110229}{doi:\nolinkurl{10.1016/j.jbiomech.2021.110229}}


\bibitem[Rutherford and Hubley-Kozey(2009)]%
        {Referencebound2}
\bibfield{author}{\bibinfo{person}{Derek~James Rutherford} {and}
  \bibinfo{person}{Cheryl Hubley-Kozey}.} \bibinfo{year}{2009}\natexlab{}.
\newblock \showarticletitle{Explaining the hip adduction moment variability
  during gait: Implications for hip abductor strengthening}.
\newblock \bibinfo{journal}{\emph{Clinical Biomechanics}} \bibinfo{volume}{24},
  \bibinfo{number}{3} (\bibinfo{year}{2009}), \bibinfo{pages}{267--273}.
\newblock
\showISSN{0268-0033}
\href{https://doi.org/10.1016/j.clinbiomech.2008.12.006}{doi:\nolinkurl{10.1016/j.clinbiomech.2008.12.006}}


\bibitem[Schwartz and Rozumalski(2008)]%
        {GDI}
\bibfield{author}{\bibinfo{person}{Michael~H. Schwartz} {and}
  \bibinfo{person}{Adam Rozumalski}.} \bibinfo{year}{2008}\natexlab{}.
\newblock \showarticletitle{The gait deviation index: A new comprehensive index
  of gait pathology}.
\newblock \bibinfo{journal}{\emph{Gait \& Posture}} \bibinfo{volume}{28},
  \bibinfo{number}{3} (\bibinfo{year}{2008}), \bibinfo{pages}{351--357}.
\newblock
\showISSN{0966-6362}
\href{https://doi.org/10.1016/j.gaitpost.2008.05.001}{doi:\nolinkurl{10.1016/j.gaitpost.2008.05.001}}


\bibitem[Simon(2004)]%
        {Quanthumanmotion}
\bibfield{author}{\bibinfo{person}{Sheldon~R. Simon}.}
  \bibinfo{year}{2004}\natexlab{}.
\newblock \showarticletitle{Quantification of human motion: gait
  analysis—benefits and limitations to its application to clinical problems}.
\newblock \bibinfo{journal}{\emph{Journal of Biomechanics}}
  \bibinfo{volume}{37}, \bibinfo{number}{12} (\bibinfo{year}{2004}),
  \bibinfo{pages}{1869--1880}.
\newblock
\showISSN{0021-9290}
\href{https://doi.org/10.1016/j.jbiomech.2004.02.047}{doi:\nolinkurl{10.1016/j.jbiomech.2004.02.047}}


\bibitem[Spallone et~al\mbox{.}(2026)]%
        {spallone2025step}
\bibfield{author}{\bibinfo{person}{Giovanni Spallone}, \bibinfo{person}{Letizia
  Mancini}, \bibinfo{person}{Arianna Carnevale}, \bibinfo{person}{Stefano
  Campi}, \bibinfo{person}{Alessandro de Sire}, \bibinfo{person}{Emiliano
  Schena}, \bibinfo{person}{Pieter D'Hooghe}, \bibinfo{person}{Michael~T
  Hirschmann}, \bibinfo{person}{Rocco Papalia}, {and}
  \bibinfo{person}{Umile~Giuseppe Longo}.} \bibinfo{year}{2026}\natexlab{}.
\newblock \showarticletitle{Step-by-step insight into gait analysis: A
  narrative review unlocking knee biomechanics}.
\newblock \bibinfo{journal}{\emph{Knee Surgery, Sports Traumatology,
  Arthroscopy}} \bibinfo{volume}{34}, \bibinfo{number}{8}
  (\bibinfo{year}{2026}), \bibinfo{pages}{2796--2813}.
\newblock
\href{https://doi.org/10.1002/ksa.70067}{doi:\nolinkurl{10.1002/ksa.70067}}


\bibitem[Team et~al\mbox{.}(2023)]%
        {gemini}
\bibfield{author}{\bibinfo{person}{Gemini Team}, \bibinfo{person}{Rohan Anil},
  \bibinfo{person}{Sebastian Borgeaud}, \bibinfo{person}{Jean-Baptiste
  Alayrac}, \bibinfo{person}{Jiahui Yu}, \bibinfo{person}{Radu Soricut},
  \bibinfo{person}{Johan Schalkwyk}, \bibinfo{person}{Andrew~M Dai},
  \bibinfo{person}{Anja Hauth}, \bibinfo{person}{Katie Millican},
  {et~al\mbox{.}}} \bibinfo{year}{2023}\natexlab{}.
\newblock \showarticletitle{Gemini: a family of highly capable multimodal
  models}.
\newblock \bibinfo{journal}{\emph{arXiv preprint arXiv:2312.11805}}
  (\bibinfo{year}{2023}).
\newblock


\bibitem[Team et~al\mbox{.}(2025)]%
        {Kimi-k2}
\bibfield{author}{\bibinfo{person}{Kimi Team}, \bibinfo{person}{Yifan Bai},
  \bibinfo{person}{Yiping Bao}, \bibinfo{person}{Guanduo Chen},
  \bibinfo{person}{Jiahao Chen}, \bibinfo{person}{Ningxin Chen},
  \bibinfo{person}{Ruijue Chen}, \bibinfo{person}{Yanru Chen},
  \bibinfo{person}{Yuankun Chen}, \bibinfo{person}{Yutian Chen},
  {et~al\mbox{.}}} \bibinfo{year}{2025}\natexlab{}.
\newblock \showarticletitle{Kimi k2: Open agentic intelligence}.
\newblock \bibinfo{journal}{\emph{arXiv preprint arXiv:2507.20534}}
  (\bibinfo{year}{2025}).
\newblock


\bibitem[van~der Krogt et~al\mbox{.}(2022)]%
        {10.3389/fnhum.2022.907565}
\bibfield{author}{\bibinfo{person}{Marjolein~M. van~der Krogt},
  \bibinfo{person}{Han Houdijk}, \bibinfo{person}{Koen Wishaupt},
  \bibinfo{person}{Kim van Hutten}, \bibinfo{person}{Sarah Dekker}, {and}
  \bibinfo{person}{Annemieke~I. Buizer}.} \bibinfo{year}{2022}\natexlab{}.
\newblock \showarticletitle{Development of a core set of gait features and
  their potential underlying impairments to assist gait data interpretation in
  children with cerebral palsy}.
\newblock \bibinfo{journal}{\emph{Frontiers in Human Neuroscience}}
  \bibinfo{volume}{Volume 16 - 2022} (\bibinfo{year}{2022}).
\newblock
\showISSN{1662-5161}
\href{https://doi.org/10.3389/fnhum.2022.907565}{doi:\nolinkurl{10.3389/fnhum.2022.907565}}


\bibitem[Van~Veen et~al\mbox{.}(2024)]%
        {van2024adapted}
\bibfield{author}{\bibinfo{person}{Dave Van~Veen}, \bibinfo{person}{Cara
  Van~Uden}, \bibinfo{person}{Louis Blankemeier}, \bibinfo{person}{Jean-Benoit
  Delbrouck}, \bibinfo{person}{Aakash Aali}, \bibinfo{person}{Christian
  Bluethgen}, \bibinfo{person}{Anuj Pareek}, \bibinfo{person}{Malte Polzin},
  {et~al\mbox{.}}} \bibinfo{year}{2024}\natexlab{}.
\newblock \showarticletitle{Adapted large language models can outperform
  medical experts in clinical text summarization}.
\newblock \bibinfo{journal}{\emph{Nature Medicine}} \bibinfo{volume}{30},
  \bibinfo{number}{4} (\bibinfo{year}{2024}), \bibinfo{pages}{1134--1142}.
\newblock
\href{https://doi.org/10.1038/s41591-024-02855-5}{doi:\nolinkurl{10.1038/s41591-024-02855-5}}


\bibitem[Vatambeti and Damera(2022)]%
        {GaitIdentification1}
\bibfield{author}{\bibinfo{person}{R Vatambeti} {and} \bibinfo{person}{VK
  Damera}.} \bibinfo{year}{2022}\natexlab{}.
\newblock \showarticletitle{Gait based person identification using deep
  learning model of generative adversarial network}.
\newblock \bibinfo{journal}{\emph{Acadlore Trans. Mach. Learn}}
  \bibinfo{volume}{1}, \bibinfo{number}{2} (\bibinfo{year}{2022}),
  \bibinfo{pages}{90--100}.
\newblock


\bibitem[Wang et~al\mbox{.}(2023)]%
        {ClinicalGPT}
\bibfield{author}{\bibinfo{person}{Guangyu Wang}, \bibinfo{person}{Guoxing
  Yang}, \bibinfo{person}{Zongxin Du}, \bibinfo{person}{Longjun Fan}, {and}
  \bibinfo{person}{Xiaohu Li}.} \bibinfo{year}{2023}\natexlab{}.
\newblock \showarticletitle{ClinicalGPT: Large Language Models Finetuned with
  Diverse Medical Data and Comprehensive Evaluation}.
\newblock \bibinfo{journal}{\emph{CoRR}}  \bibinfo{volume}{abs/2306.09968}
  (\bibinfo{year}{2023}).
\newblock
\showeprint[arXiv]{2306.09968}
\href{https://doi.org/10.48550/ARXIV.2306.09968}{doi:\nolinkurl{10.48550/ARXIV.2306.09968}}


\bibitem[Wang et~al\mbox{.}(2025)]%
        {GPT-5}
\bibfield{author}{\bibinfo{person}{Shansong Wang}, \bibinfo{person}{Mingzhe
  Hu}, \bibinfo{person}{Qiang Li}, \bibinfo{person}{Mojtaba Safari}, {and}
  \bibinfo{person}{Xiaofeng Yang}.} \bibinfo{year}{2025}\natexlab{}.
\newblock \showarticletitle{Capabilities of gpt-5 on multimodal medical
  reasoning}.
\newblock \bibinfo{journal}{\emph{arXiv preprint arXiv:2508.08224}}
  (\bibinfo{year}{2025}).
\newblock


\bibitem[Waskito et~al\mbox{.}(2023)]%
        {Otter}
\bibfield{author}{\bibinfo{person}{Steven Waskito}, \bibinfo{person}{Kai~Jie
  Leow}, \bibinfo{person}{Pramuka Medaranga}, \bibinfo{person}{Tejas Gupta},
  \bibinfo{person}{Shantanu Chakrabarty}, \bibinfo{person}{Manoj Gulati}, {and}
  \bibinfo{person}{Ambuj Varshney}.} \bibinfo{year}{2023}\natexlab{}.
\newblock \showarticletitle{Otter: Simplifying Embedded Sensor Data Collection
  and Analysis using Large Language Models}. In
  \bibinfo{booktitle}{\emph{Proceedings of the 29th Annual International
  Conference on Mobile Computing and Networking}} (Madrid, Spain)
  \emph{(\bibinfo{series}{ACM MobiCom '23})}. \bibinfo{publisher}{Association
  for Computing Machinery}, \bibinfo{address}{New York, NY, USA}, Article
  \bibinfo{articleno}{152}, \bibinfo{numpages}{3}~pages.
\newblock
\showISBNx{9781450399906}
\href{https://doi.org/10.1145/3570361.3615759}{doi:\nolinkurl{10.1145/3570361.3615759}}


\bibitem[Wei et~al\mbox{.}(2022)]%
        {CoTPrompt}
\bibfield{author}{\bibinfo{person}{Jason Wei}, \bibinfo{person}{Xuezhi Wang},
  \bibinfo{person}{Dale Schuurmans}, \bibinfo{person}{Maarten Bosma},
  \bibinfo{person}{brian ichter}, \bibinfo{person}{Fei Xia},
  \bibinfo{person}{Ed Chi}, \bibinfo{person}{Quoc~V Le}, {and}
  \bibinfo{person}{Denny Zhou}.} \bibinfo{year}{2022}\natexlab{}.
\newblock \showarticletitle{Chain-of-Thought Prompting Elicits Reasoning in
  Large Language Models}. In \bibinfo{booktitle}{\emph{Advances in Neural
  Information Processing Systems}},
  \bibfield{editor}{\bibinfo{person}{S.~Koyejo}, \bibinfo{person}{S.~Mohamed},
  \bibinfo{person}{A.~Agarwal}, \bibinfo{person}{D.~Belgrave},
  \bibinfo{person}{K.~Cho}, {and} \bibinfo{person}{A.~Oh}} (Eds.),
  Vol.~\bibinfo{volume}{35}. \bibinfo{publisher}{Curran Associates, Inc.},
  \bibinfo{pages}{24824--24837}.
\newblock
\urldef\tempurl%
\url{https://proceedings.neurips.cc/paper_files/paper/2022/file/9d5609613524ecf4f15af0f7b31abca4-Paper-Conference.pdf}
\showURL{%
\tempurl}


\bibitem[Wei et~al\mbox{.}(2025)]%
        {OneModeltoFit}
\bibfield{author}{\bibinfo{person}{Qingxin Wei}, \bibinfo{person}{Jiaming
  Huang}, \bibinfo{person}{Yi Gao}, {and} \bibinfo{person}{Wei Dong}.}
  \bibinfo{year}{2025}\natexlab{}.
\newblock \showarticletitle{One Model to Fit Them All: Universal IMU-based
  Human Activity Recognition with LLM-assisted Cross-dataset Representation}.
\newblock \bibinfo{journal}{\emph{Proc. ACM Interact. Mob. Wearable Ubiquitous
  Technol.}} \bibinfo{volume}{9}, \bibinfo{number}{3}, Article
  \bibinfo{articleno}{139} (\bibinfo{date}{Sept.} \bibinfo{year}{2025}),
  \bibinfo{numpages}{22}~pages.
\newblock
\href{https://doi.org/10.1145/3749509}{doi:\nolinkurl{10.1145/3749509}}


\bibitem[Whittle(1996)]%
        {GaitReview1}
\bibfield{author}{\bibinfo{person}{Michael~W Whittle}.}
  \bibinfo{year}{1996}\natexlab{}.
\newblock \showarticletitle{Clinical gait analysis: A review}.
\newblock \bibinfo{journal}{\emph{Human movement science}}
  \bibinfo{volume}{15}, \bibinfo{number}{3} (\bibinfo{year}{1996}),
  \bibinfo{pages}{369--387}.
\newblock


\bibitem[Wren et~al\mbox{.}(2011)]%
        {Gaitanalysisreview}
\bibfield{author}{\bibinfo{person}{Tishya~A.L. Wren},
  \bibinfo{person}{George~E. Gorton}, \bibinfo{person}{Sylvia Õunpuu}, {and}
  \bibinfo{person}{Carole~A. Tucker}.} \bibinfo{year}{2011}\natexlab{}.
\newblock \showarticletitle{Efficacy of clinical gait analysis: A systematic
  review}.
\newblock \bibinfo{journal}{\emph{Gait \& Posture}} \bibinfo{volume}{34},
  \bibinfo{number}{2} (\bibinfo{year}{2011}), \bibinfo{pages}{149--153}.
\newblock
\showISSN{0966-6362}
\href{https://doi.org/10.1016/j.gaitpost.2011.03.027}{doi:\nolinkurl{10.1016/j.gaitpost.2011.03.027}}


\bibitem[Wu et~al\mbox{.}(2025)]%
        {WirelessSensingGait}
\bibfield{author}{\bibinfo{person}{Patrick Wu}, \bibinfo{person}{Yiwen Dong},
  \bibinfo{person}{Chenhan Xu}, \bibinfo{person}{Mark Geil},
  \bibinfo{person}{Modupe Akintomide}, \bibinfo{person}{Xinyue Zhang}, {and}
  \bibinfo{person}{Zongxing Xie}.} \bibinfo{year}{2025}\natexlab{}.
\newblock \showarticletitle{Wireless Sensing of Gait for Neurodegenerative
  Disease Assessment: A Scoping Review}. In
  \bibinfo{booktitle}{\emph{Proceedings of the 3rd International Workshop on
  Human-Centered Sensing, Modeling, and Intelligent Systems}} (Irvine, CA, USA)
  \emph{(\bibinfo{series}{HumanSys '25})}. \bibinfo{publisher}{Association for
  Computing Machinery}, \bibinfo{address}{New York, NY, USA},
  \bibinfo{pages}{58–63}.
\newblock
\showISBNx{9798400716096}
\href{https://doi.org/10.1145/3722570.3726893}{doi:\nolinkurl{10.1145/3722570.3726893}}


\bibitem[Wu et~al\mbox{.}(2023)]%
        {SelfadaptiveICL}
\bibfield{author}{\bibinfo{person}{Zhiyong Wu}, \bibinfo{person}{Yaoxiang
  Wang}, \bibinfo{person}{Jiacheng Ye}, {and} \bibinfo{person}{Lingpeng Kong}.}
  \bibinfo{year}{2023}\natexlab{}.
\newblock \showarticletitle{Self-Adaptive In-Context Learning: An Information
  Compression Perspective for In-Context Example Selection and Ordering}. In
  \bibinfo{booktitle}{\emph{Proceedings of the 61st Annual Meeting of the
  Association for Computational Linguistics (Volume 1: Long Papers)}},
  \bibfield{editor}{\bibinfo{person}{Anna Rogers}, \bibinfo{person}{Jordan
  Boyd-Graber}, {and} \bibinfo{person}{Naoaki Okazaki}} (Eds.).
  \bibinfo{publisher}{Association for Computational Linguistics},
  \bibinfo{address}{Toronto, Canada}, \bibinfo{pages}{1423--1436}.
\newblock
\href{https://doi.org/10.18653/v1/2023.acl-long.79}{doi:\nolinkurl{10.18653/v1/2023.acl-long.79}}


\bibitem[Xia et~al\mbox{.}(2025)]%
        {MentalHealthAI}
\bibfield{author}{\bibinfo{person}{Wuyue Xia}, \bibinfo{person}{Hanya Shao},
  \bibinfo{person}{Ningxin Kong}, \bibinfo{person}{Yuang Fan}, {and}
  \bibinfo{person}{Jingping Nie}.} \bibinfo{year}{2025}\natexlab{}.
\newblock \showarticletitle{The Convergence of Mental Health and AI: A
  Cross-Disciplinary Survey of Ubiquitous Sensing, LLMs, and Clinical
  Alignment}.
\newblock \bibinfo{journal}{\emph{Authorea Preprints}} (\bibinfo{year}{2025}).
\newblock


\bibitem[Yang et~al\mbox{.}(2025b)]%
        {Qwen3}
\bibfield{author}{\bibinfo{person}{An Yang}, \bibinfo{person}{Anfeng Li},
  \bibinfo{person}{Baosong Yang}, \bibinfo{person}{Beichen Zhang},
  \bibinfo{person}{Binyuan Hui}, \bibinfo{person}{Bo Zheng},
  \bibinfo{person}{Bowen Yu}, \bibinfo{person}{Chang Gao},
  \bibinfo{person}{Chengen Huang}, \bibinfo{person}{Chenxu Lv},
  {et~al\mbox{.}}} \bibinfo{year}{2025}\natexlab{b}.
\newblock \showarticletitle{Qwen3 technical report}.
\newblock \bibinfo{journal}{\emph{arXiv preprint arXiv:2505.09388}}
  (\bibinfo{year}{2025}).
\newblock


\bibitem[Yang et~al\mbox{.}(2025a)]%
        {SocialMind}
\bibfield{author}{\bibinfo{person}{Bufang Yang}, \bibinfo{person}{Yunqi Guo},
  \bibinfo{person}{Lilin Xu}, \bibinfo{person}{Zhenyu Yan},
  \bibinfo{person}{Hongkai Chen}, \bibinfo{person}{Guoliang Xing}, {and}
  \bibinfo{person}{Xiaofan Jiang}.} \bibinfo{year}{2025}\natexlab{a}.
\newblock \showarticletitle{SocialMind: LLM-based Proactive AR Social Assistive
  System with Human-like Perception for In-situ Live Interactions}.
\newblock \bibinfo{journal}{\emph{Proc. ACM Interact. Mob. Wearable Ubiquitous
  Technol.}} \bibinfo{volume}{9}, \bibinfo{number}{1}, Article
  \bibinfo{articleno}{23} (\bibinfo{date}{March} \bibinfo{year}{2025}),
  \bibinfo{numpages}{30}~pages.
\newblock
\href{https://doi.org/10.1145/3712286}{doi:\nolinkurl{10.1145/3712286}}


\bibitem[Yang et~al\mbox{.}(2024)]%
        {DrHouse}
\bibfield{author}{\bibinfo{person}{Bufang Yang}, \bibinfo{person}{Siyang
  Jiang}, \bibinfo{person}{Lilin Xu}, \bibinfo{person}{Kaiwei Liu},
  \bibinfo{person}{Hai Li}, \bibinfo{person}{Guoliang Xing},
  \bibinfo{person}{Hongkai Chen}, \bibinfo{person}{Xiaofan Jiang}, {and}
  \bibinfo{person}{Zhenyu Yan}.} \bibinfo{year}{2024}\natexlab{}.
\newblock \showarticletitle{DrHouse: An LLM-empowered Diagnostic Reasoning
  System through Harnessing Outcomes from Sensor Data and Expert Knowledge}.
\newblock \bibinfo{journal}{\emph{Proc. ACM Interact. Mob. Wearable Ubiquitous
  Technol.}} \bibinfo{volume}{8}, \bibinfo{number}{4}, Article
  \bibinfo{articleno}{153} (\bibinfo{date}{Nov.} \bibinfo{year}{2024}),
  \bibinfo{numpages}{29}~pages.
\newblock
\href{https://doi.org/10.1145/3699765}{doi:\nolinkurl{10.1145/3699765}}


\bibitem[Yang et~al\mbox{.}(2026)]%
        {biomechgpt}
\bibfield{author}{\bibinfo{person}{Ruize Yang}, \bibinfo{person}{Ann Kennedy},
  {and} \bibinfo{person}{R.~James Cotton}.} \bibinfo{year}{2026}\natexlab{}.
\newblock \bibinfo{title}{BiomechGPT: Extending Motion-Language Models to
  Clinical Motion Understanding}.
\newblock
\showeprint[arxiv]{2505.18465}~[cs.CV]
\urldef\tempurl%
\url{https://arxiv.org/abs/2505.18465}
\showURL{%
\tempurl}


\bibitem[Yang et~al\mbox{.}(2025c)]%
        {Yang_2025_CVPR}
\bibfield{author}{\bibinfo{person}{Shaopeng Yang}, \bibinfo{person}{Jilong
  Wang}, \bibinfo{person}{Saihui Hou}, \bibinfo{person}{Xu Liu},
  \bibinfo{person}{Chunshui Cao}, \bibinfo{person}{Liang Wang}, {and}
  \bibinfo{person}{Yongzhen Huang}.} \bibinfo{year}{2025}\natexlab{c}.
\newblock \showarticletitle{Bridging Gait Recognition and Large Language Models
  Sequence Modeling}. In \bibinfo{booktitle}{\emph{Proceedings of the IEEE/CVF
  Conference on Computer Vision and Pattern Recognition (CVPR)}}.
  \bibinfo{pages}{3460--3469}.
\newblock


\bibitem[Yim et~al\mbox{.}(2023)]%
        {MEDCON}
\bibfield{author}{\bibinfo{person}{Wen-wai Yim}, \bibinfo{person}{Yujuan Fu},
  \bibinfo{person}{Asma Ben~Abacha}, \bibinfo{person}{Neal Snider},
  \bibinfo{person}{Thomas Lin}, {and} \bibinfo{person}{Meliha Yetisgen}.}
  \bibinfo{year}{2023}\natexlab{}.
\newblock \showarticletitle{Aci-bench: a novel ambient clinical intelligence
  dataset for benchmarking automatic visit note generation}.
\newblock \bibinfo{journal}{\emph{Scientific data}} \bibinfo{volume}{10},
  \bibinfo{number}{1} (\bibinfo{year}{2023}), \bibinfo{pages}{586}.
\newblock


\bibitem[Zeni et~al\mbox{.}(2008)]%
        {Twosimplemethods}
\bibfield{author}{\bibinfo{person}{J.A. Zeni}, \bibinfo{person}{J.G. Richards},
  {and} \bibinfo{person}{J.S. Higginson}.} \bibinfo{year}{2008}\natexlab{}.
\newblock \showarticletitle{Two simple methods for determining gait events
  during treadmill and overground walking using kinematic data}.
\newblock \bibinfo{journal}{\emph{Gait \& Posture}} \bibinfo{volume}{27},
  \bibinfo{number}{4} (\bibinfo{year}{2008}), \bibinfo{pages}{710--714}.
\newblock
\showISSN{0966-6362}
\href{https://doi.org/10.1016/j.gaitpost.2007.07.007}{doi:\nolinkurl{10.1016/j.gaitpost.2007.07.007}}


\bibitem[Zhang et~al\mbox{.}(2020)]%
        {BERTScore}
\bibfield{author}{\bibinfo{person}{Tianyi Zhang}, \bibinfo{person}{Varsha
  Kishore}, \bibinfo{person}{Felix Wu}, \bibinfo{person}{Kilian~Q. Weinberger},
  {and} \bibinfo{person}{Yoav Artzi}.} \bibinfo{year}{2020}\natexlab{}.
\newblock \showarticletitle{BERTScore: Evaluating Text Generation with {BERT}}.
  In \bibinfo{booktitle}{\emph{8th International Conference on Learning
  Representations, {ICLR} 2020, Addis Ababa, Ethiopia, April 26-30, 2020}}.
  \bibinfo{publisher}{OpenReview.net}.
\newblock
\urldef\tempurl%
\url{https://openreview.net/forum?id=SkeHuCVFDr}
\showURL{%
\tempurl}


\bibitem[Zhang et~al\mbox{.}(2024b)]%
        {LeveragingLLMs}
\bibfield{author}{\bibinfo{person}{Tianyi Zhang}, \bibinfo{person}{Songyan
  Teng}, \bibinfo{person}{Hong Jia}, {and} \bibinfo{person}{Simon D'Alfonso}.}
  \bibinfo{year}{2024}\natexlab{b}.
\newblock \showarticletitle{Leveraging LLMs to Predict Affective States via
  Smartphone Sensor Features}. In \bibinfo{booktitle}{\emph{Companion of the
  2024 on ACM International Joint Conference on Pervasive and Ubiquitous
  Computing}} (Melbourne VIC, Australia) \emph{(\bibinfo{series}{UbiComp
  '24})}. \bibinfo{publisher}{Association for Computing Machinery},
  \bibinfo{address}{New York, NY, USA}, \bibinfo{pages}{709–716}.
\newblock
\showISBNx{9798400710582}
\href{https://doi.org/10.1145/3675094.3678420}{doi:\nolinkurl{10.1145/3675094.3678420}}


\bibitem[Zhang et~al\mbox{.}(2024a)]%
        {DBLP:conf/ijcai/0001C0S24}
\bibfield{author}{\bibinfo{person}{Xiyuan Zhang}, \bibinfo{person}{Ranak~Roy
  Chowdhury}, \bibinfo{person}{Rajesh~K. Gupta}, {and} \bibinfo{person}{Jingbo
  Shang}.} \bibinfo{year}{2024}\natexlab{a}.
\newblock \showarticletitle{Large Language Models for Time Series: {A} Survey}.
  In \bibinfo{booktitle}{\emph{Proceedings of the Thirty-Third International
  Joint Conference on Artificial Intelligence, {IJCAI} 2024, Jeju, South Korea,
  August 3-9, 2024}}. \bibinfo{publisher}{ijcai.org},
  \bibinfo{pages}{8335--8343}.
\newblock
\urldef\tempurl%
\url{https://www.ijcai.org/proceedings/2024/921}
\showURL{%
\tempurl}


\bibitem[Zhang et~al\mbox{.}(2025)]%
        {SensorLM}
\bibfield{author}{\bibinfo{person}{Yuwei Zhang}, \bibinfo{person}{Kumar Ayush},
  \bibinfo{person}{Siyuan Qiao}, \bibinfo{person}{A.~Ali Heydari},
  \bibinfo{person}{Girish Narayanswamy}, \bibinfo{person}{Maxwell~A. Xu},
  \bibinfo{person}{Ahmed~A. Metwally}, \bibinfo{person}{Shawn Xu},
  \bibinfo{person}{Jake Garrison}, \bibinfo{person}{Xuhai Xu},
  \bibinfo{person}{Tim Althoff}, \bibinfo{person}{Yun Liu},
  \bibinfo{person}{Pushmeet Kohli}, \bibinfo{person}{Jiening Zhan},
  \bibinfo{person}{Mark Malhotra}, \bibinfo{person}{Shwetak~N. Patel},
  \bibinfo{person}{Cecilia Mascolo}, \bibinfo{person}{Xin Liu},
  \bibinfo{person}{Daniel McDuff}, {and} \bibinfo{person}{Yuzhe Yang}.}
  \bibinfo{year}{2025}\natexlab{}.
\newblock \showarticletitle{SensorLM: Learning the Language of Wearable
  Sensors}.
\newblock \bibinfo{journal}{\emph{CoRR}}  \bibinfo{volume}{abs/2506.09108}
  (\bibinfo{year}{2025}).
\newblock
\showeprint[arXiv]{2506.09108}
\href{https://doi.org/10.48550/ARXIV.2506.09108}{doi:\nolinkurl{10.48550/ARXIV.2506.09108}}


\bibitem[Zhao et~al\mbox{.}(2024)]%
        {DBLP:conf/naacl/ZhaoM0A24}
\bibfield{author}{\bibinfo{person}{Zheng Zhao}, \bibinfo{person}{Emilio Monti},
  \bibinfo{person}{Jens Lehmann}, {and} \bibinfo{person}{Haytham Assem}.}
  \bibinfo{year}{2024}\natexlab{}.
\newblock \showarticletitle{Enhancing Contextual Understanding in Large
  Language Models through Contrastive Decoding}. In
  \bibinfo{booktitle}{\emph{Proceedings of the 2024 Conference of the North
  American Chapter of the Association for Computational Linguistics: Human
  Language Technologies (Volume 1: Long Papers), {NAACL} 2024, Mexico City,
  Mexico, June 16-21, 2024}}, \bibfield{editor}{\bibinfo{person}{Kevin Duh},
  \bibinfo{person}{Helena G{\'{o}}mez{-}Adorno}, {and} \bibinfo{person}{Steven
  Bethard}} (Eds.). \bibinfo{publisher}{Association for Computational
  Linguistics}, \bibinfo{pages}{4225--4237}.
\newblock
\href{https://doi.org/10.18653/V1/2024.NAACL-LONG.237}{doi:\nolinkurl{10.18653/V1/2024.NAACL-LONG.237}}


\bibitem[Zhou et~al\mbox{.}(2024)]%
        {NEURIPS2024_e41efb03}
\bibfield{author}{\bibinfo{person}{Pei Zhou}, \bibinfo{person}{Jay Pujara},
  \bibinfo{person}{Xiang Ren}, \bibinfo{person}{Xinyun Chen},
  \bibinfo{person}{Heng-Tze Cheng}, \bibinfo{person}{Quoc~V. Le},
  \bibinfo{person}{Ed H.}, \bibinfo{person}{Denny Zhou},
  \bibinfo{person}{Swaroop Mishra}, {and} \bibinfo{person}{Huaixiu~Steven
  Zheng}.} \bibinfo{year}{2024}\natexlab{}.
\newblock \showarticletitle{SELF-DISCOVER: Large Language Models Self-Compose
  Reasoning Structures}. In \bibinfo{booktitle}{\emph{Advances in Neural
  Information Processing Systems}},
  \bibfield{editor}{\bibinfo{person}{A.~Globerson},
  \bibinfo{person}{L.~Mackey}, \bibinfo{person}{D.~Belgrave},
  \bibinfo{person}{A.~Fan}, \bibinfo{person}{U.~Paquet},
  \bibinfo{person}{J.~Tomczak}, {and} \bibinfo{person}{C.~Zhang}} (Eds.),
  Vol.~\bibinfo{volume}{37}. \bibinfo{publisher}{Curran Associates, Inc.},
  \bibinfo{pages}{126032--126058}.
\newblock
\href{https://doi.org/10.52202/079017-4004}{doi:\nolinkurl{10.52202/079017-4004}}


\end{thebibliography}

\appendix

\section{Gait Term Library} \label{appendix:gaitterms}

The proposed Gait Term Library in \textit{MEDCON-Gait} consists of domain-specific terms covering anatomical structures, motion attributes, gait phases/events, clinical status descriptors, and commonly used abbreviations.

\noindent\textbf{Anatomical Structures:} \\
Foot, Ankle, Knee, Hip, Pelvis, Trunk, Arm, Shoulder, Upper extremity, Lower extremity, Leg, Thigh, Toe, Hindfoot, Forefoot.

\noindent\textbf{Motion Attributes:} \\
Flexion, Extension, Dorsiflexion, Plantarflexion, Abduction, Adduction, Rotation, Internal rotation, External rotation, Valgus, Varus, Tilt, Obliquity, Circumduction, Moment, Power, Velocity, Angle, Alignment, Rate.

\noindent\textbf{Gait Phases and Events:} \\
Initial Contact, Loading Response, Mid Stance, Terminal Stance, Pre-Swing, Swing, Stance, Single Limb Support, Double Limb Support, Heel Strike, Toe Off.

\noindent\textbf{Clinical Status Descriptors:} \\
Increased, Decreased, Reduced, Minimal, Frequent, Occasional, Absent, Normal, Improved, Posterior, Anterior, Ipsilateral, Contralateral, Level, Downward, Upward, Neutral.

\noindent\textbf{Abbreviations:} \\
WNL, DF, PF, IR, ER, FPA, GDI, HS, SLS, IC.

\section{CoT Prompt Template for Baseline LLMs} \label{appendix:CoT}
\begin{figure}[htbp]
    \centering
    \includegraphics[width=1.0\linewidth]{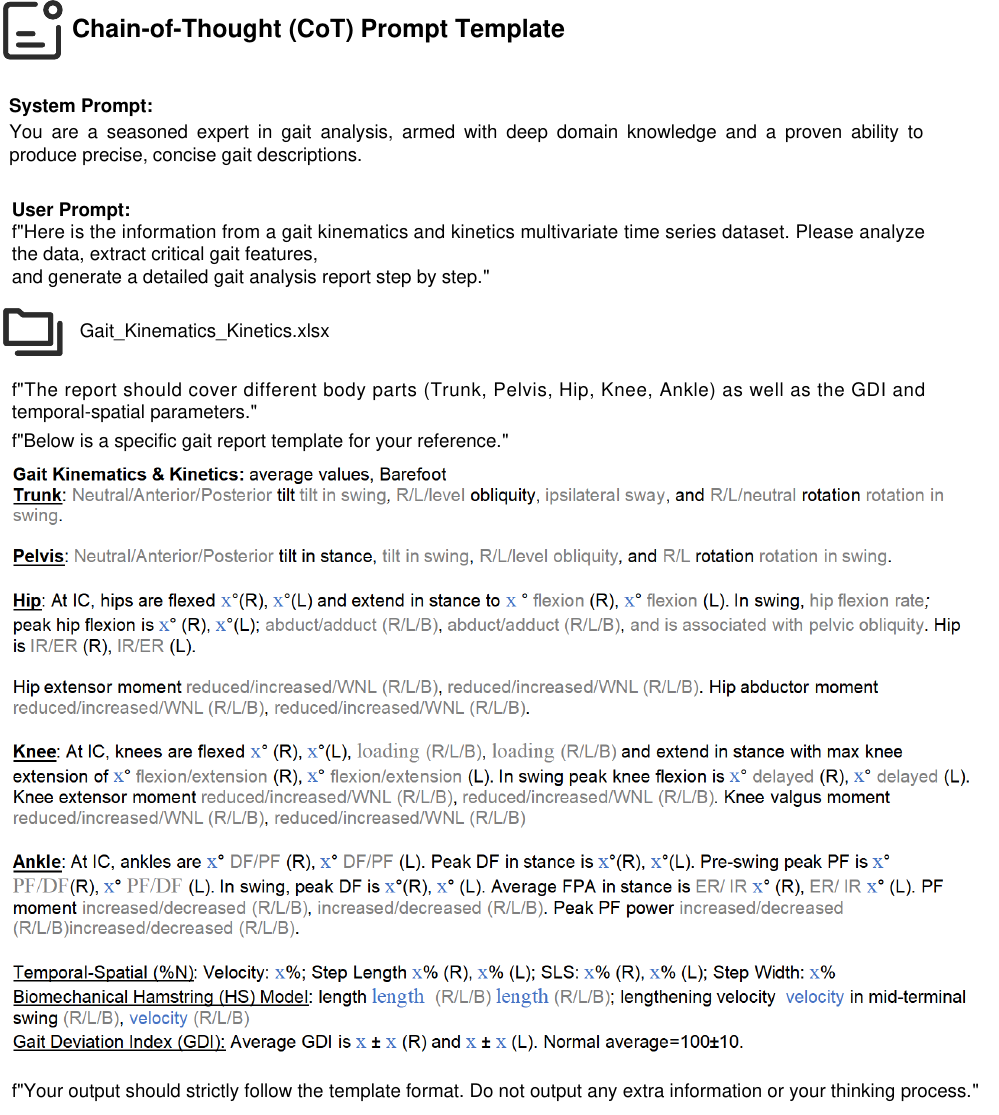}
    \caption{CoT Prompt Template for Baseline LLMs.}
    \label{Fig:Appendix1}
\end{figure}

We provide the CoT prompt template used in baseline LLMs, as shown in Fig. ~\ref{Fig:Appendix1}.

\end{document}